\documentclass[useAMS,usenatbib,usedcolumn]{mn2e}
\usepackage[figuresright]{rotating}
\usepackage{lscape}
\usepackage{graphics}
\usepackage{epsfig}
\usepackage{multirow}
\usepackage{bigdelim}
\usepackage{bigstrut}
\usepackage{amsmath}
\usepackage{amssymb}
\usepackage{lscape}
\usepackage{supertabular}
\usepackage{times}
\usepackage[T1]{fontenc}
\usepackage{aecompl}
\usepackage{subcaption}
\usepackage{afterpage}
\DeclareMathOperator{\sgn}{sgn}
\DeclareMathOperator{\sech}{sech}
\title[Weighing the disk: simulations]{Weighing the Galactic disk using the Jeans equation:\\ lessons from simulations}
\author[G.~N. Candlish, R. Smith, C. Moni Bidin, B.K. Gibson]{G.~N. Candlish$^{1,2}$\thanks{E-mail: gcandlish@astro-udec.cl}, R. Smith$^{3}$, C. Moni Bidin$^{4}$, B.K. Gibson$^{5}$\\
$^{1}$Departamento de Astronom\'ia, Universidad de Chile, Casilla 36-D, Santiago, Chile\\
$^{2}$Departamento de Astronom\'ia, Universidad de Concepci\'on, Concepci\'on, Chile\\
$^{3}$Department of Astronomy, Yonsei University, Seoul 120-749, Korea\\
$^{4}$Instituto de Astronom\'ia, Universidad Cat\'olica del Norte, Av. Angamos 0610, Antofagasta, Chile\\
$^{5}$E.A. Milne Centre for Astrophysics, Dept. of Physics and Mathematics, University of Hull, Hull HU6 7RX, UK}
\begin{document}

\maketitle

\begin{abstract}
Using three-dimensional stellar kinematic data from simulated galaxies, we examine the efficacy of a Jeans equation analysis in reconstructing the total disk surface density, including the dark matter, at the ``Solar'' radius. Our simulation dataset includes galaxies formed in a cosmological context using state-of-the-art high resolution cosmological zoom simulations, and other idealised models. The cosmologically formed galaxies have been demonstrated to lie on many of the observed scaling relations for late-type spirals, and thus offer an interesting surrogate for real galaxies with the obvious advantage that all the kinematical data are known perfectly. We show that the vertical velocity dispersion is typically the dominant kinematic quantity in the analysis, and that the traditional method of using only the vertical force is reasonably effective at low heights above the disk plane. At higher heights the inclusion of the radial force becomes increasingly important. We also show that the method is sensitive to uncertainties in the measured disk parameters, particularly the scale lengths of the assumed double exponential density distribution, and the scale length of the radial velocity dispersion. In addition, we show that disk structure and low number statistics can lead to significant errors in the calculated surface densities. Finally we examine the implications of our results for previous studies of this sort, suggesting that more accurate measurements of the scale lengths may help reconcile conflicting estimates of the local dark matter density in the literature.
\end{abstract}


\section{Introduction}
\label{intro}
Attempting to determine the density of the material near the plane of the Galactic disk in the solar neighbourhood is an old endeavour, dating back to the pioneering work of \cite{kapteyn} and \cite{oort}, with many subsequent studies over the years (e.g. \citealp{kuijken,creze,holmberg,korchagin,dejong,salucci,nesti}). The most direct method that may be employed is one-dimensional (i.e. it ignores the radial force) with the vertical velocity dispersion used in the Jeans equation to determine the vertical force acting on the stars. This is typically justified by arguing for the dominance of these terms over all the other terms, including the so-called `tilt' term in the vertical force expression, which depends on $\overline{UW}$ (the mean product of the radial and vertical velocities). Then, using the Poisson equation, one is able to recover the matter density.

These kinds of studies are restricted to near the plane of the Galaxy, i.e. the thin disk region, and have been undertaken using ever more complete datasets with varying estimates for the dark matter density in the Solar neighbourhood. As the stellar surface density near the Galactic plane may be ascertained by direct observations, any mass discrepancy with the dynamical estimate using the tracer stellar populations is attributed to the presence of dark matter. While there is considerable uncertainty, both from observational errors, as well as difficulties in separating the dynamically old stellar populations, generally the estimates\footnote{For comparison the estimated total volume density within $\pm 50$~pc is approximately $0.1-0.11 M_{\odot} \text{pc}^{-3}$, using old red giant stars from the \emph{Hipparcos} catalog in the disk \citep{korchagin}.} are on the order of $\rho_{\text{DM}} \sim 0.01 M_{\odot} \text{pc}^{-3}$. Such measurements of the Galactic disk surface density therefore offer a means of constraining the properties of dark matter. For a very recent discussion of observational difficulties related to these kinds of studies, see \cite{hessman}.

As the data available for the stellar dynamics in the Milky Way continue to improve, it is interesting to contemplate going beyond the thin disk, inferring both vertical and radial forces from the Jeans equation and using the full Poisson equation, without restricting to the vertical dynamics alone. A recent study along these lines was undertaken by \cite{mb12} (hereafter referred to as MB12), where the dark matter density was determined to be significantly lower than previous estimates ($\rho_{\text{DM}} \approx 0-1 mM_{\odot} \text{pc}^{-3}$), with the caveat that multiple approximations and assumptions were required in the analysis. Subsequent to this, the study by \cite{bt12} (hereafter referred to as BT12) proposed that this result could be explained by modifying one of the assumptions made by MB12 concerning the dependence of the average azimuthal velocity on the radius. In BT12 the assumption of $\partial_R \overline{V} = 0$, as used in MB12, was discarded as unphysical, and the radial force term was assumed negligible, leading to a lower limit on the local dark matter density from a purely vertical analysis. This apparently restored $\rho_{\text{DM}}$ to a value similar to those previously reported in the literature.

The analysis of BT12 has, however, been shown to be incorrect in \cite{mb15} (hereafter referred to as MB15), as a vanishing radial force term (using the MB12 data) requires that the vertical gradient of $\partial_R \overline{V}$ is too steep, disagreeing with observations both of the Milky Way and of external galaxies. Furthermore, although the assumption about $\overline{V}$ used in MB12 is indeed unphysical, relaxing this assumption and using the sparse observational data available in the analysis results in a negligible change in the calculated dark matter density, as demonstrated in MB15. Hence, further investigation is needed to determine the reason for the unusual results of MB12.

In this work we will use the full three-dimensional kinematic data from simulations in an attempt to demonstrate the efficacy or otherwise of this approach, to analyse which parameters are most crucial to the analysis, and to provide some insight into possible reasons for the results of MB12. A study investigating the one-dimensional method using simulations was undertaken in \cite{garbari}, using a thin disk galaxy model. Here we will be concerned with the full three-dimensional formulation, considering model data of galaxies with thick disks. We will also utilise state-of-the-art simulations in which disk galaxies are formed in a cosmological context, including many associated physical processes, both environmental and secular, rather than limiting ourselves only to idealised simulations.

The structure of the paper is as follows: in Section~\ref{JeansEquation} we derive the form of the Jeans equation that we will use, carefully stating the assumptions used to simplify the terms; in Section~\ref{data} we will summarise the simulated data that we use for this work and how we measure the various quantities required for the Jeans equation; in Section~\ref{results} we show the surface densities calculated from the kinematics of our various simulation datasets to test how well the method works in an ideal case of perfect information and investigate which parameters are most crucial to an accurate measurement, as well as discussing the potential impact of our results on previous observational studies of this sort; and finally we conclude in Section~\ref{conclusions}.

\section{The Jeans equation}
\label{JeansEquation}
The Jeans equations, in cylindrical coordinates $(R,z,\phi)$, for a tracer stellar population of an axisymmetric system in a steady-state are:
\begin{equation}
\label{jeansforces}
\begin{split}
F_z &= \frac{1}{R\rho_*} \frac{\partial (R\rho_* \overline{UW} )}{\partial R} + \frac{1}{\rho_*} \frac{ \partial (\rho_* \overline{W^2} )}{\partial z}, \\
F_R &= \frac{1}{\rho_*} \frac{\partial (\rho_* \overline{U^2} )}{\partial R} + \frac{1}{\rho_*} \frac{ \partial (\rho_* \overline{UW})}{\partial z} + \frac{\overline{U^2} - \overline{V^2}}{R}
\end{split}
\end{equation}
where $F_R = -\partial_R \Phi, F_z = -\partial_z \Phi$ and $\Phi$ is the \emph{total} gravitational potential, $\rho_*$ is the stellar density distribution, and $U, V$ and $W$ are the radial, azimuthal and vertical velocities respectively. All these kinematic quantities refer to the kinematics of the test stellar sample used to probe the Galactic potential. The surface density $\Sigma$ is then calculated from the kinematics by using these expressions for the forces in the Poisson equation, integrated in the vertical direction:
\begin{equation}
\label{poissonvertical}
\int_{0}^{+z} -\frac{2}{R} \frac{\partial }{\partial R} (RF_R) dz - 2F_z \Big|_{0}^{+z} = 4\pi G \Sigma(z),
\end{equation}
where $\Sigma(z)$ is the surface density as a function of height. As all terms in the expression for the forces are assumed symmetric in $z$ (which makes physical sense for a disk galaxy) we integrate over half the symmetric interval of integration and multiply by 2 all terms on the left hand side. Note that we will evaluate these terms at the radius $R=8.2$~kpc, roughly corresponding to the solar Galacto-centric radius. In what follows, we will refer to a directly ``measured'' surface density in a simulation as the ``true'' surface density, denoted $\Sigma_{\text{true}}(z)$ and the surface density calculated \emph{via} the Jeans and Poisson equations as $\Sigma^C(z)$.

\subsection{Simplifying the equation}

\subsubsection{The density}
At this stage we begin to make some assumptions regarding the form of the terms in the Jeans equation, in a similar manner to those made in MB12. Note that, in general, it is only the test tracer population that need satisfy the following assumptions, not the disk as a whole. The disks of our simulated galaxies, however, have only a single component, thus we apply these assumptions to the whole disk.\\

\noindent
\textbf{Assumption 1} - the density has an exponential drop-off in both the vertical and radial directions: $\rho(R,z) \sim \exp(-R/h_R) \exp(-|z|/h_z)$. We furthermore assume that $h_R$ and $h_z$ are constants, i.e. there is no dependence on height or radius for these values.\\

\noindent
Checking the density profiles of the simulation data directly this assumption is seen to be generally valid, with the caveat that the radial scale length $h_R$ (and that of the radial velocity dispersion $h_U$) typically increases somewhat with height, violating the second part of assumption 1. We will limit ourselves to the kind of analysis utilised in MB12 and so we choose to measure the radial density profile at a single height given by the vertical bin closest to the plane of the disk.

Given this assumption for the density, we can rewrite Eq.~\ref{jeansforces} as:
\begin{equation}
\label{jeansforces_vertical}
F_z = \left( \frac{1}{R} - \frac{1}{h_R} \right) \overline{UW} + \frac{\partial \overline{UW}}{\partial R} - \frac{1}{h_z} \overline{W^2} + \frac{ \partial \overline{W^2} }{\partial z}
\end{equation}
for the vertical force and
\begin{equation}
\label{jeansforces_radial}
\begin{split}
\frac{1}{R} \frac{\partial }{\partial R} (RF_R) &= -\frac{1}{Rh_R} \overline{U^2} + \left( \frac{2}{R} - \frac{1}{h_R} \right) \frac{ \partial \overline{U^2}}{\partial R}  + \frac{\partial^2 \overline{U^2}}{\partial R^2} \\
&- \frac{1}{Rh_z} \overline{UW} - \frac{1}{h_z} \frac{ \partial \overline{UW} }{\partial R} + \frac{1}{R} \frac{ \partial \overline{UW} }{\partial z} + \frac{ \partial^2 \overline{UW} }{\partial z \partial R} \\
&- \frac{1}{R} \frac{\partial \overline{V^2}}{\partial R},
\end{split}
\end{equation}
where we have written the left hand side as it appears in Eq.~\ref{poissonvertical}, the vertical integration of the Poisson equation. This integral over $z$ means we need not calculate the derivatives with respect to $z$ appearing in Eq.~\ref{jeansforces_radial}, but the $z$ derivative in the fourth term of Eq. \ref{jeansforces_vertical} remains. These expressions are strictly only valid for positive values of $z$. The more general expression, valid for positive and negative $z$, would include occurrences of the $\sgn(z)$ function which arises due to taking a derivative of $|z|$ in $\rho(R,z)$.

Note that in MB12 the authors incorrectly assumed that the fourth and fifth terms on the right hand side of Eq. \ref{jeansforces_radial} would vanish from the integration over $z$ due to the antisymmetry of $\overline{UW}$. To see that this is not true, we can restore the $\sgn$ function to those terms:
\begin{equation}
\label{restoredterms}
\frac{1}{R}\frac{\partial}{\partial R} (R F_R) = \ldots - \frac{1}{Rh_z} \overline{UW} \sgn(z) - \frac{1}{h_z} \frac{\partial \overline{UW}}{\partial R} \sgn(z) \ldots
\end{equation}
The antisymmetry in $z$ of $\overline{UW}$ (and $\partial_R \overline{UW}$) combined with the antisymmetry of $\sgn(z)$ ensure that these terms are symmetric and thus do not vanish upon integration over a symmetric interval in $z$. This is also to be expected based on the symmetry in $z$ of the radial force, as we now demonstrate. The second term in the expression for $F_R$ in Eq. \ref{jeansforces} gives rise to the two terms highlighted in Eq. \ref{restoredterms}. The symmetry of $\rho_*(R,z)$ in $z$ due to the dependence on $|z|$ combined with the antisymmetry of $\overline{UW}(z)$ means the function within the derivative is antisymmetric in $z$. When differentiated with respect to $z$ this function becomes symmetric in $z$, and thus this whole term is symmetric in $z$ as anticipated for the radial force.

\subsubsection{The velocity dispersions}
The analysis in MB12 used the radial and vertical velocity \emph{dispersions} rather than the mean squared radial velocities. The two are of course related by
\begin{equation}
\sigma_U^2 = \overline{U^2} - \overline{U}^2, \quad \sigma_W^2 = \overline{W^2} - \overline{W}^2.
\end{equation}
Both $\overline{U}$ and $\overline{W}$ are assumed in MB12 to be zero, as expected for a system in a steady-state equilibrium, and so $\sigma_U^2 = \overline{U^2}$ and $\sigma_W^2 = \overline{W^2}$. We can then replace all occurrences of the mean squared radial and vertical velocities with the velocity dispersion. The simulation data allow us to work directly with the mean squared velocities, although we have checked that, in all simulations, the assumption of zero mean radial and vertical motions is a very accurate one, and we could just as easily have worked with the dispersions instead. For the azimuthal velocities we will replace $\overline{V^2}$ in Eq. \ref{jeansforces_radial} using
\begin{equation}
\label{aziveldisp}
\overline{V^2} = \sigma_V^2 + \overline{V}^2.
\end{equation}

Our second and third assumptions, again verified by checking with the simulations and shown to be in excellent agreement with the data, are\\

\noindent
\textbf{Assumption 2} - the squared mean radial velocity (or the squared radial velocity dispersion) is assumed to have an exponential drop-off with radius: $\overline{U^2} \sim \sigma_U^2 \sim \exp(-R/h_U)$.\\

\noindent
\textbf{Assumption 3} - the squared azimuthal velocity dispersion is assumed to have an exponential drop-off with radius: $\sigma_V^2 \sim \exp(-R/h_V)$.\\

\noindent
Note that we do \emph{not} restrict either $h_V$ or $h_U$ to be equal to $h_R$, as done in MB12. We will see later that imposing $h_U = h_R$ can lead to significant problems for some parameter sets, potentially resulting in a large negative $\Sigma^C(z)$ at large heights, an unphysical result.

With assumptions 2 and 3, we may now write the terms contributing to the surface density as follows. We start with the terms that arise from $F_z$:
\begin{equation}
\label{verticalforceterm}
\begin{split}
\Sigma_{F_z} &= -\frac{1}{4\pi G} \left\{ \left( \frac{2}{R} - \frac{2}{h_R} \right) \overline{UW} \Big|_0^z + 2\frac{\partial \overline{UW}}{\partial R} \Big|_0^z - \frac{2}{h_z} \overline{W^2} \Big|_0^z \right. \\
&\left.+ 2\frac{\partial \overline{W^2}}{\partial z}\Big|_0^z \right\}
\end{split}
\end{equation}
while the terms coming from $F_R$ give:
\begin{equation}
\label{radialforceterm}
\begin{split}
\Sigma_{F_R} &= -\frac{1}{4\pi G} \left\{ \frac{2}{R} \overline{UW} \Big|_0^z + 2 \frac{\partial \overline{UW}}{\partial R} \Big|_0^z \right. \\
& \left. + \left(\frac{2}{h_Rh_U} - \frac{2}{Rh_R} - \frac{4}{Rh_U} + \frac{2}{h_U^2} \right) \int_0^z \overline{U^2} dz' \right. \\
& \left. - \frac{2}{Rh_z} \int_0^z \overline{UW} dz' - \frac{2}{h_z} \int_0^z \frac{\partial \overline{UW}}{\partial R} dz' \right. \\
& \left. + \frac{2}{Rh_V} \int_0^z \sigma_V^2 dz' - \frac{2}{R} \int_0^z \frac{\partial \overline{V}^2}{\partial R} dz' \right\},
\end{split}
\end{equation}
where we have used Eq. \ref{aziveldisp}.

\subsection{Summary of the equation}
\label{listofterms}
At this stage it is useful to enumerate the terms in the equation for the surface density. We will refer to these numbers when we compare with plots of the simulation data. The total list of terms, where we have made our three additional assumptions beyond reflection symmetry in the $z$-axis (exponential drop-off with radius and height for $\rho$, and exponential drop-off with radius for $\overline{U^2}$ and $\sigma_V^2$), is given below:
\begin{equation*}
\label{jeansfinal}
\begin{split}
\text{(i)}& \quad \left(  \frac{1}{h_R} - \frac{2}{R} \right) \overline{UW} \Big|_0^z \quad \quad \text{(ii)} \quad \frac{1}{Rh_z} \int_0^z \overline{UW} dz'\\
\text{(iii)}& \quad -2\frac{\partial \overline{UW}}{\partial R} \Big|_0^z \quad \quad \text{(iv)} \quad \frac{1}{h_z} \int_0^z \frac{ \partial \overline{UW}}{\partial R} dz'\\
\text{(v)}& \quad \left(\frac{2}{Rh_U} + \frac{1}{Rh_R} - \frac{1}{h_Uh_R} - \frac{1}{h_U^2} \right) \int_0^z \overline{U^2} dz'\\
\text{(vi)}& \quad -\frac{1}{Rh_V} \int_0^z \sigma_V^2 dz' \quad \quad \text{(vii)} \quad \frac{1}{R} \int_0^z \frac{ \partial \overline{V}^2}{\partial R} dz'\\
\text{(viii)}& \quad \frac{1}{h_z} \overline{W^2} \Big|_0^z \quad \quad \text{(ix)} \quad -\frac{\partial \overline{W^2}}{\partial z} \Big|_0^z
\end{split}
\end{equation*}
Summing all these terms, and multiplying the result by $1/2\pi G$, gives the surface density $\Sigma^C(z)$ at radius $R$. As discussed earlier terms (ii) and (iv) were incorrectly omitted from the analysis of MB12.

\subsection{Analytic exponential disk models}
\label{analytic}
In simplifying the Jeans equation we have assumed that the density distribution of the stellar disk is that of a double exponential profile:
\begin{equation}
\label{analyticexp}
\rho_d(R,z) = \frac{\Sigma_0}{2h_z} \exp(-R/h_R) \exp(-|z|/h_z),
\end{equation}
where $\Sigma_0$ is the central surface density and $h_R$ and $h_z$ are the scale length and scale height of the disk. To give us some guidance in the expected behaviour of the Jeans equation terms, we will now calculate the radial and vertical forces for the density distribution of Eq. \ref{analyticexp}, using the Poisson solver of the \textsc{RAMSES} N-body/hydrodynamics code \citep{teyssier}. Thus we are simply numerically solving the Poisson equation using a specified analytic density distribution. Note that MB15 provides examples of analytic Miyamoto-Nagai disk models and an NFW halo, and considers the contribution of the radial force term in these cases. Our choice of an exponential disk allows a more direct comparison with the analysis undertaken in this study.

To demonstrate the effect of a dark matter halo on the vertical and radial forces, we also add a logarithmic halo density distribution:
\begin{equation}
\rho_h = \frac{v_0^2}{4\pi Gq^2} \frac{(2q^2 + 1)R_c^2 + R^2 + (2-q^{-2})z^2}{(R_c^2 + R^2 + z^2q^{-2})^2}
\end{equation}
where $R_c = 12$~kpc, $v_0 = 200$~km/s and $q$ is the halo flattening, with $q=1$ for a spherical halo, and $q=0.7$ for a very flattened halo. After numerically determining the vertical and radial forces, we can then use these directly in Eq. \ref{poissonvertical} to recover the surface density at a specific radius $R$. Doing this at $R=8$~kpc, we find the dependence on height $z$ for the radial and vertical terms in Eq. \ref{poissonvertical} for various choices of disk parameters. We choose stellar disks with scale lengths of $1.25$, $2$, $3.8$ and $5$~kpc. We also add a $q=1$ dark matter halo to the disks with scale lengths $2$ and $3.8$~kpc, and a $q=0.7$ dark matter halo to the disk with $R_d = 3.8$~kpc, for a total of 7 models. All disks have $h_z = 0.9$~kpc, i.e. they are all thick disk models.

\begin{figure*}
\centering
\begin{tabular}{cc}
\includegraphics[width=7.0cm]{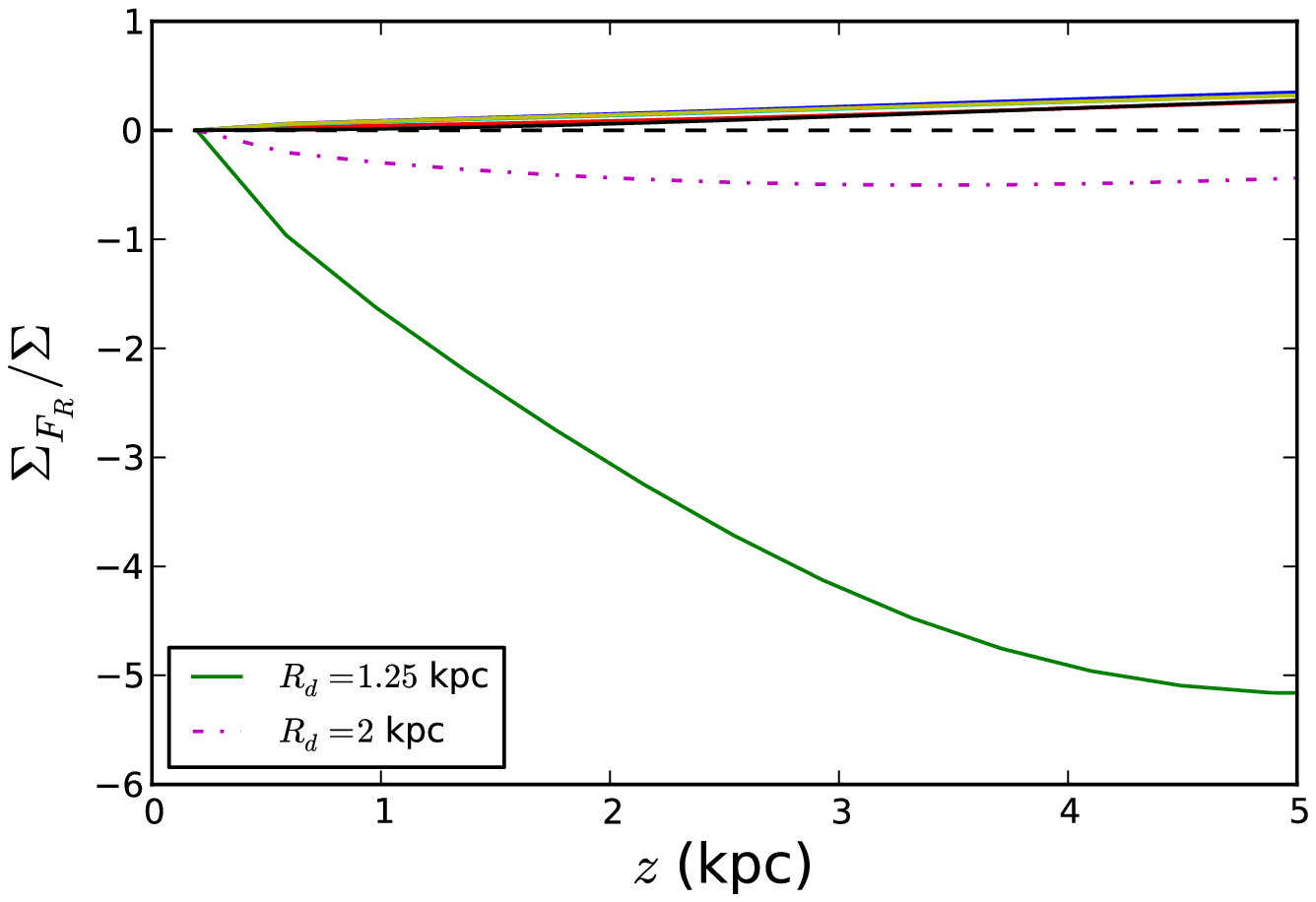} & \includegraphics[width=7.0cm]{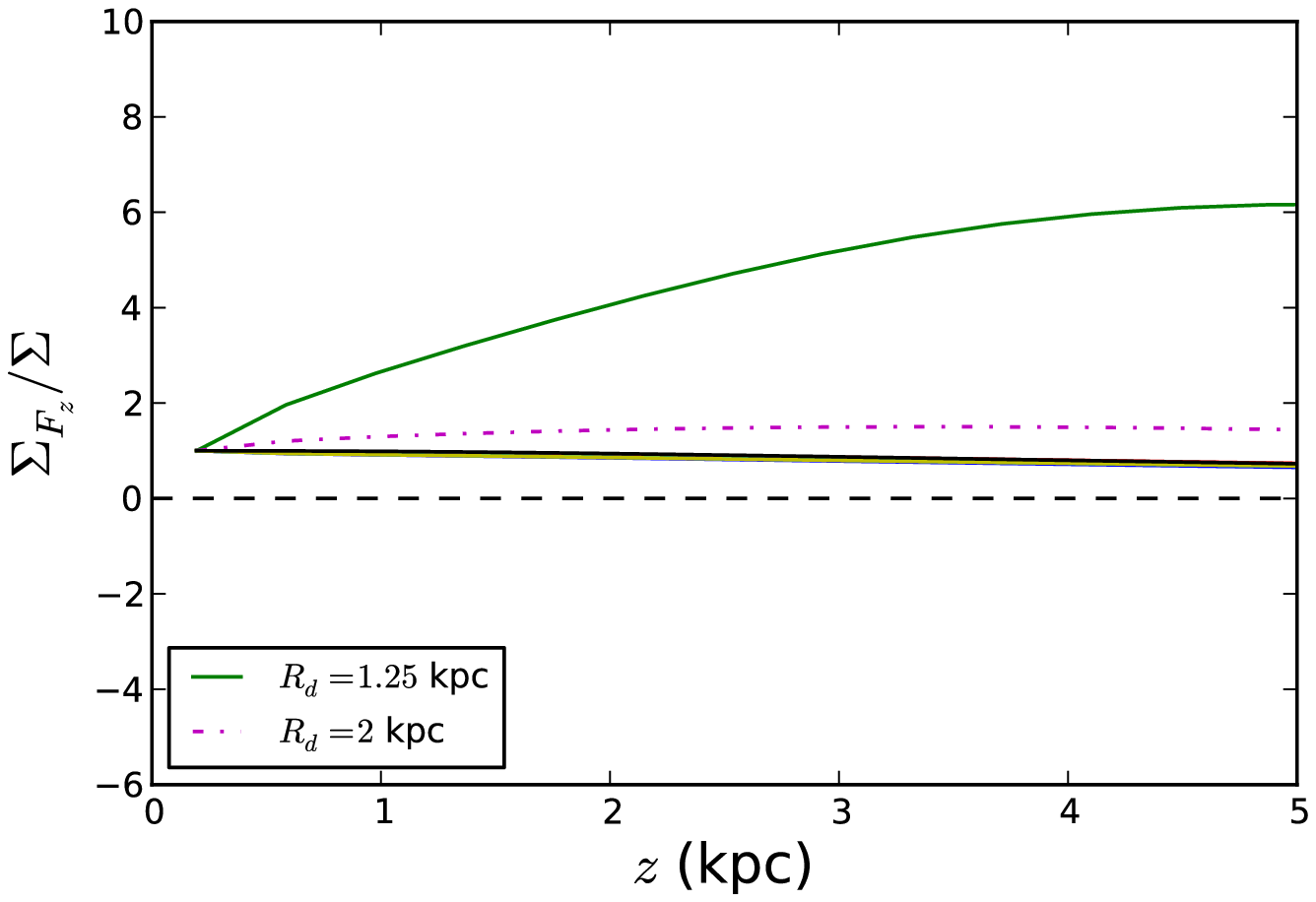}
\end{tabular}
\caption{The left panel shows the radial force term of Eq. \ref{poissonvertical} divided by the total surface density, and the right panel shows the vertical force term of Eq. \ref{poissonvertical} divided by the total surface density.}
\label{expdisk_analytic}
\end{figure*}

In the left panel of Fig. \ref{expdisk_analytic} we show the radial force term of Eq. \ref{poissonvertical} divided by the total surface density in order to assess their relative importance. Clearly the scale length controls the size of the contribution from the radial term, which may be considerable and is also not guaranteed to be positive. In fact, we can only distinguish the lines associated to the models with very short scale lengths, i.e. $R_d = 1.25$~kpc and $R_d = 2$~kpc. All other models are essentially indistinguishable in these plots. The ratio of the vertical force term to the total surface density is shown in the right panel of Fig. \ref{expdisk_analytic}. Again, only for the small scale length disks (small relative to the scale height) can we distinguish different behaviour. Although the vertical term is always a larger contribution to the calculated surface density than the radial term, we can see that for very short scale length disks the radial term is important, especially at higher heights. For larger scale lengths the radial term becomes increasingly positive, again contributing more at higher heights, but it is always dominated by the contribution of the vertical term. The inclusion of a dark matter halo similarly increases both radial and vertical terms (although this is not evident in Fig.~\ref{expdisk_analytic} as the change is too small) with more weighting on the vertical term for the case of the flattened $q=0.7$ halo. Note that a logarithmic halo always increases both the radial and vertical contributions to the surface density, regardless of the parameter choices.

MB15 reached similar conclusions when looking at the force terms for an analytic Miyamoto-Nagai disk: for a longer scale length disk the contribution of the radial term becomes increasingly positive. The inclusion of an NFW dark matter halo also leads to an increased positive contribution from the radial term.

We thus see that, if the total stellar distribution is well-described by an exponential disk, only for a short scale length disk with negligible dark matter\footnote{One could choose a dark matter halo distribution such that the circular velocity decreases with radius, accentuating the negative contribution to the surface density coming from the radial force term in Eq. \ref{poissonvertical}. Such a dark matter configuration is, of course, at odds with expectations from observations and simulations.} at $R=8$~kpc do we expect to see a negative contribution from the radial force term. Furthermore, only in this case do the radial and vertical terms become comparable: all other models show that the radial term is subdominant, except at large heights above the disk. 

The stellar disks of our simulation datasets are all well described by a single-component exponential disk mass distribution (except for the dark matter-free model). Thus these results provide us with some intuition regarding our Jeans equation analysis applied to the simulations, independent of the details of the stellar kinematics. It is important to remember, however, that when applied to our Galaxy, only the tracer stellar distribution need satisfy the double exponential distribution, not the whole disk. As such, our analytic results here do not imply that the radial force term for measurements at the solar radius in our Galaxy must be negligible or positive.

\section{Simulation data}
\label{data}
Now we will investigate the behaviour of the terms in the Jeans equation using a total of eight simulated disk galaxies, four of which were recently utilised in \cite{gibson2013}. Two of these are based on g1536 and g15784 from the McMaster Unbiased Galaxy Simulations (MUGS) project \citep{stinson2010}, using the star formation and feedback recipes described in \cite{stinson2010} and \cite{pilkington2012a}, while the other two use the same merger histories as the MUGS galaxies, but instead use the star formation and feedback recipes of the Making Galaxies In a Cosmological Context (MaGICC) project \citep{brook2011,brook2012a,brook2012b,brook2012c,pilkington2012b,stinson2012,stinson2013}. We also use one simulated galaxy from the RAMSES Disk Environment Survey (RaDES) \citep{few2012}, for a total of five cosmologically-formed disk galaxy models.

The last three simulated disks are idealised models. These are the models GD1 and GD2, drawn from the suite described by \cite{huntkawatamartel}, each realised with the smoothed particle hydrodynamics code GCD+ \citep{kawatagibson2003}\footnote{http://astrowiki.ph.surrey.ac.uk/dokuwiki/doku.php} and one more simulation performed for this study using the \textsc{RAMSES} adaptive mesh refinement code.

\subsection{Cosmological simulations}
Our simulated galaxy sample from the cosmological simulations are those referred to as MUGS g1536, MUGS g15784, MaGICC g1536, MaGICC g15784 and the ``Selene'' simulation from RaDES. The stellar disks of all these galaxies are single thick disks. The only differences between the MUGS and MaGICC galaxies are the star formation and feedback recipes employed. The main galaxy properties are summarised in Table~\ref{galaxysample}.

\begin{table*}
\begin{center}
\begin{tabular}{ c | c | c | c | c }
  \hline
  Model & $M_*$ & $M_g$ & $M_\text{DM}$ & $h_z$ (kpc) \\
  \hline
  Cosmological & & & & \\ 
  \hline                      
  MUGS g1536    & $5.5 \times 10^{10}$ & $9.7 \times 10^{9}$ & $9.3 \times 10^{10}$ & 2.1\\
  MUGS g15784   & $8.6 \times 10^{10}$ & $6.3 \times 10^{9}$ & $1.5 \times 10^{11}$ & 1.7 \\
  MaGICC g1536  & $2.3 \times 10^{10}$ & $1.5 \times 10^{10}$ & $8.9 \times 10^{10}$ & 1.3 \\
  MaGICC g15784 & $8.2 \times 10^{10}$ & $2.3 \times 10^{10}$ & $1.6 \times 10^{11}$ & 0.9 \\
  RaDES ``Selene'' & $6.1 \times 10^{10}$ & $1.2 \times 10^{10}$ & $1.0 \times 10^{11}$ & 1.5 \\
\hline
  Idealised & & & & \\
\hline
  GD1 (smooth) & $3 \times 10^{10}$ & - & $1.75 \times 10^{12} (M_{200}) $ & 0.2 \\
  GD2 (barred) & $5 \times 10^{10}$ & - & $2 \times 10^{12} (M_{200}) $ & 0.4 \\
  Idealised dark matter-free disk & $1.24 \times 10^{11}$ & - & - & 7.8 \\
  \hline  
\end{tabular}
\end{center}
\caption{Basic physical properties of the simulated galaxies. All these values are calculated within the region $R \leq 20$~kpc, $0 \leq z \leq 5$~kpc for the cosmological and idealised models, while the GD1 and GD2 models use a vertical range of $0 \leq z \leq 1.75$~kpc and $0 \leq z \leq 2.5$~kpc respectively. $M_*$ denotes the total stellar mass within this region, $M_g$ is the total gas mass, while $M_\text{DM}$ is the mass in dark matter. We also note the disk vertical scale height $h_z$: these are all single thick disk models except for the GD1 and GD2 models, which are single thin disks. The halo mass for the GD1 and GD2 models is quoted as the $M_{200}$ (in solar masses) of the analytic NFW halo used. The model without a DM halo thickens substantially and behaves more as an isothermal slab distribution, hence the very large ``scale height'' in this case.}
\label{galaxysample}
\end{table*}

The MUGS and MaGICC galaxies were run using the N-body/SPH code \textsc{GASOLINE}, while the RaDES galaxy was simulated with the AMR grid code \textsc{RAMSES}. All these galaxies are dark matter dominated. The MUGS and MaGICC merger histories are considered to be `quiet', given that their last major merger takes place prior to $z=3$ \citep{stinson2010}, allowing them to form large disks. The RaDES ``Selene'' model is a field galaxy that undergoes only a few minor merger events at high redshift (i.e. at a lookback time of $\gtrsim 10$~Gyr) and therefore also has a largely uneventful star formation history.

In all cases, the lack of any recent merger activity ensures that the disks in these models are not strongly disrupted and therefore are more likely to be dynamically settled, consistent with the steady-state assumption used for the Jeans equation analysis of this study.

\subsection{Idealised models}
The smooth GD1 and barred GD2 idealised disks are live galaxy models inside analytic NFW dark matter halo profiles. It is clear from the vertical scale heights given in Table \ref{galaxysample} that these disks are single-component thin disk models, rather than thick disks as in the cosmologically simulated galaxies.

To test the Jeans equation analysis in the absence of dark matter we also model, using \textsc{RAMSES}, a single-component stellar disk (no bulge, no gas and no dark matter) which is set up using the GalactICs initial conditions generation code \citep{dubinski}. This code generates a dark matter halo as well (albeit chosen to have a negligible mass) which we remove before running the model. The initial conditions have $Q > 1.92$ at all radii, where $Q$ is the Toomre stability parameter. We then reset $w = 0$ for all particles, i.e. there are no vertical motions at the start of the simulation. The model is then evolved for a long period of time in an attempt to equilibriate the stellar dynamics: slightly more than $11.1$~Gyr. The absence of a dark matter halo causes the disk to thicken substantially over time, until it more closely resembles an isothermal slab distribution, with very low rotation velocity at all radii. Thus this evolution timescale corresponds to roughly $6$ full rotations of the disk.

\subsection{Determining the kinematical quantities from the simulations}
We consider the particles (stars for all the models, plus dark matter and gas for the non-idealised models) within a radius of $20$~kpc in the galactic plane and a height (above and below the disk) of $5$~kpc for the cosmological and dark matter-free models, $1.75$~kpc for the GD1 model and $2.5$~kpc for the GD2 model. Only the star particles are used to determine the kinematics in the disk. The dark matter and gas particles are only used to determine the total disk density.

This region is then binned into $50$ bins in the $z$ and $R$ directions. The bin width in the $R$ direction is $0.2$~kpc for all models, and in the $z$ direction it is $0.2$~kpc for the cosmological models, $0.07$~kpc for the GD1 model, and $0.1$~kpc for the GD2 model. The following quantities are determined in each bin:
\begin{itemize}
\item $\overline{U^2}, \sigma_U$: the mean squared radial velocity and the radial velocity dispersion of the star particles.
\item $\overline{V}, \sigma_V$: the mean azimuthal velocity and the azimuthal velocity dispersion of the star particles.
\item $\overline{W^2}, \sigma_W$: the mean squared vertical velocity and the vertical velocity dispersion of the star particles.
\item $\overline{UW}$: the mean product of the radial and vertical velocities of the star particles.
\item $\rho_*$: the stellar density (the total star particle mass divided by the bin volume).
\item $\rho_g$: the gas density (the total gas particle mass, or gas densities in all the grid cells within our bin, divided by the bin volume).
\item $\rho_{DM}$: the dark matter density (the total DM particle mass divided by the bin volume).
\end{itemize}
These quantities, as functions of $R$ and $z$, are then used in the Jeans equation to calculate the surface density and to compare with the true surface density determined by direct measurement in the simulation. Note that, for now, we consider all particles within an annulus at a specific radius and height, in order to use as many particles as possible, we do not select only particles from particular wedges in the disk. In Section~\ref{azimuth} we will see how the results vary for each disk quadrant. Clearly at very high heights above the disk the number of star particles becomes low, introducing noise into the results. In addition, the low particle resolution at these heights in the \textsc{GASOLINE} SPH simulations implies a large hydrodynamical smoothing length, and correspondingly poor resolution limiting the ability to accurately capture drag effects in the gas. This will then have an impact on the kinematics of any thick disk stars formed \emph{in-situ} at high heights. In the \textsc{RAMSES} simulations, a low resolution grid at high heights implies poor sampling of the gravitational potential, also leading to inaccurate stellar dynamics. These concerns do not apply at lower heights where the particle and grid resolution is high.

The various radial scale lengths $h_R, h_U$ and $h_V$ are measured at a single height close to the disk plane: $z=0.1$~kpc for the cosmological models and the dark matter-free disk; $z=0.05$~kpc and $z=0.035$~kpc for the thin disk GD1 and GD2 models respectively. The vertical scale height $h_z$ is measured for the radial bin at $R=8.2$~kpc for all models.

\subsection{Vertical velocity dispersions}
\label{subsec:vertveldisp}
\begin{figure*}
\centering
\begin{tabular}{cc}
\includegraphics[width=6.0cm]{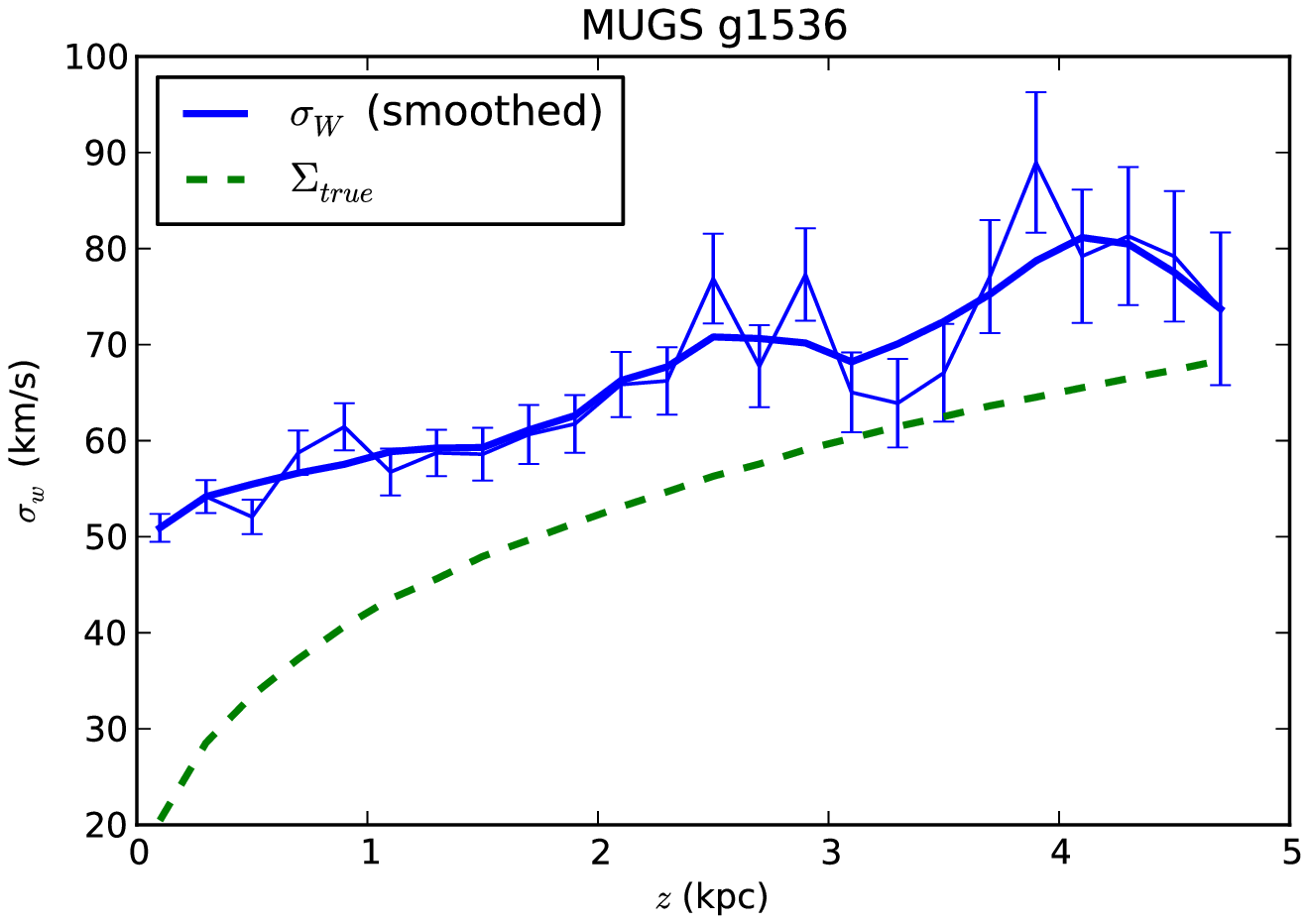} & \includegraphics[width=6.0cm]{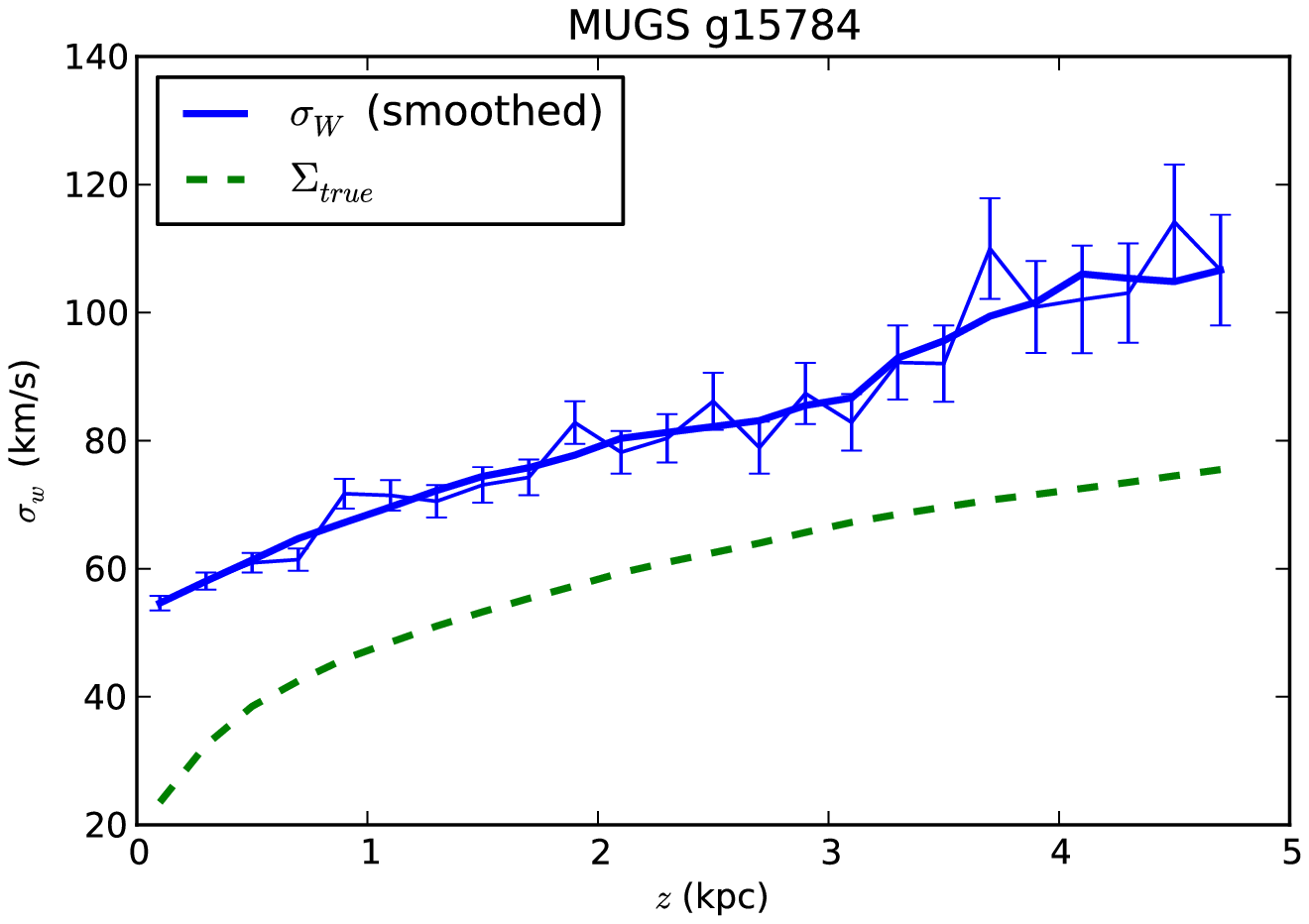} \\
\includegraphics[width=6.0cm]{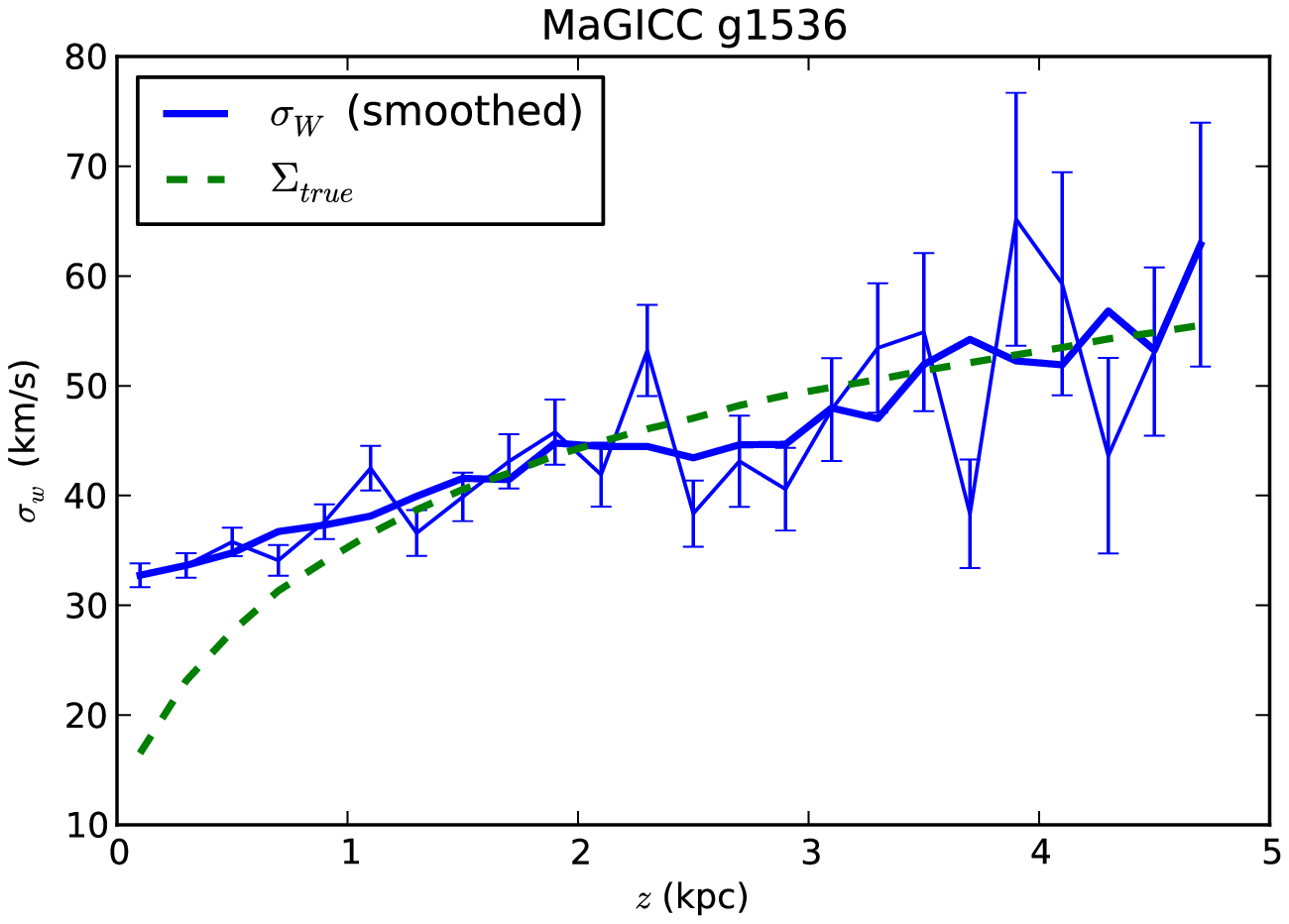} & \includegraphics[width=6.0cm]{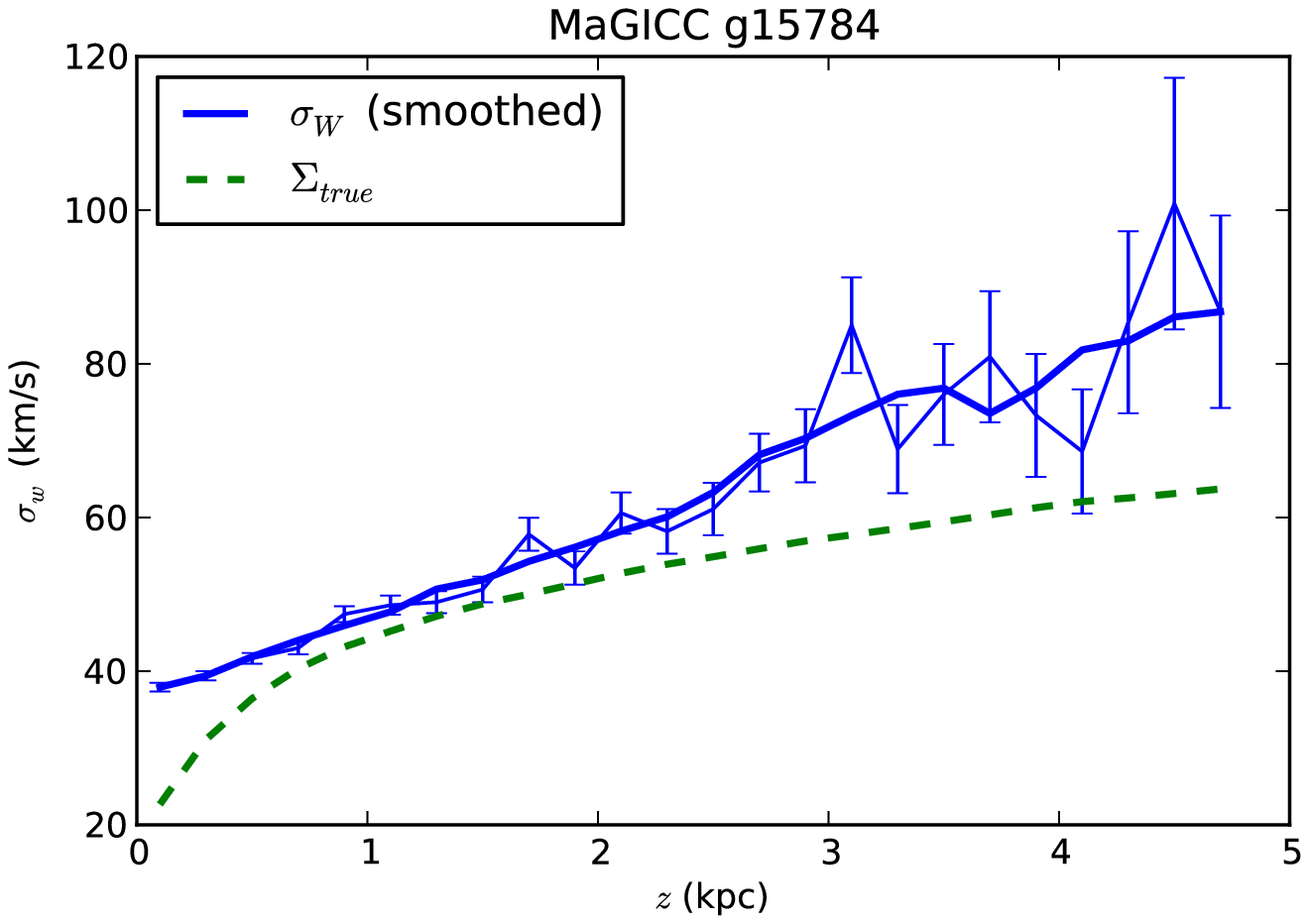} \\
\includegraphics[width=6.0cm]{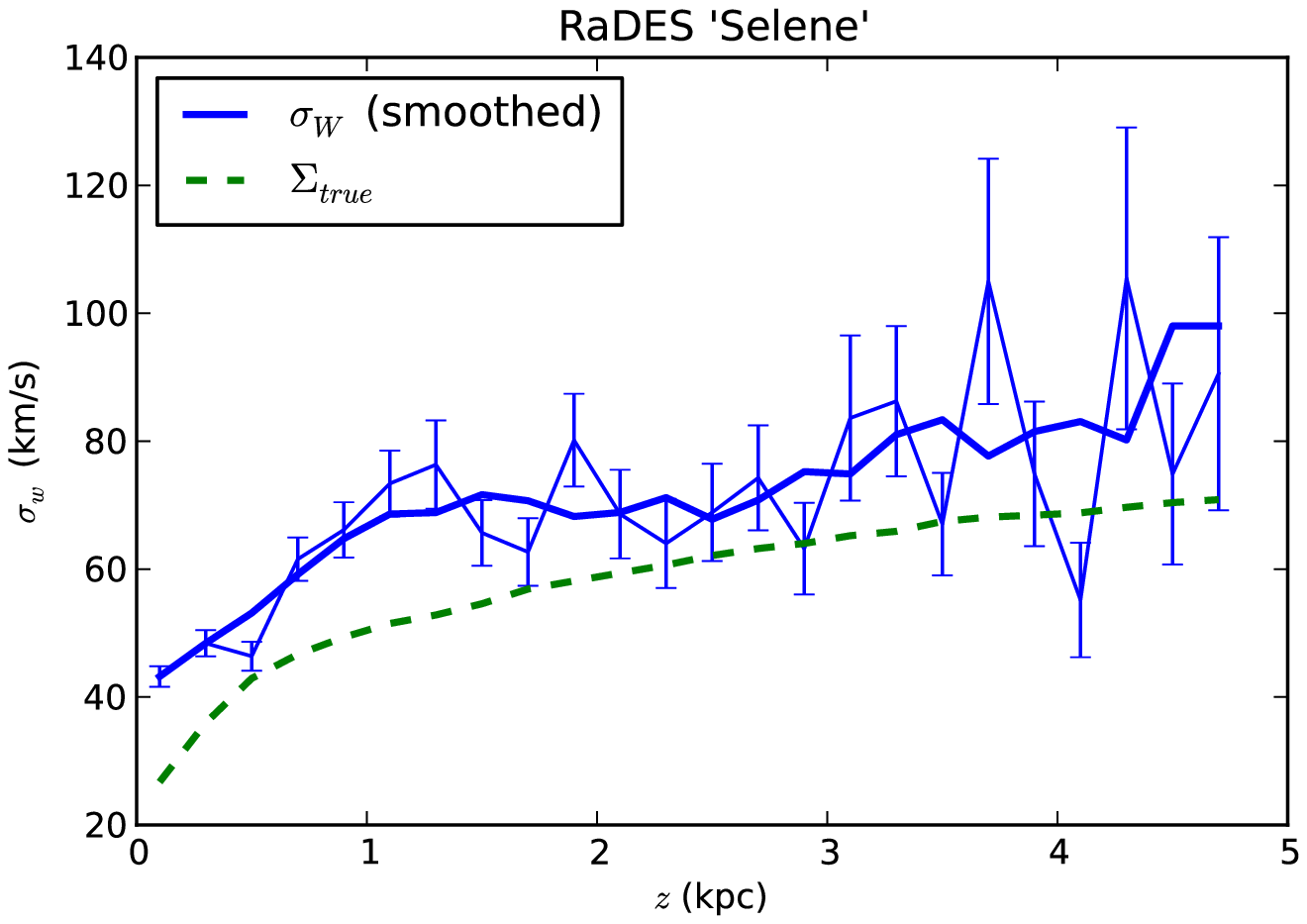} & \includegraphics[width=6.0cm]{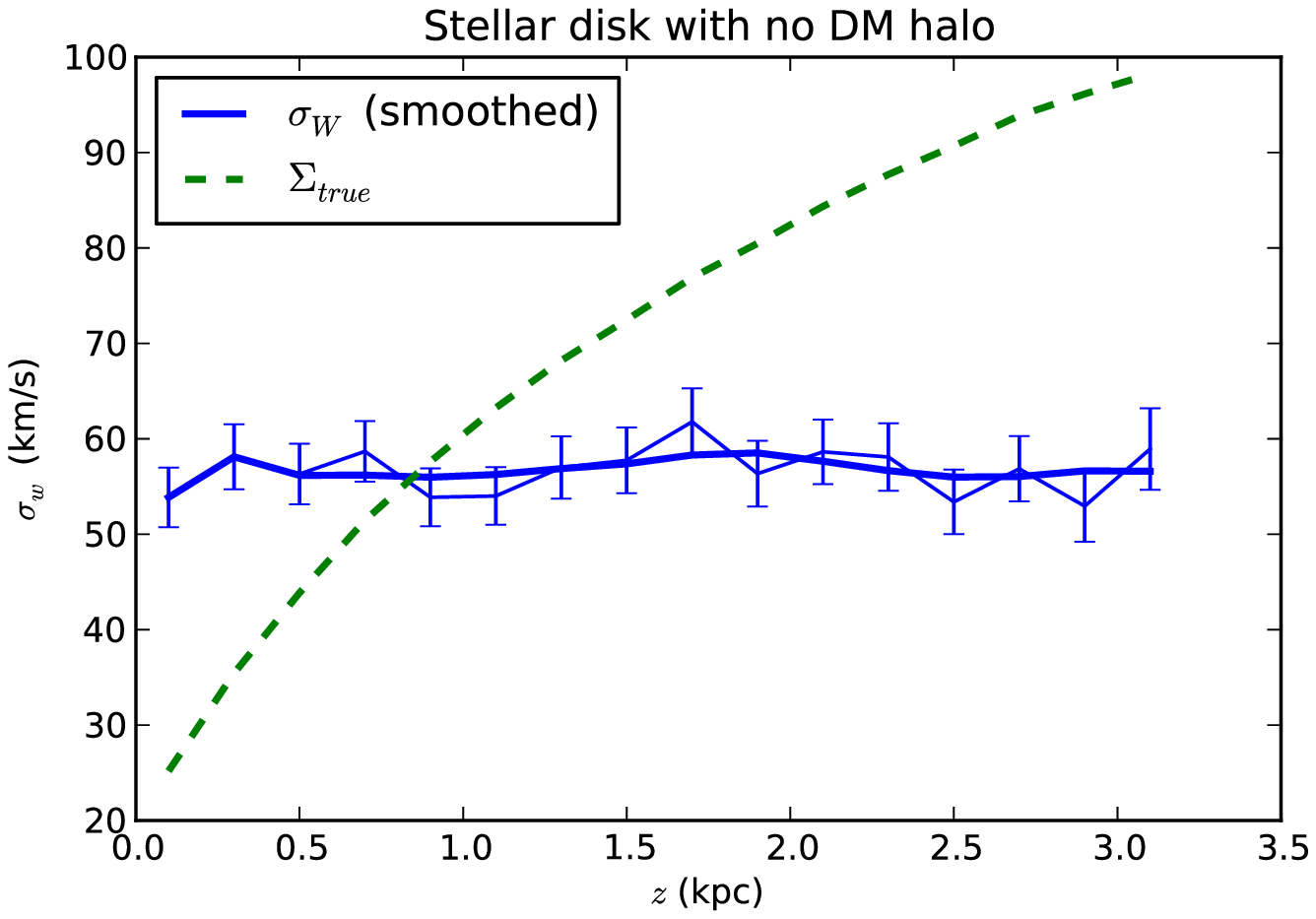} \\
\includegraphics[width=6.0cm]{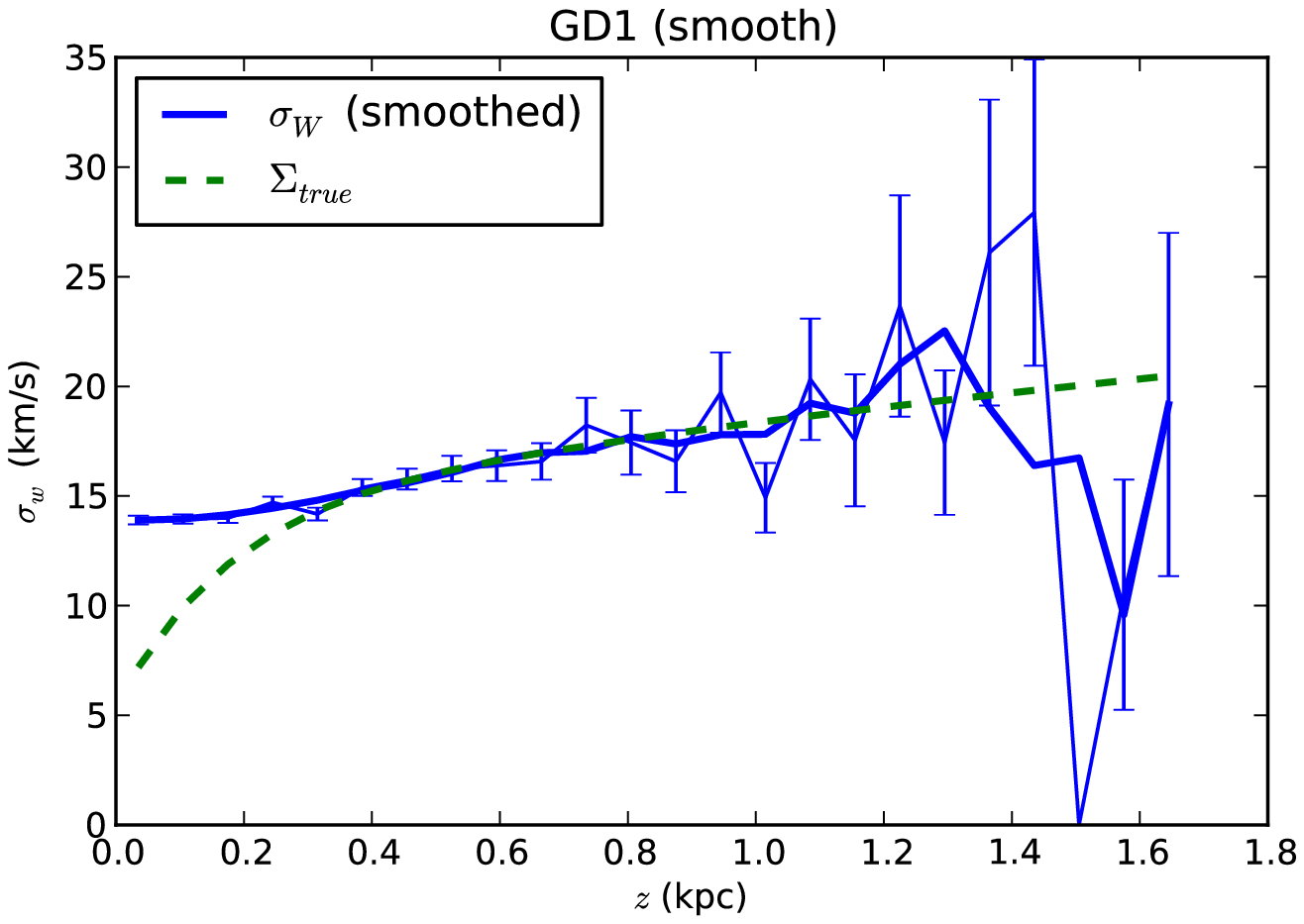} & \includegraphics[width=6.0cm]{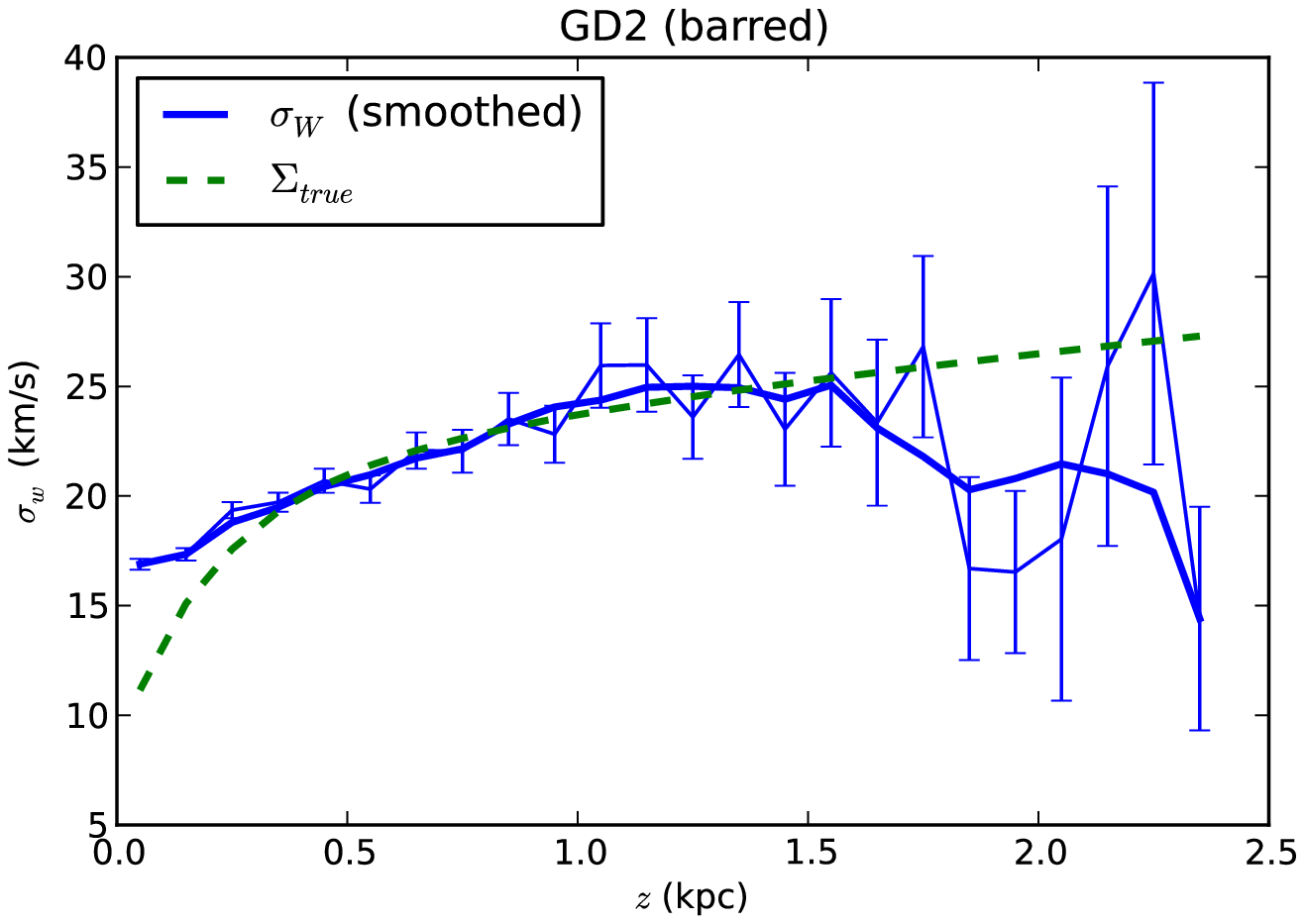}
\end{tabular}
\caption{$\sigma_w$ as a function of $z$ at $R=8.2$~kpc. The thin blue line denotes the values determined in each vertical bin, while the error bars are derived from the sample size in each bin (see text). The thick blue line is the 5-point moving average, extrapolated for low and high $z$. Clearly the noise and error increases as we move further from the disk, where there are fewer stars. Note that the idealised models exhibit low number statistics at lower $z$ than the cosmological models, hence the lower cut-off in height in these plots. The thick green line in all plots is the ``predicted'' vertical velocity dispersion from the actual surface density in the model (see text).}
\label{sigmaw}
\end{figure*}

In Fig.~\ref{sigmaw} the vertical velocity dispersion as a function of height $\sigma_w(z)$ is shown for each model. The blue lines in the various panels are the values obtained for that model in the vertical bins above the disk plane at a radius of $R=8.2$~kpc. The error in each bin $\epsilon(z)$ is calculated as follows \citep{statistics}:
\begin{equation}
\epsilon(z) = \frac{\sigma_w(z)}{\sqrt{2(N(z)-1)}}
\end{equation}
where $N(z)$ is the number of star particles in that $z$-bin. The length of each symmetric error bar is then $2\epsilon(z)$. We smooth the values of $\sigma_w$ with a 5-point running average (extrapolating at low and high $z$) which is shown as a thick blue line in Fig.~\ref{sigmaw}. Our calculation of the surface density uses the smoothed $\sigma_w$, and so all future references to the vertical velocity dispersion refer to the smoothed values. The thick green line in these plots is the ``predicted'' vertical velocity dispersion from the true surface density of the model (see Section~\ref{comparison1d3d}).

\subsection{Applying the equation to the simulation data}
We must decide how to numerically evaluate the derivatives appearing in the equation for the surface density. One approach would be to try using the data directly, by calculating derivatives with a finite difference approximation. The results with this approach are, however, extremely noisy due to the non-smoothness of the data. An alternative procedure is utilised in \cite{garbari} where an SPH-style smoothing kernel is applied to the particle data, and then a polynomial reconstruction at each point in the particle ``fluid'' is used to determine gradients. Our strategy is to employ a simpler approach, more along the lines of the analysis undertaken in MB12, where we use function fitting for those terms that involve a derivative. Note that we do \emph{not} use these fits for terms that do not involve a derivative, only for terms (iii), (iv), (vii) and (ix). Terms (i), (ii), (v), (vi) and (viii) will use the binned simulation data.

To proceed with our function fitting, we therefore apply further assumptions:\\

\noindent
\textbf{Assumption 4} - the squared mean azimuthal velocity is assumed to have a \emph{linear} dependence on radius: $\overline{V}^2 \sim \alpha_V R + \beta_V$, with $\alpha_V$ the slope of the fit line and $\beta_V$ the intercept, at least within a chosen range in $R$ (given later) that includes our ``solar radius'' of $R=8.2$~kpc. Although it is a fit parameter, the intercept $\beta_V$ is never needed in our analysis. We fit the slope at all heights, due to the vertical dependence of this term (see BT12 and MB15). This term was ignored in MB12, equivalent to setting $\alpha_V = 0$. It is demonstrated in MB15 that this assumption, while incorrect, makes little difference to their final value of $\Sigma_{\text{calc}}(z)$. The integral of the slope values over $z$ required for term (vii) is then calculated numerically.\\

\noindent
\textbf{Assumption 5} - the mean product of the radial and vertical velocities is assumed to have a \emph{linear} dependence on radius: $\overline{UW} \sim \alpha_{UW} R + \beta_{UW}$, with $\alpha_{UW}$ the slope of the fit line and $\beta_{UW}$ the intercept, again within a chosen range in $R$. The intercept value, although fitted, is unnecessary for our analysis. For this quantity we fit the slope at the height $z=0.1$~kpc only. Adding a height dependence makes a negligible difference to our results. This linear fit shows that $\overline{UW}$ has a very weak dependence on $R$ (although it is noisy) for all of our models. Contrary to our assumption, in MB12 this quantity was assumed to have an exponential dependence on $R$, with a scale length equal to the density. If we calculate the surface density using the terms in Section \ref{listofterms} (including terms (ii) and (iv) that were incorrectly omitted from the analysis of MB12) with the data used in MB12 using an exponential radial dependence in $\overline{UW}$ we find a negative $\Sigma_{\text{calc}}(z)$. It is nonetheless generally assumed in the literature, based on analytical approximations, that $\overline{UW}$ has an exponential dependence on $R$, although the scale length may not be equal to $h_R$.  A linear fit may therefore be taken as a reasonable assumption, at least locally, if the scale length is substantially larger than $h_R$. In any case, the simulation data are in reasonable agreement with a linear fit, given the noise, and we will see that the $\overline{UW}$ terms are subdominant in the analysis.\\

\noindent
\textbf{Assumption 6} - the vertical velocity dispersion has a linear dependence on $z$: $\sigma_{W} = \alpha_W z + \beta_W$. This assumption is also used in MB12, given the data of \cite{mb12other}, and in Fig. \ref{sigmaw} we can see that it is approximately true for our simulation data. For large heights above the disk, beyond the region where the mass density changes significantly with height, this is a well-motivated assumption. As one approaches the disk plane, however, the height dependence of $\sigma_W$ will vary from linearity, bending down towards zero. Thus this assumption leads to poor estimates of the surface density close to the disk plane. \\

Our use of the simulation data in the Jeans equation therefore begins with determining the radial and vertical scale lengths of the density, and the radial scale lengths of $\overline{U^2}$ and $\sigma_V^2$. We use a linear regression routine in NumPy to calculate linear fits to $\log \rho(R,z)$, $\log \overline{U^2}(R)$, $\log \sigma_V^2(R)$, $\overline{V}^2(R)$, $\overline{UW}(R)$ and $\sigma_W(z)$, where $R=8.2$~kpc when the $z$-dependence is being considered, and in the $z$-bin closest to the disk plane when the $R$-dependence is being considered, except for the measurements of $\overline{V}^2(R)$ which we make at all heights. For these fits we use data in a radial range $4.6 \leq R \leq 11.8$~kpc, roughly centered on our chosen ``Galactocentric solar radius'' of $8.2$~kpc, for most of the models. For the MUGS g1536 model we restrict the radius range used to measure $h_U$ to $4.6 \leq R \leq 9.4$ due to substantial noise in the data at larger $R$ leading to a poor fit. For the GD2 model the radial range is restricted to $6.2 \leq R \leq 11.8$~kpc due to complications arising from the bar in this galaxy. The vertical range used for measuring the fits of all models except the idealised models is $0.1 \leq z \leq 4.7$~kpc. The thin disks of the GD1 and GD2 models limit our useful vertical range to $0.05 \leq z \leq 1.0$ and $0.035 \leq z \leq 1.225$~kpc respectively. For the dark matter-free model low number statistics force us to restrict the vertical range to $0.1 \leq z \leq 3.1$~kpc.

The integrals in the terms in Section \ref{listofterms} are performed numerically by summing up values in each bin, multiplying by the bin width. We reiterate that terms (i), (ii), (v), (vi) and (viii) use simulation data, while terms (iii), (iv), (vii) and (ix) use the function fits, with the term (vii) fits calculated at each height. The slope of $\overline{V}^2$ has considerable variation across models (and for different heights for each model) but, as we will see later, the azimuthal velocity term is a subdominant contribution to the calculated surface density. 

It is important to note that all function fits are, of course, checked against the simulation data and are not arbitrary assumptions purely to simplify the Jeans equation. In particular, $\rho(R,z)$, $\overline{U^2}(R)$ and $\sigma^2_V(R)$ all show very clear exponential behaviour, as shown by the generally high $R^2$ values of the fits in Table~\ref{fitparams}.

We will also consider two simplified forms of the equation for the surface density. The first one uses only the vertical terms (viii) and (ix):
\begin{equation}
\Sigma^C_{Vert} = \frac{1}{4\pi G} \left\{\frac{2}{h_z} \overline{W^2} \Big|_0^z - 2\frac{\partial \overline{W^2}}{\partial z}\Big|_0^z \right\}.
\end{equation}
The other includes all terms in the expression for the vertical force, i.e. Eq.~\ref{verticalforceterm}. These simplified forms correspond respectively to: 1) the traditional method of using only the vertical kinematics to determine the surface density at the solar radius, assuming that the `tilt' term $(1/(R\rho_*))\partial_R (R \rho_* \overline{UW})$ is negligible, and 2) the method employed in BT12 that includes this term.

\begin{figure*}
\centering
\begin{tabular}{cc}
\includegraphics[width=7.0cm]{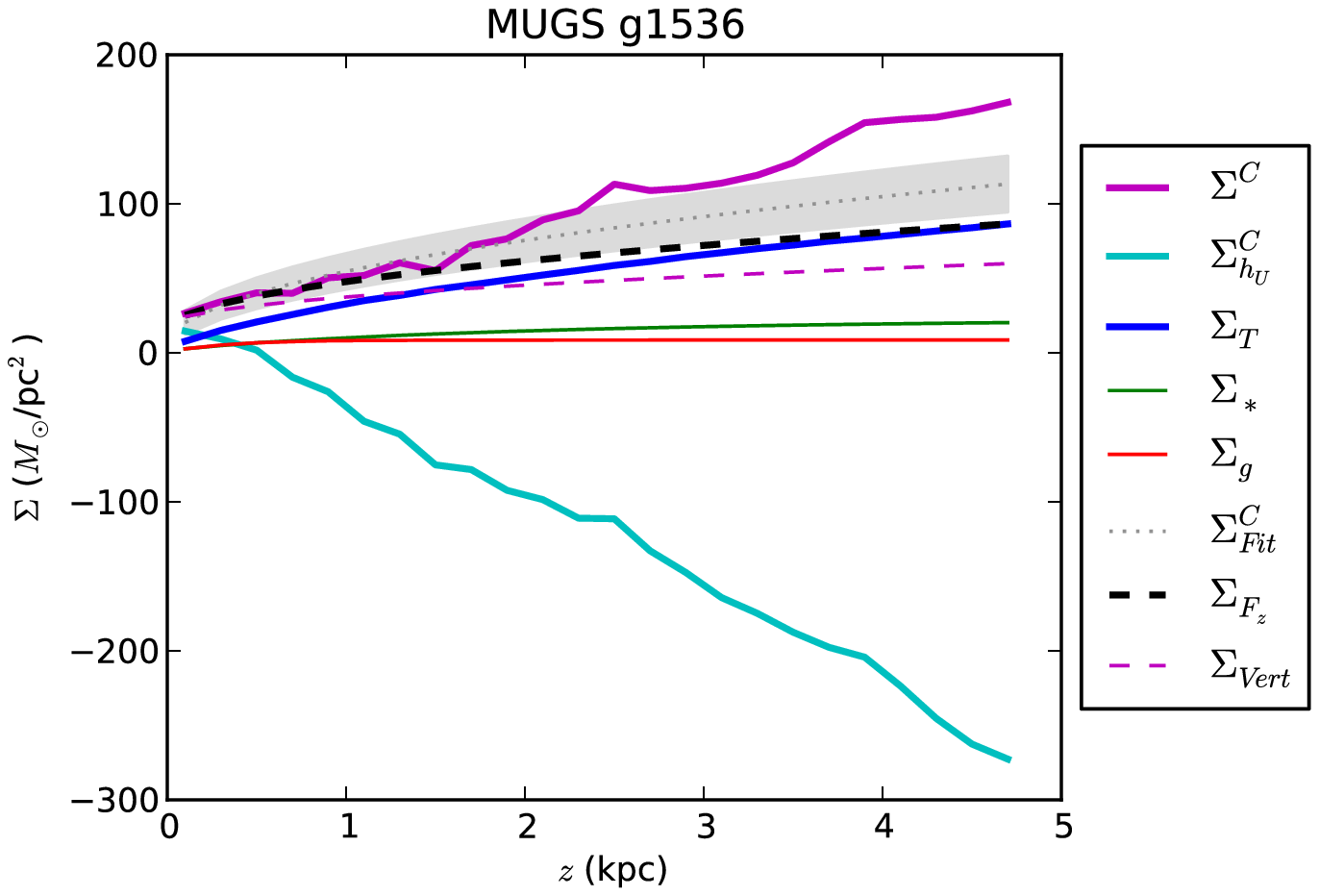} & \includegraphics[width=7.0cm]{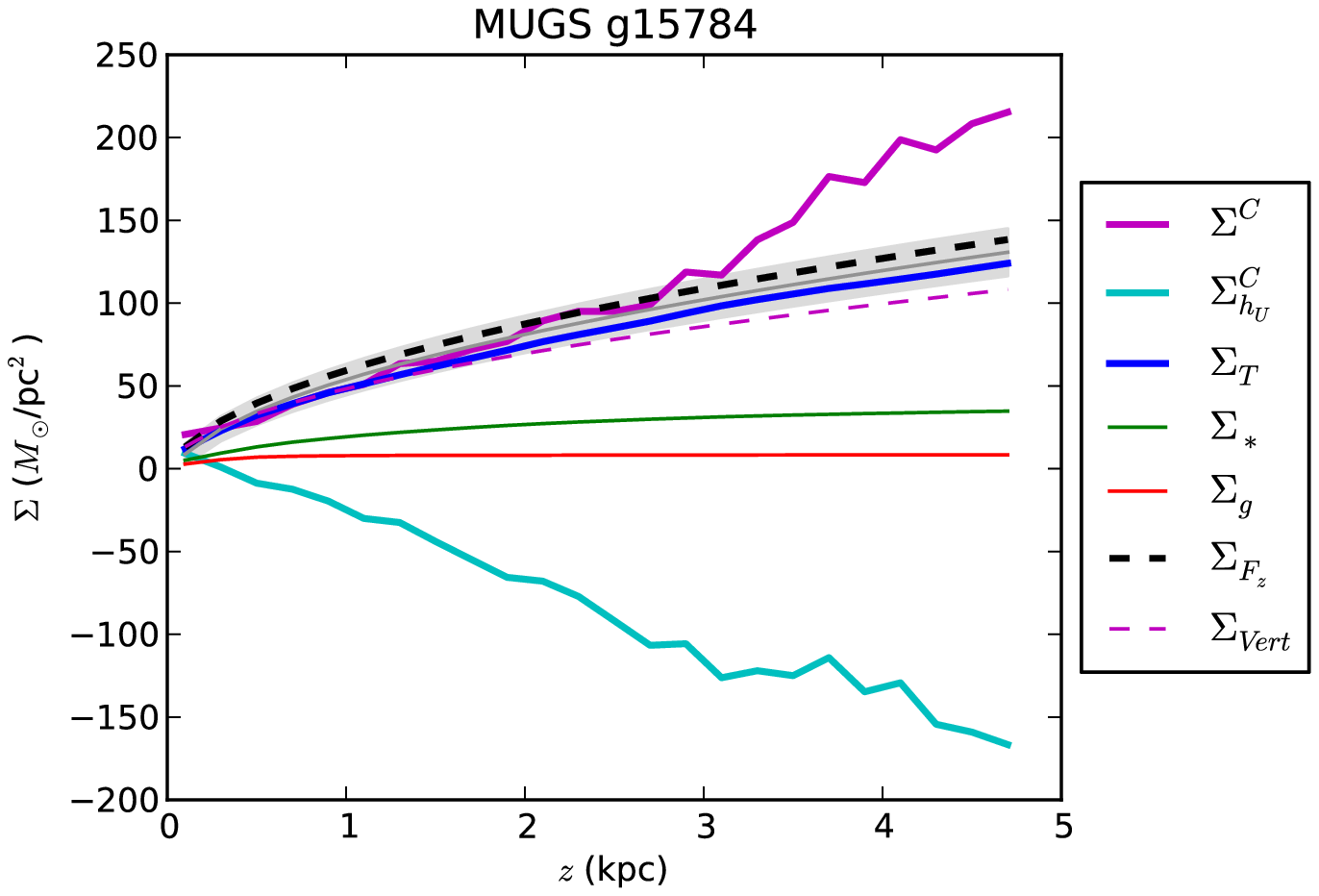} \\
\includegraphics[width=7.0cm]{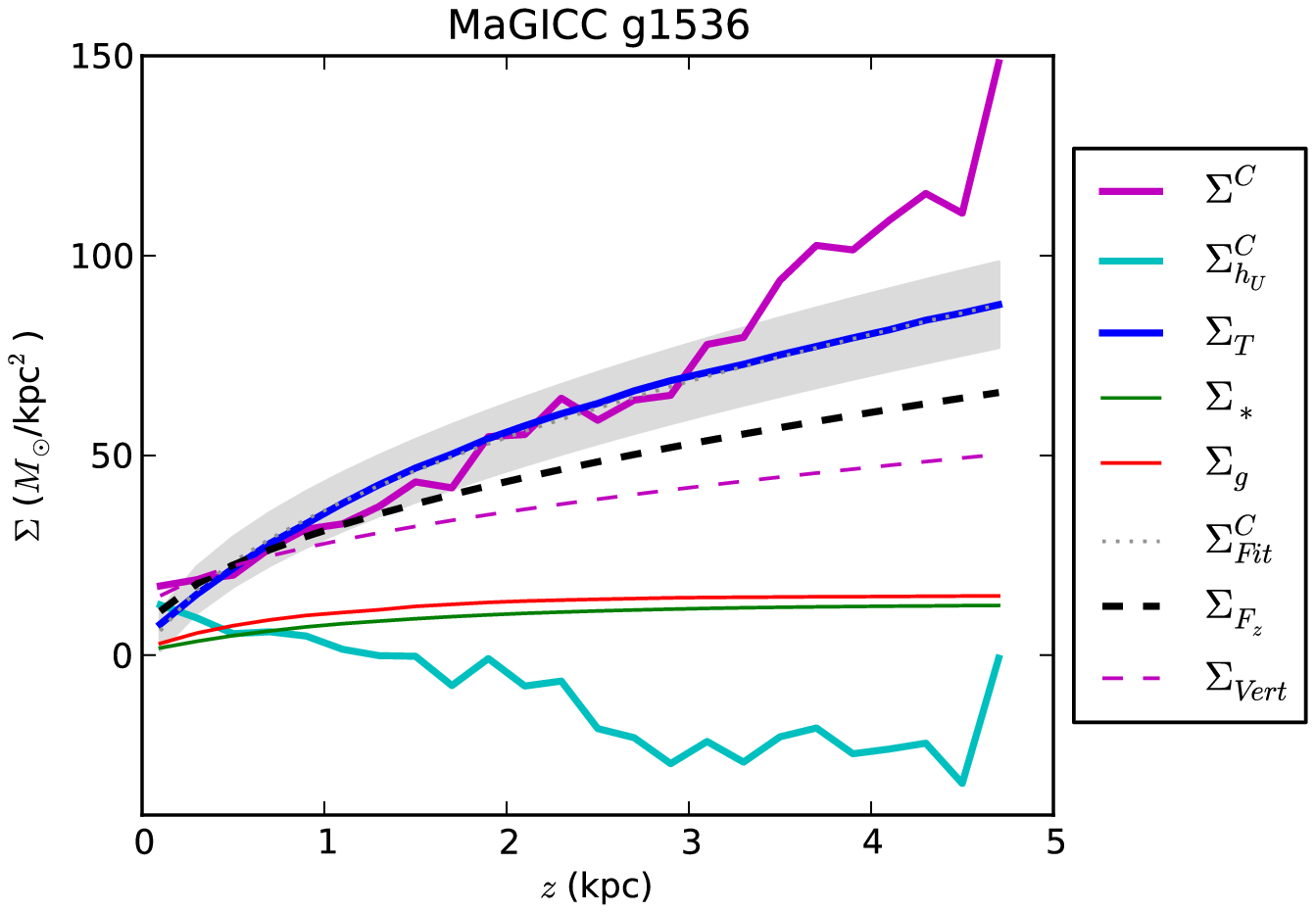} & \includegraphics[width=7.0cm]{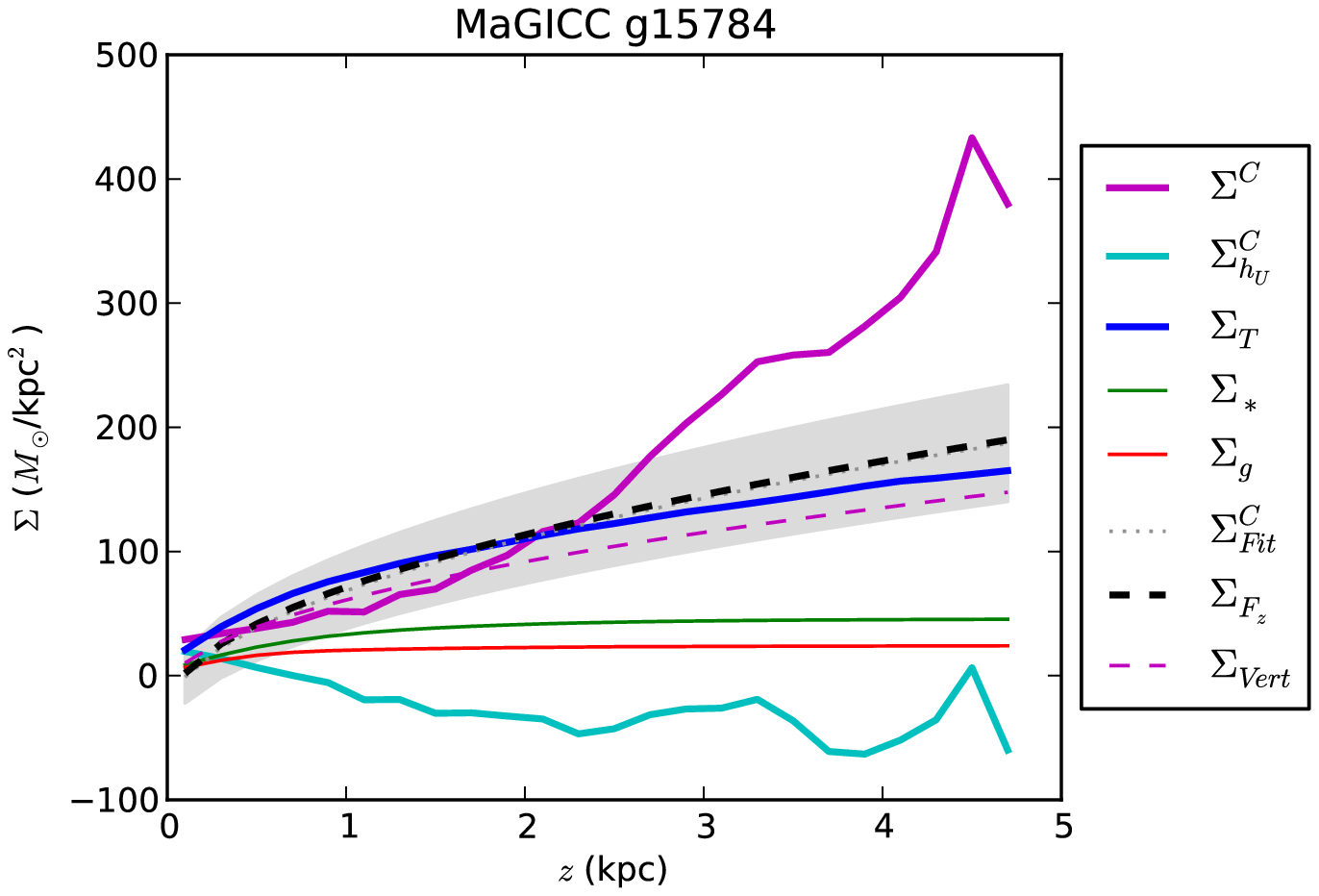} \\
\includegraphics[width=7.0cm]{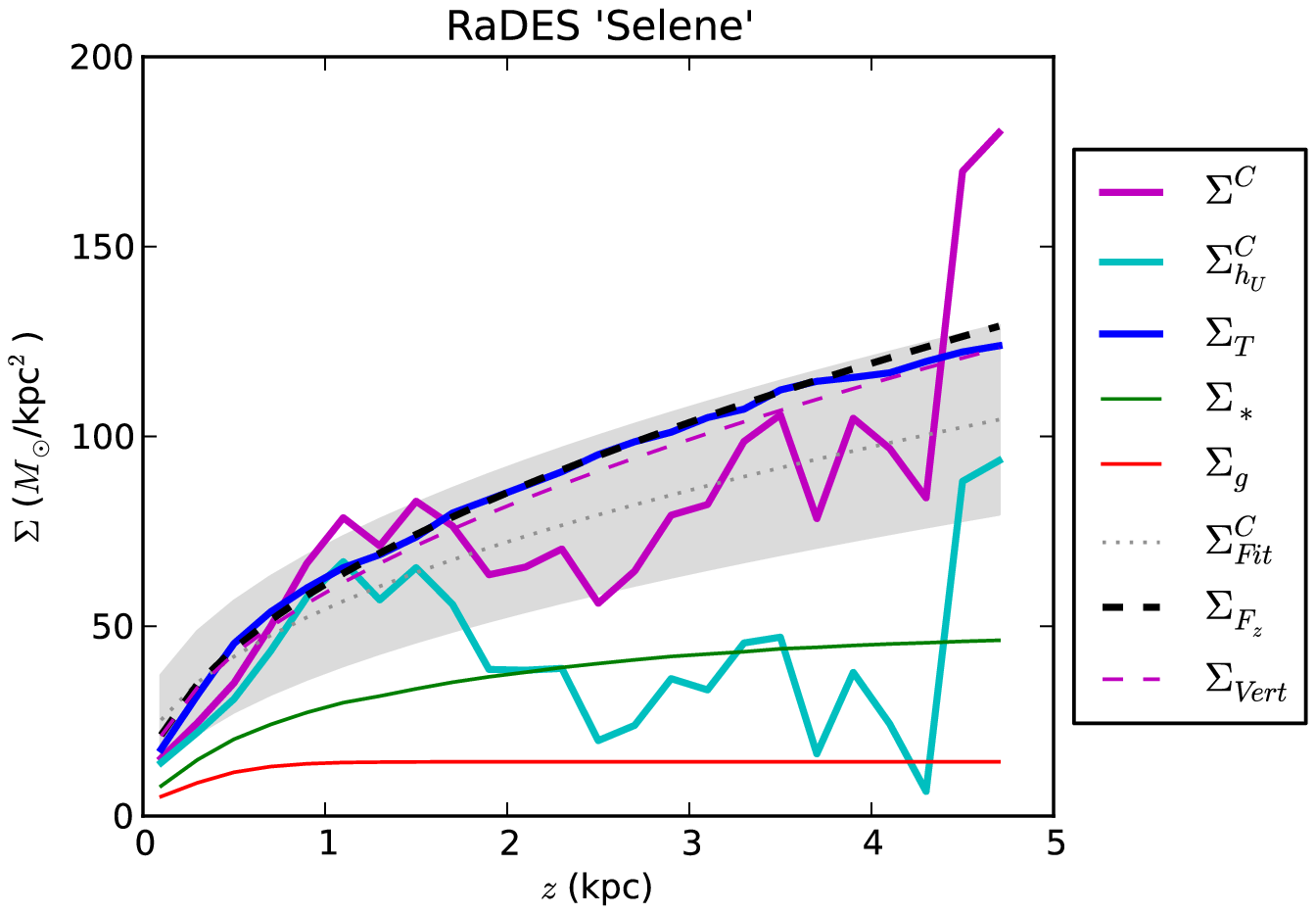} & \includegraphics[width=7.0cm]{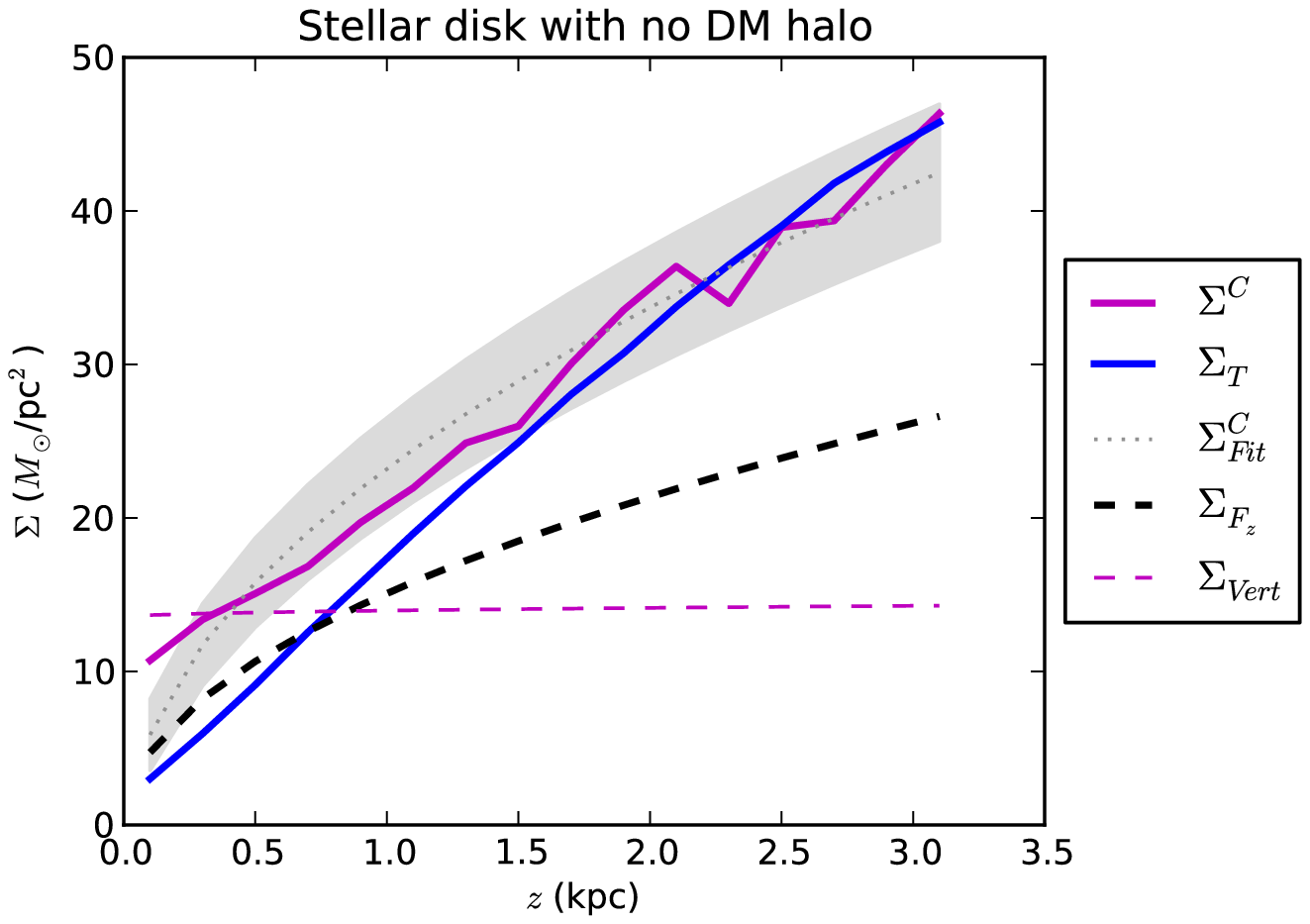} \\
\includegraphics[width=7.0cm]{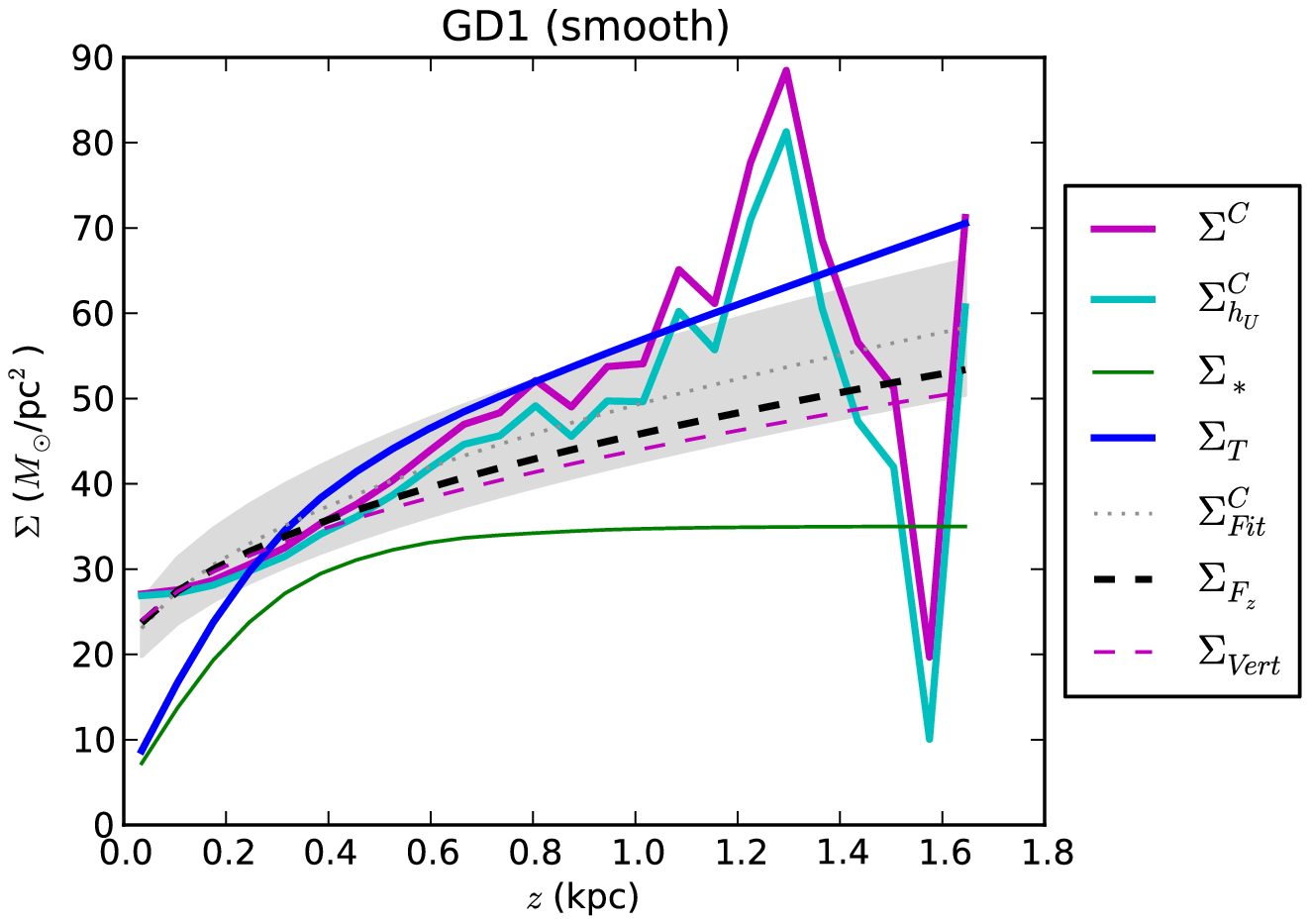} & \includegraphics[width=7.0cm]{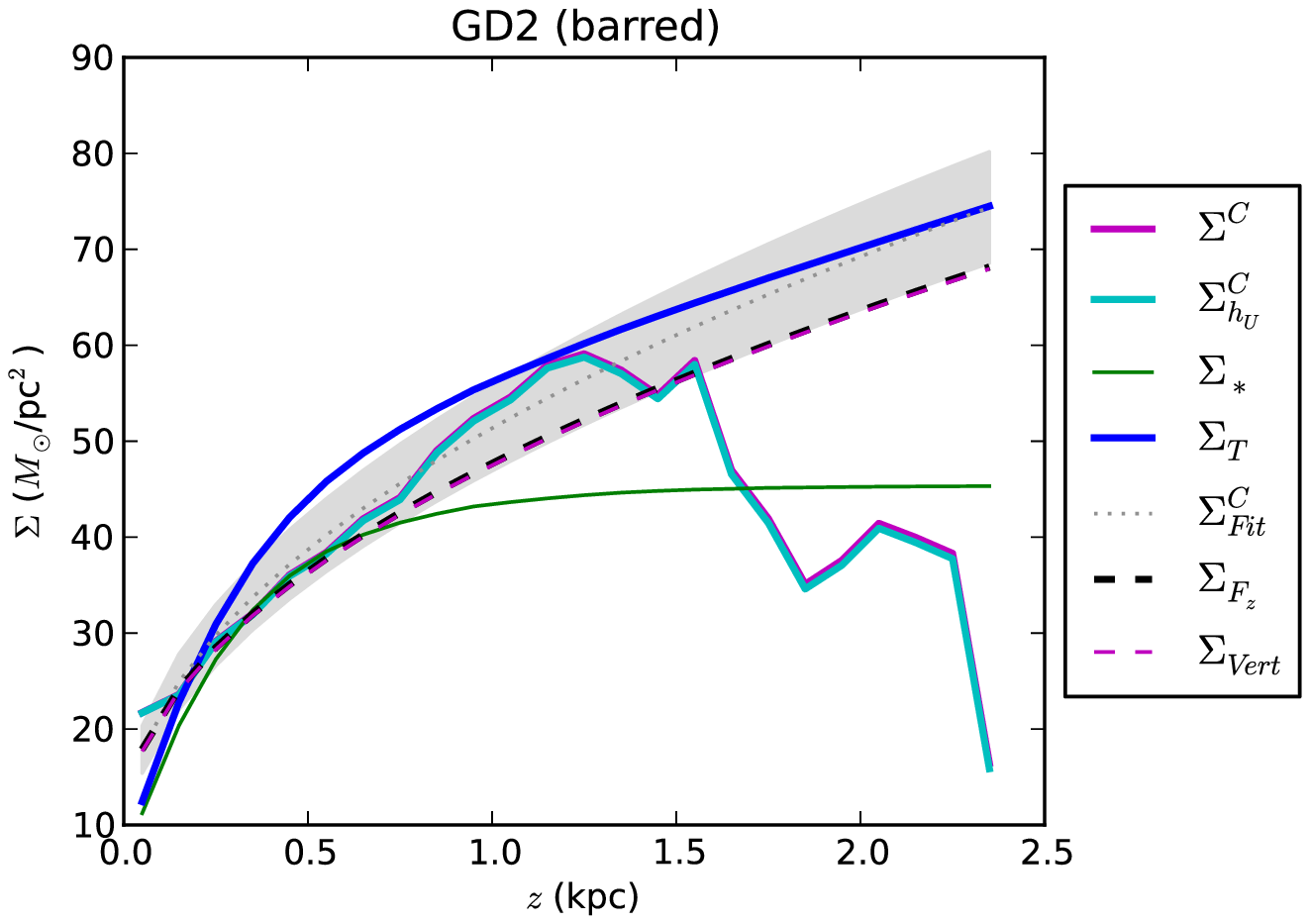}
\end{tabular}
\caption{Total surface density calculation compared to actual surface density. The $\Sigma^C$ line includes all terms, the $\Sigma^C_{Vert}$ is the vertical-only version, and the $\Sigma^C_{F_z}$ line uses the full vertical force of Eq.~\ref{verticalforceterm} including the 'tilt' terms that depend on $\overline{UW}$. The $\Sigma^C_{h_U}$ line (where shown) uses the $h_U = h_R$ condition. The total surface density (including gas if present) in the simulation is $\Sigma_T$, the stellar surface density is $\Sigma_*$ and the gas surface density is $\Sigma_g$. The dotted line and grey bands are the result of the function fit described in the text. The idealised dark matter-free model only contains stars, and so the total surface density in this case is the stellar surface density. Note that for the GD2 (barred) model the calculated surface density using the measured $h_U$ and that using the $h_U = h_R$ condition are coincident.}
\label{surfdensresults}
\end{figure*}

\begin{figure*}
\centering
\begin{tabular}{cc}
\includegraphics[width=7.0cm]{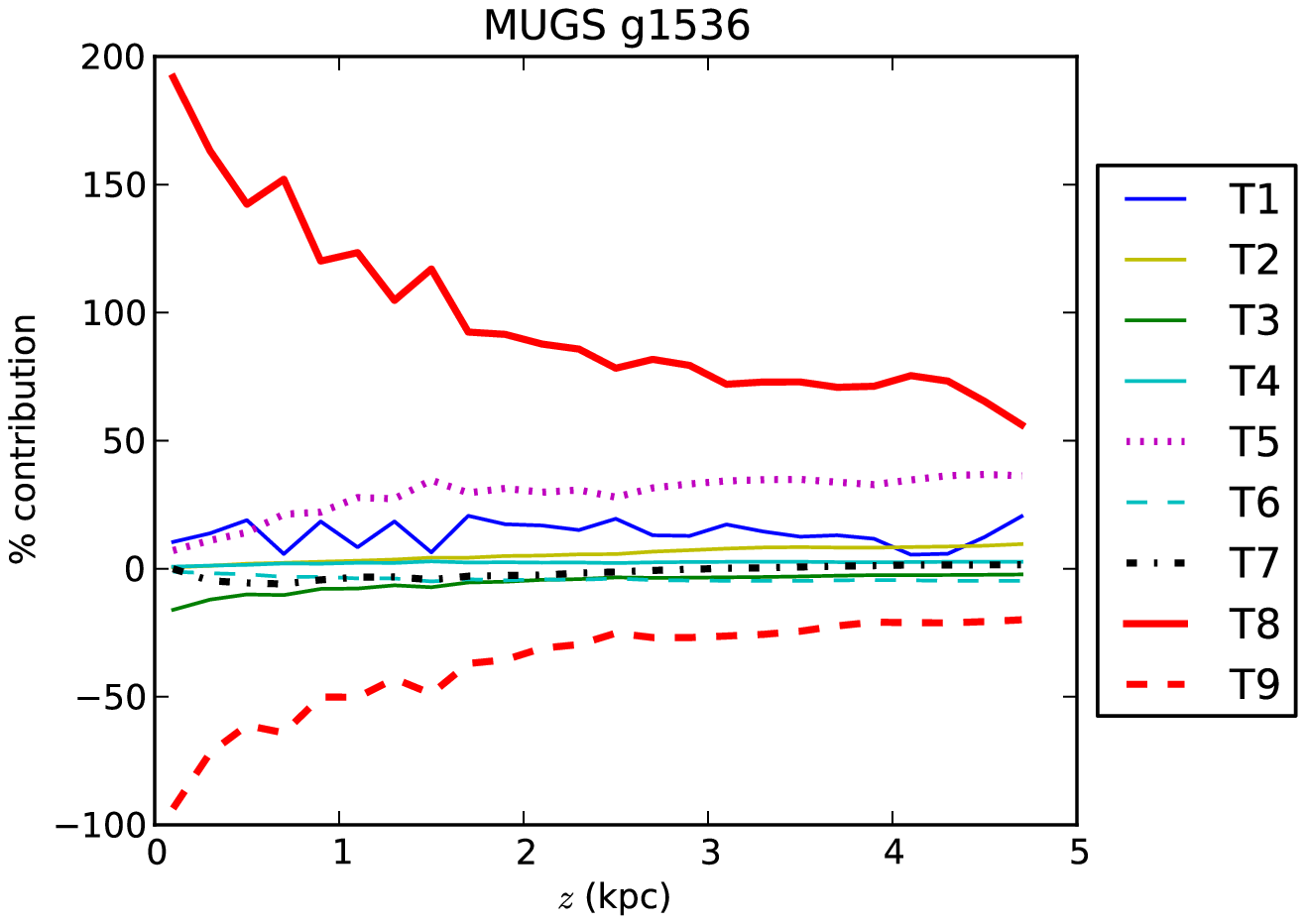} & \includegraphics[width=7.0cm]{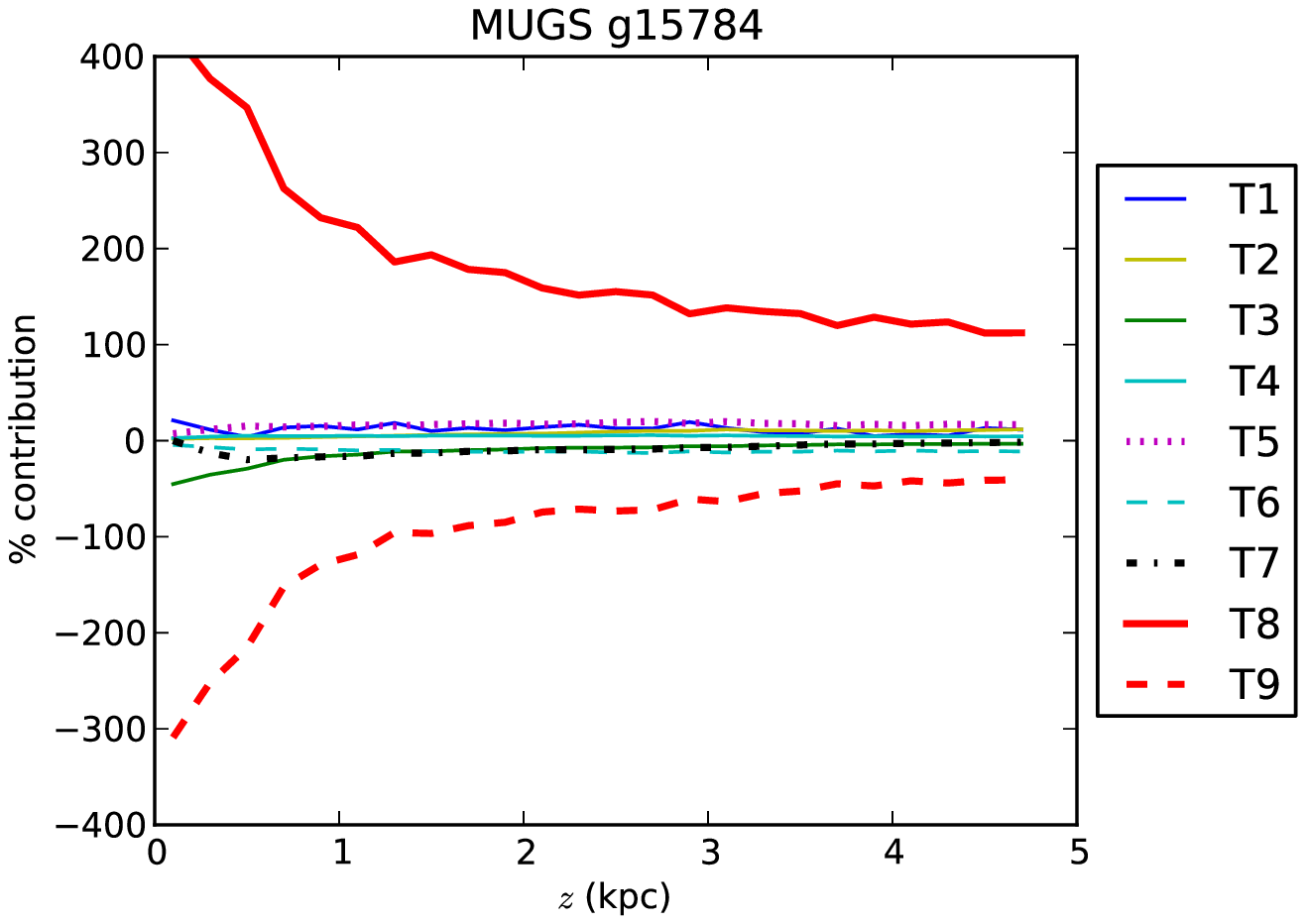} \\
\includegraphics[width=7.0cm]{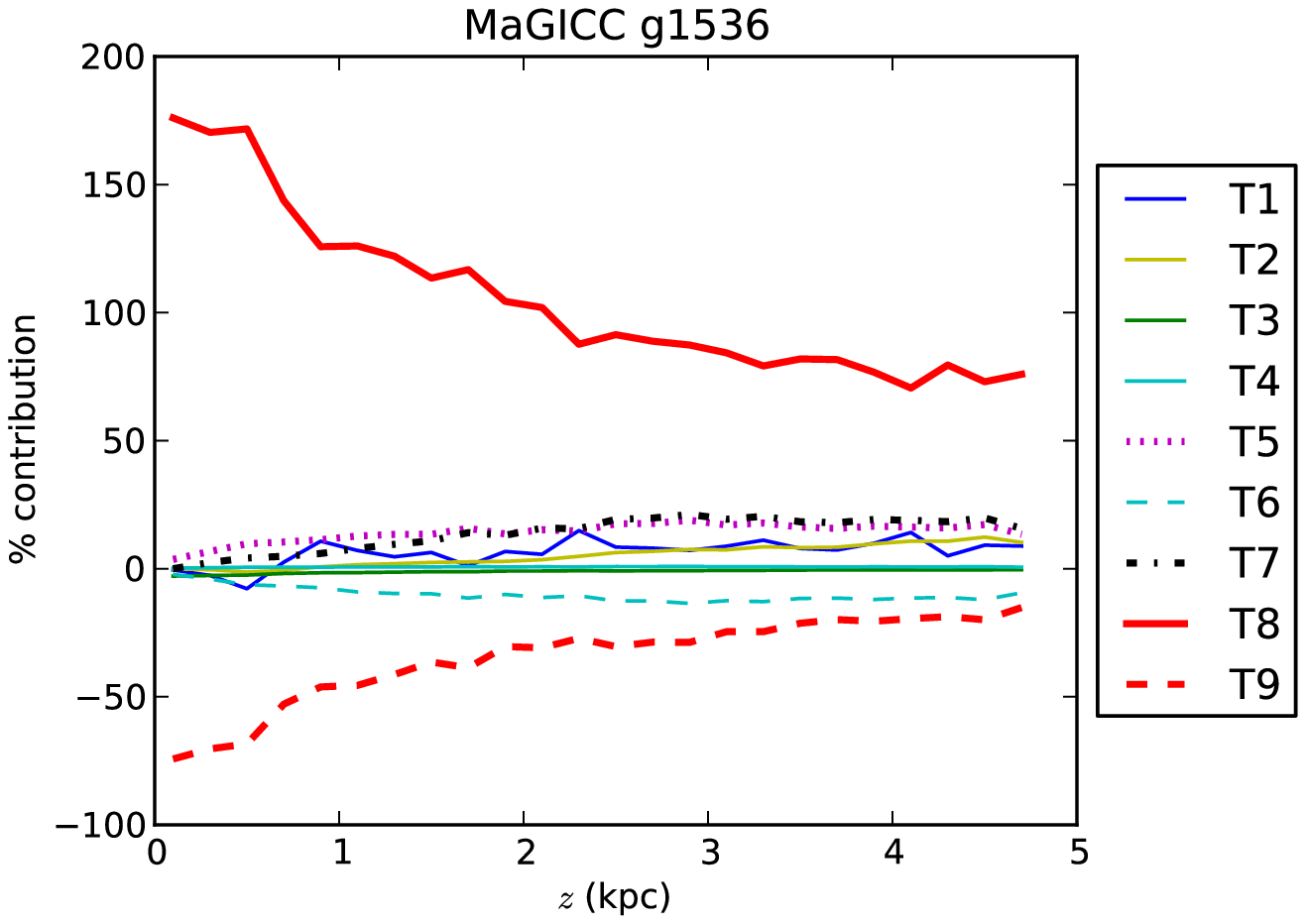} & \includegraphics[width=7.0cm]{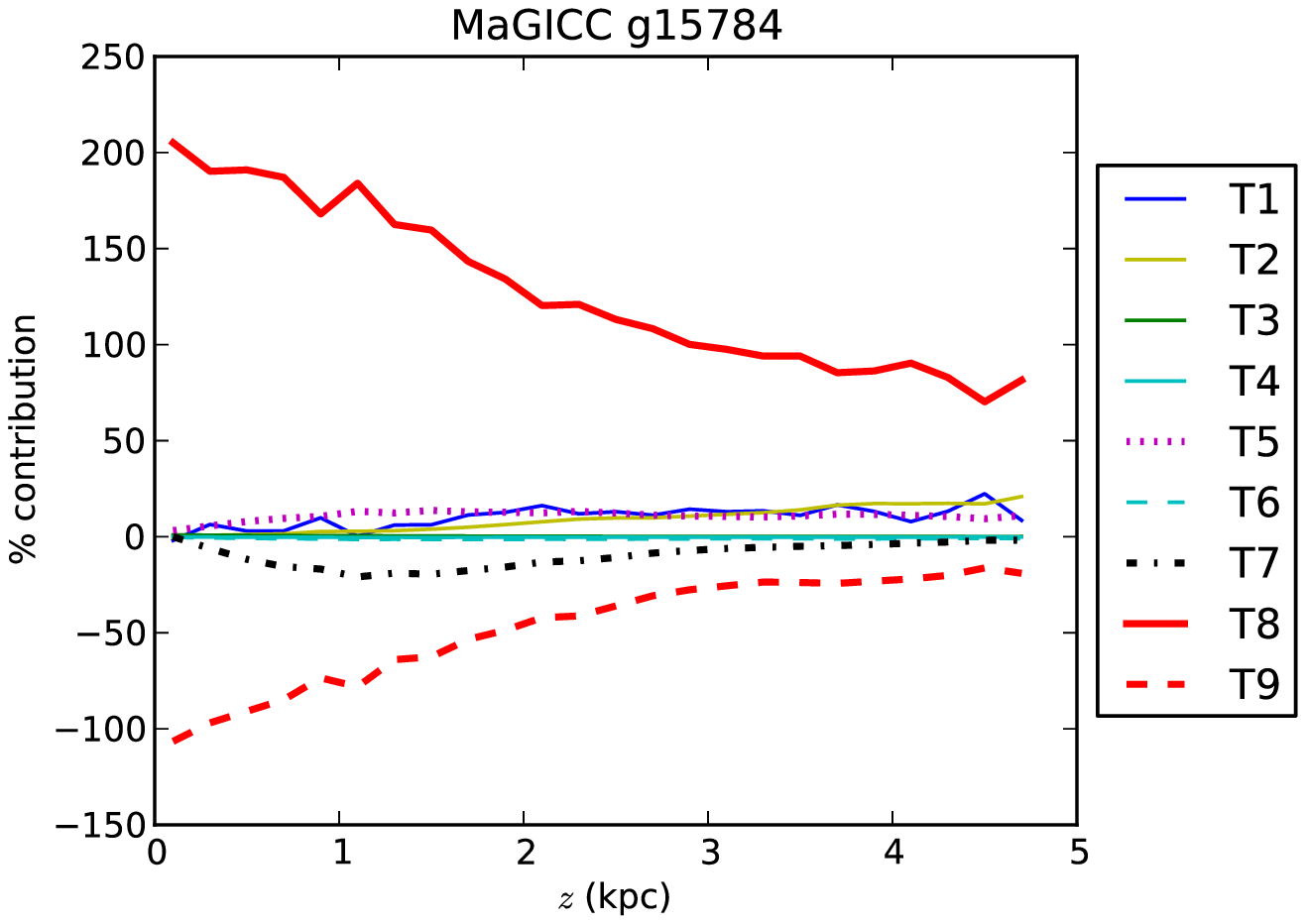} \\
\includegraphics[width=7.0cm]{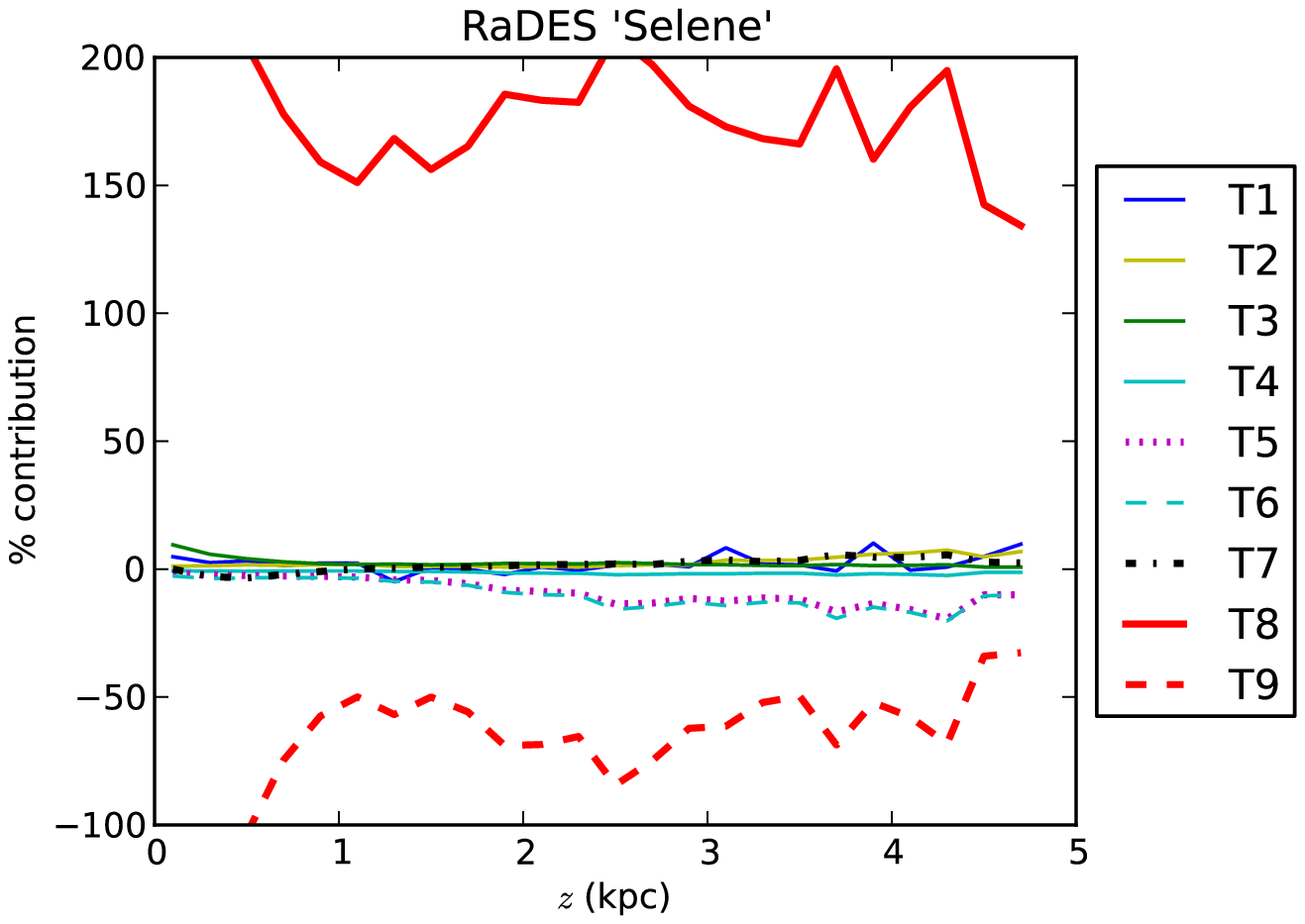} & \includegraphics[width=7.0cm]{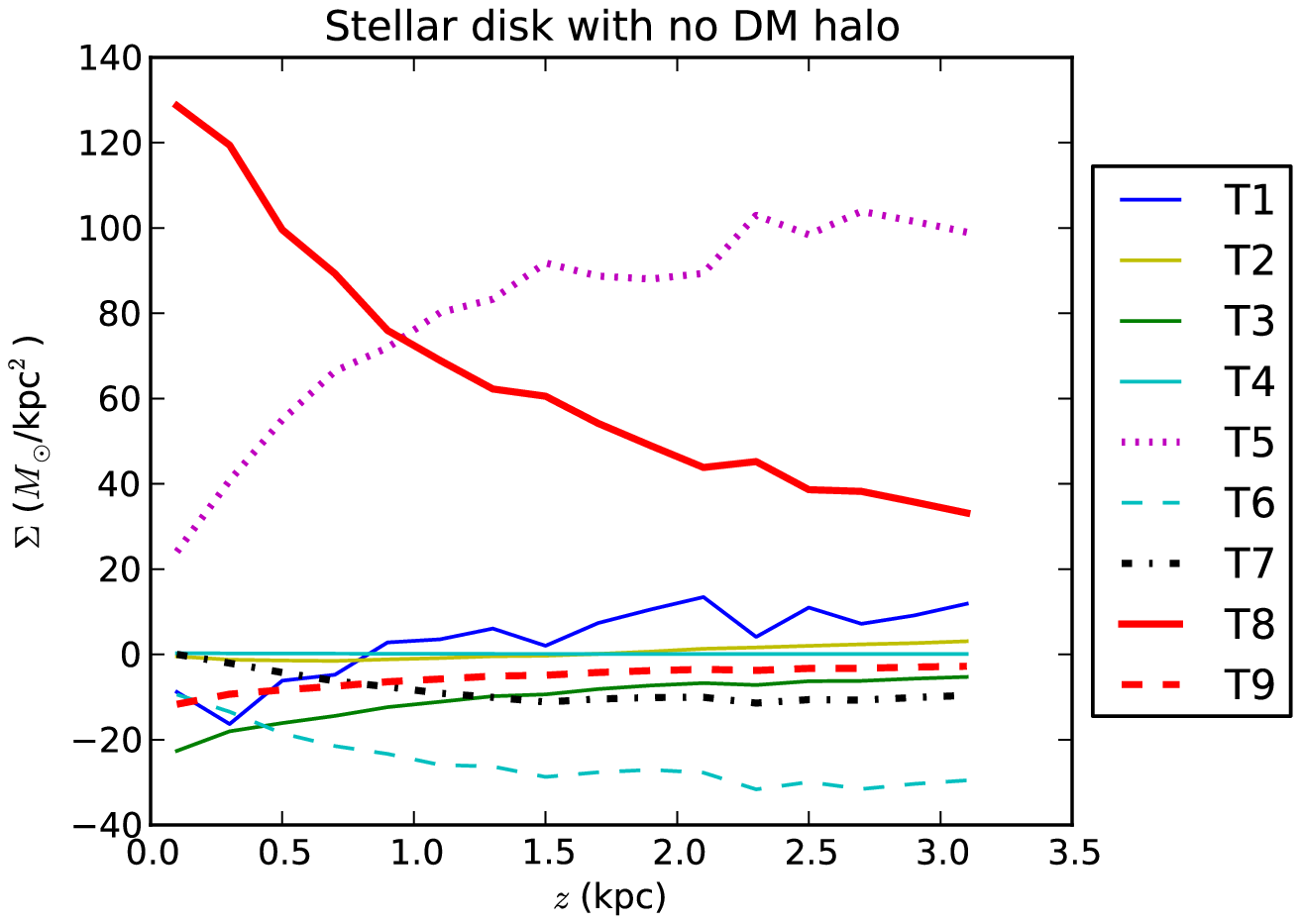} \\
\includegraphics[width=7.0cm]{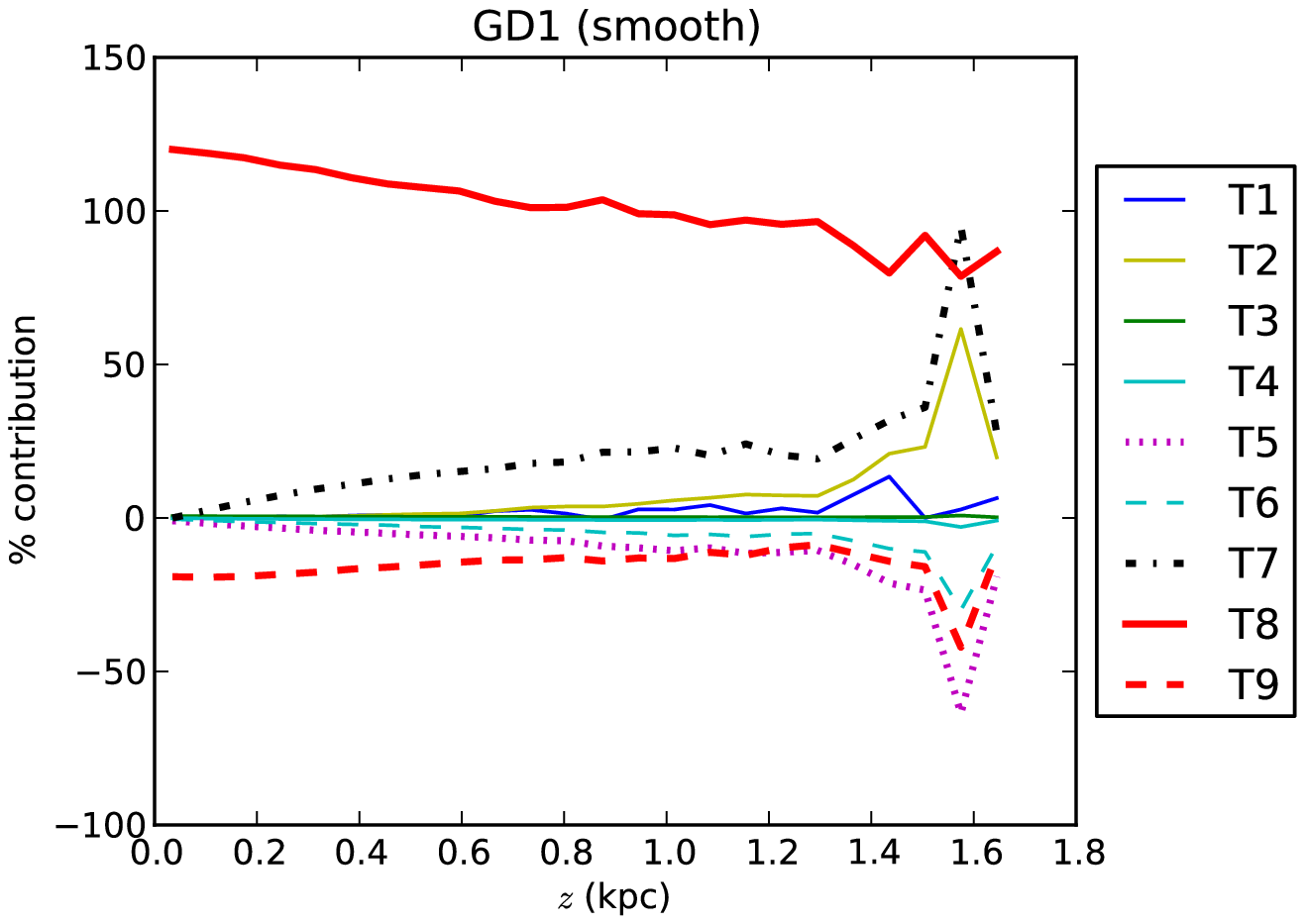} & \includegraphics[width=7.0cm]{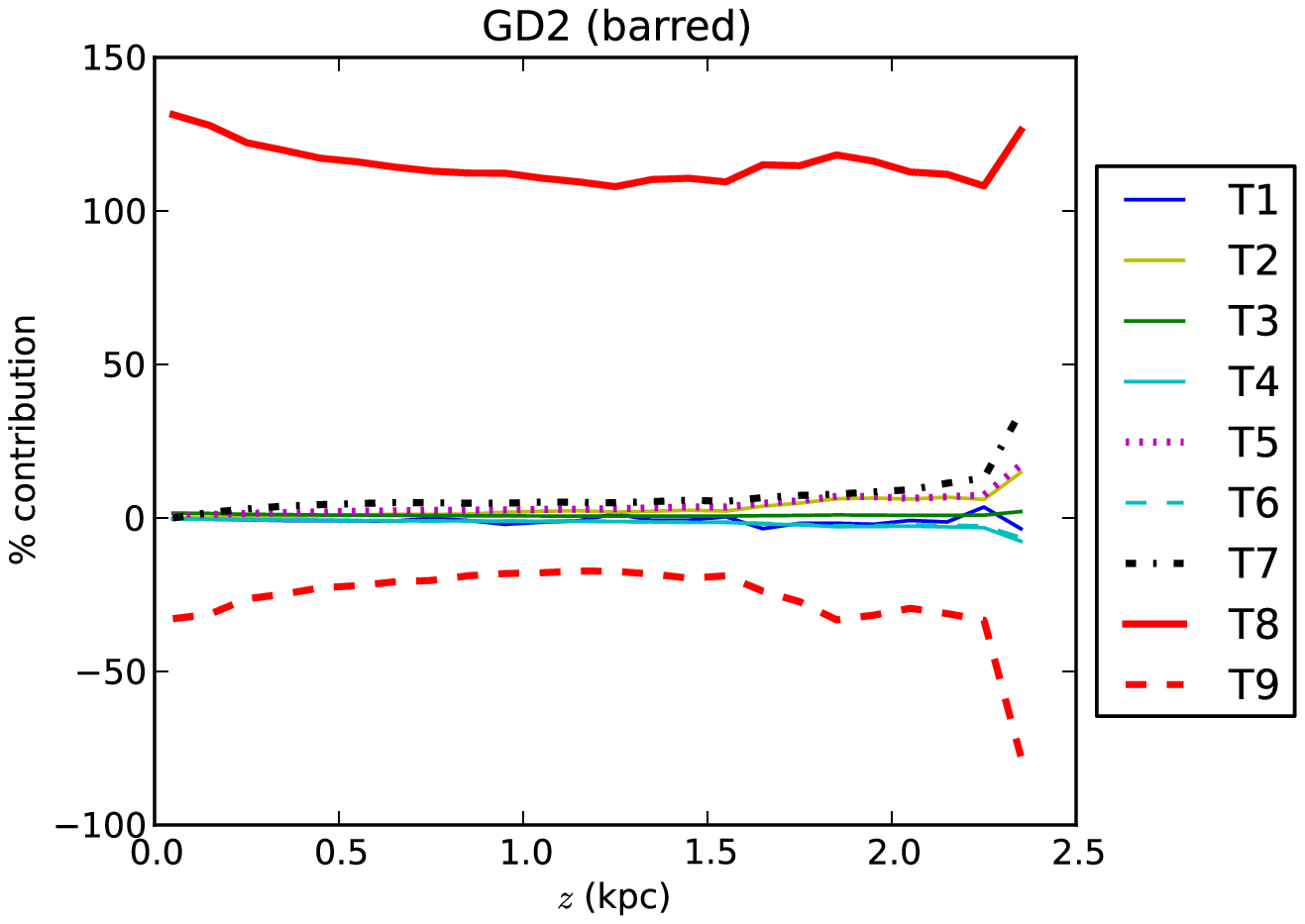}
\end{tabular}
\caption{Contributions of each term to the total surface density calculation, as enumerated in Section \ref{jeansfinal}.}
\label{surfdenscontributions}
\end{figure*}

\begin{figure*}
\centering
\begin{tabular}{cc}
\includegraphics[width=7.0cm]{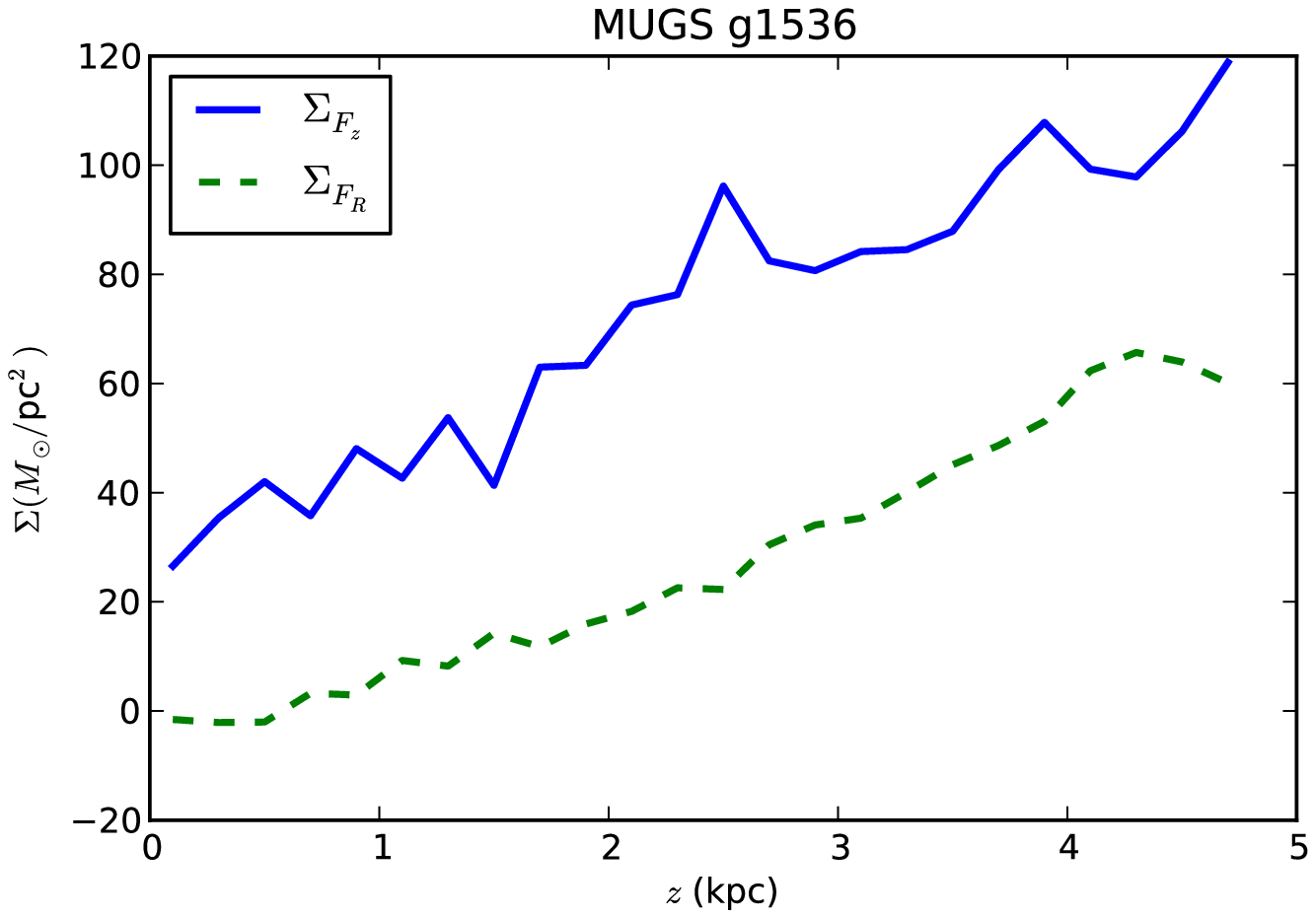} & \includegraphics[width=7.0cm]{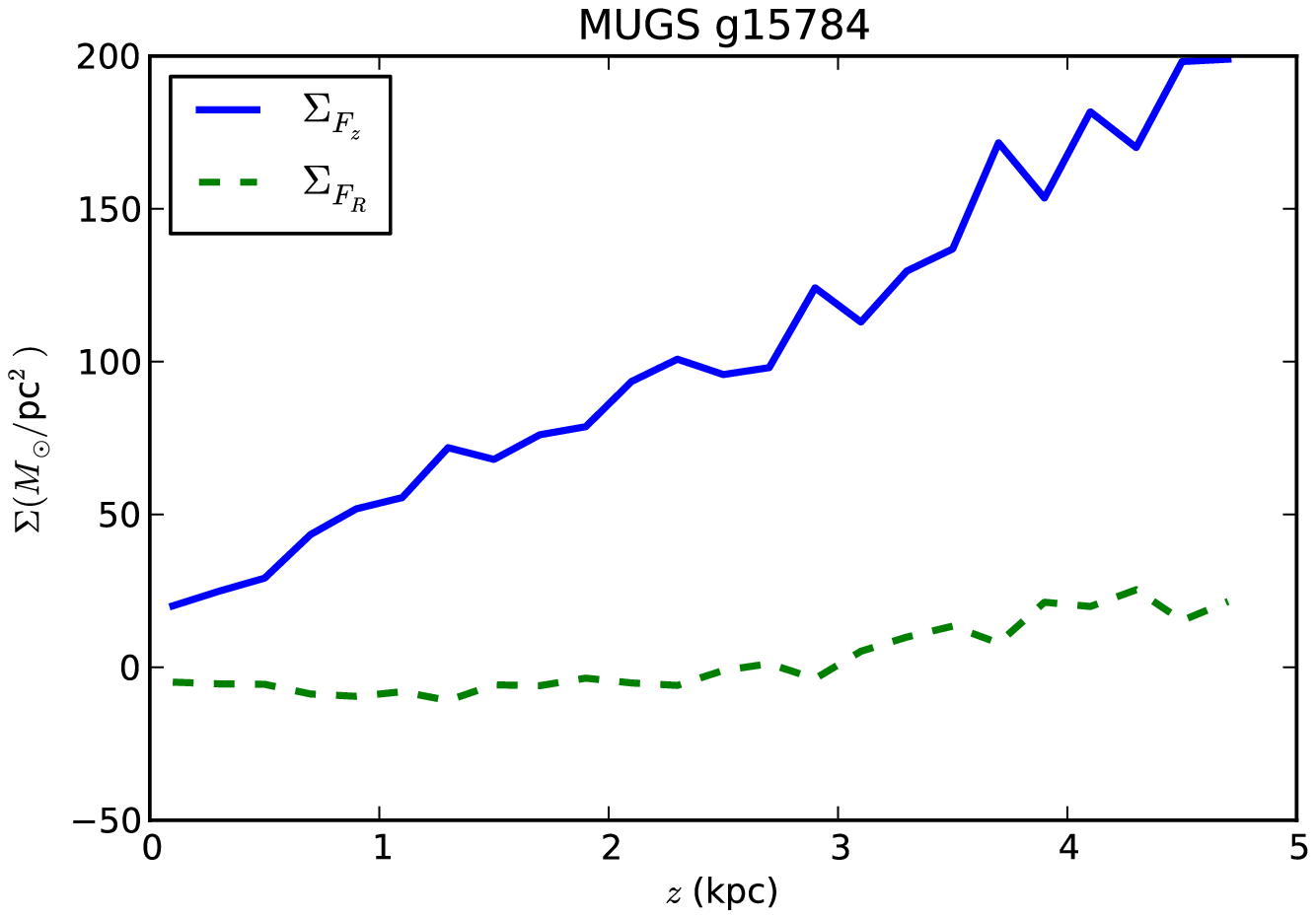} \\
\includegraphics[width=7.0cm]{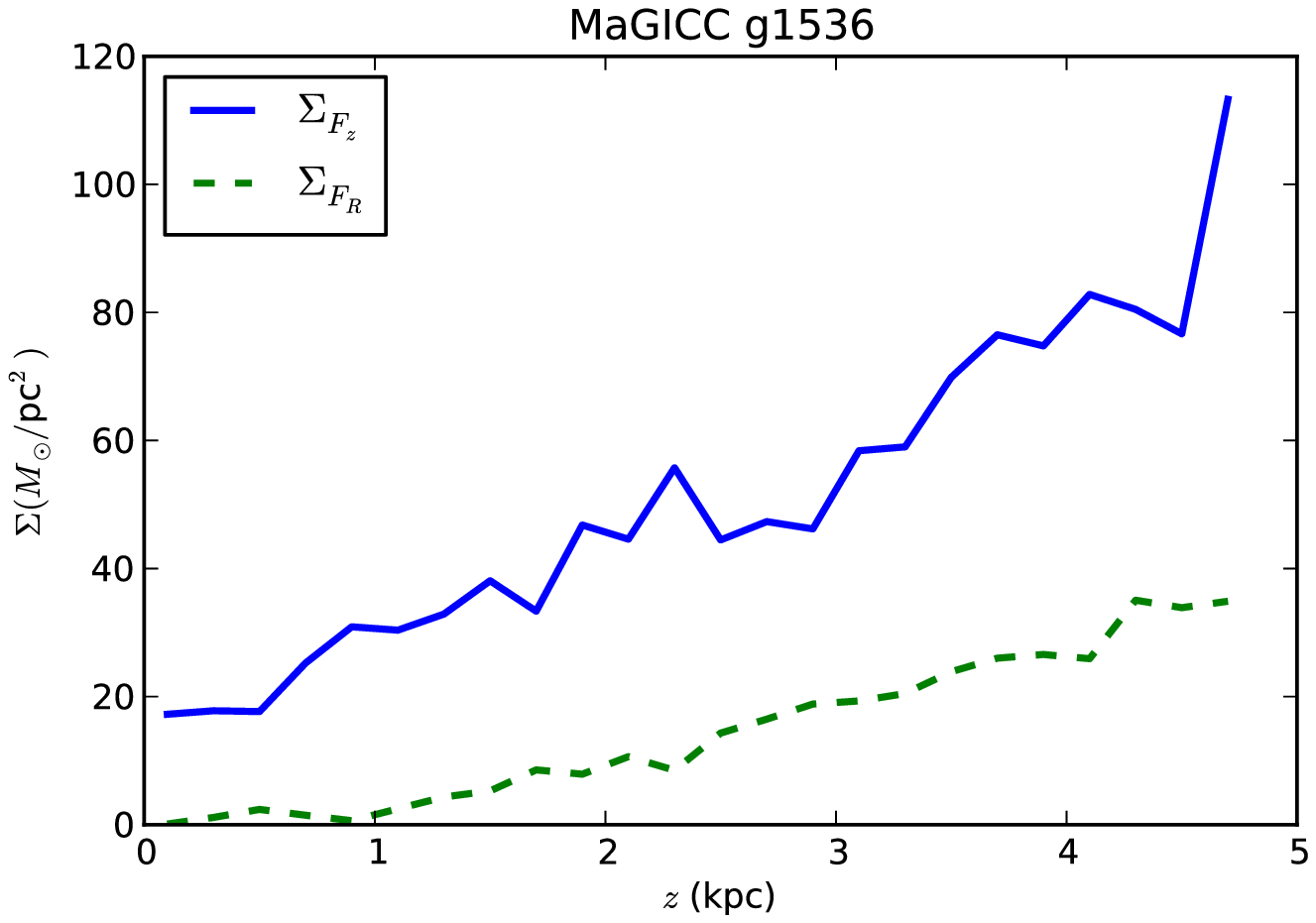} & \includegraphics[width=7.0cm]{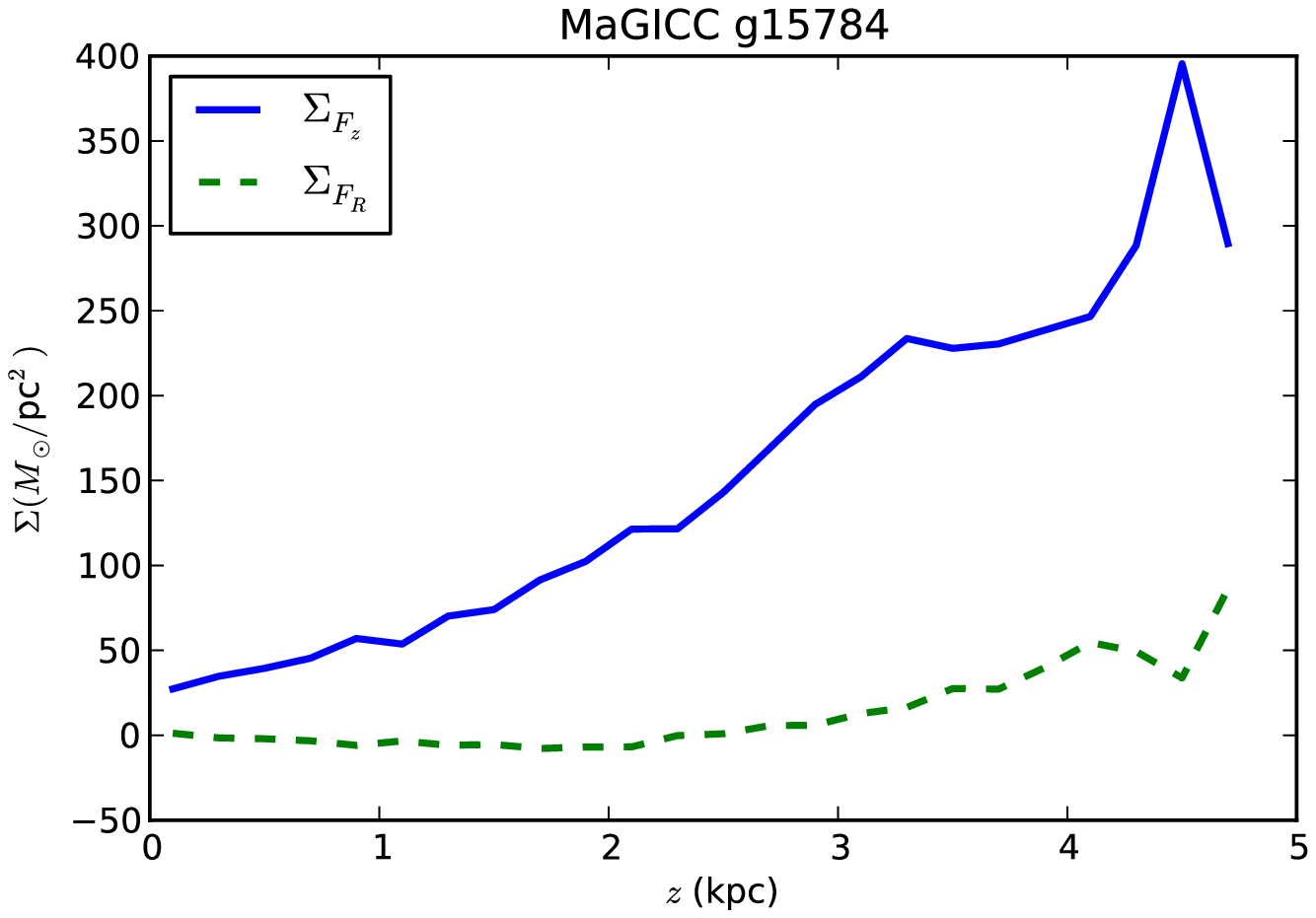} \\
\includegraphics[width=7.0cm]{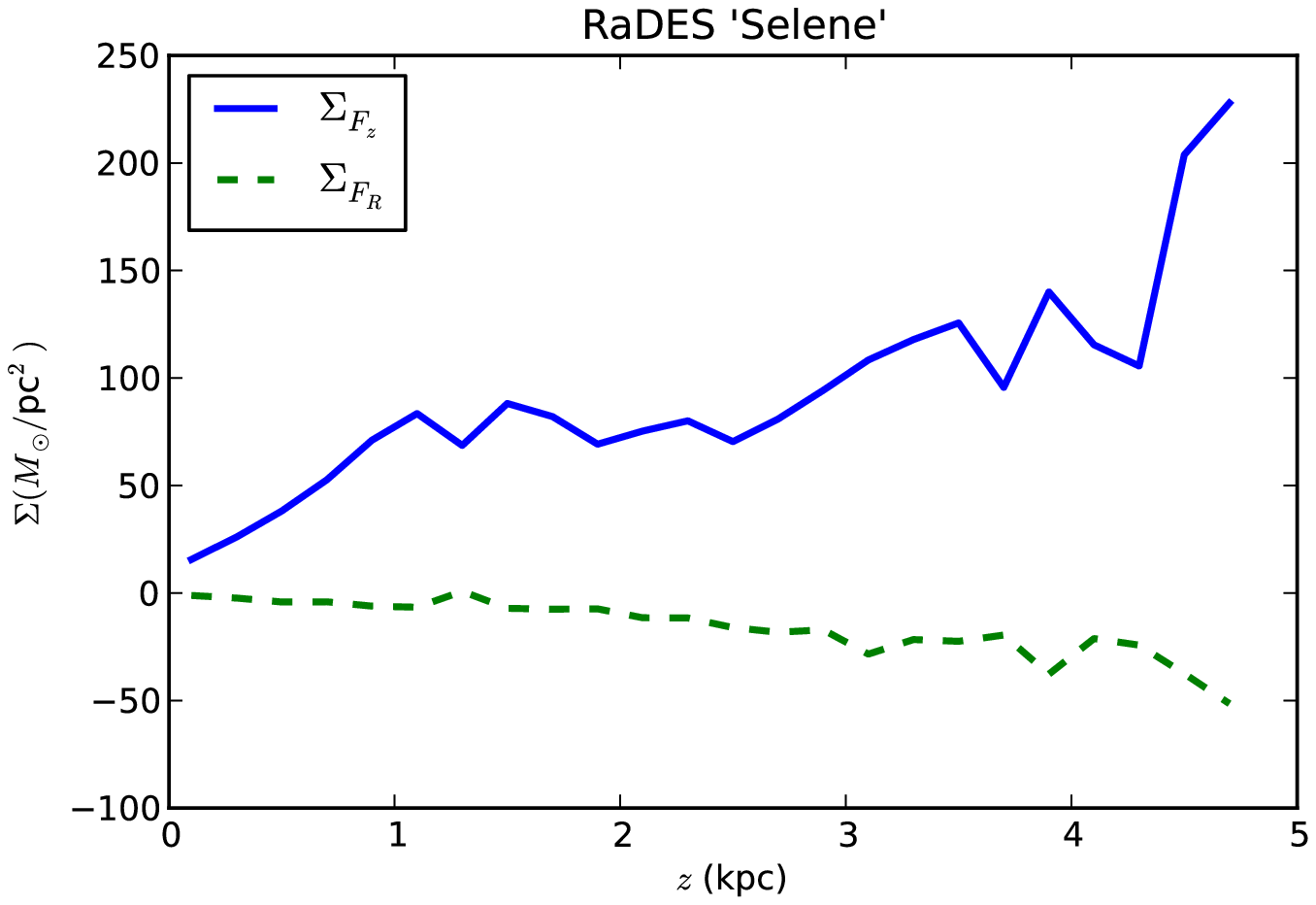} & \includegraphics[width=7.0cm]{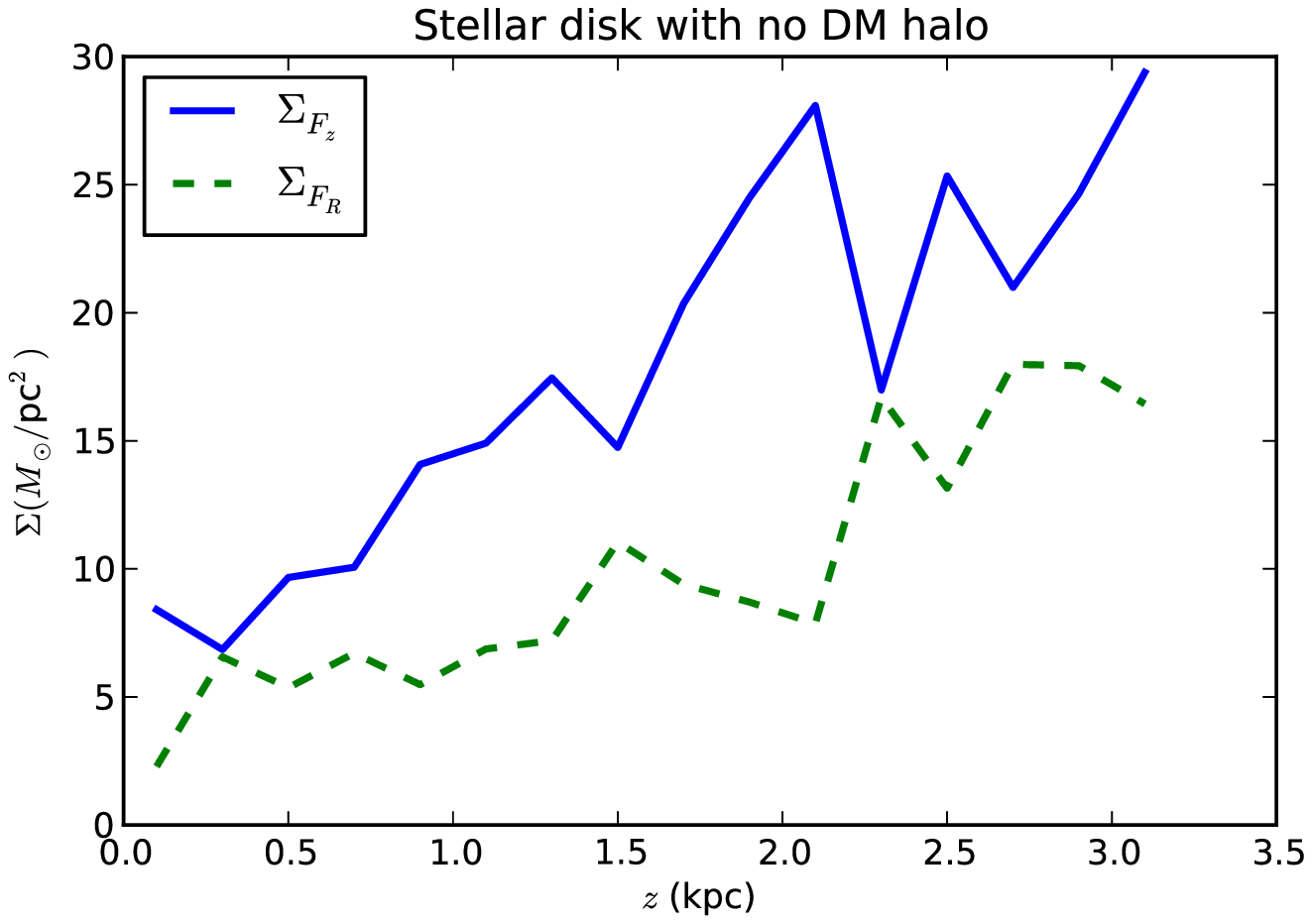} \\
\includegraphics[width=7.0cm]{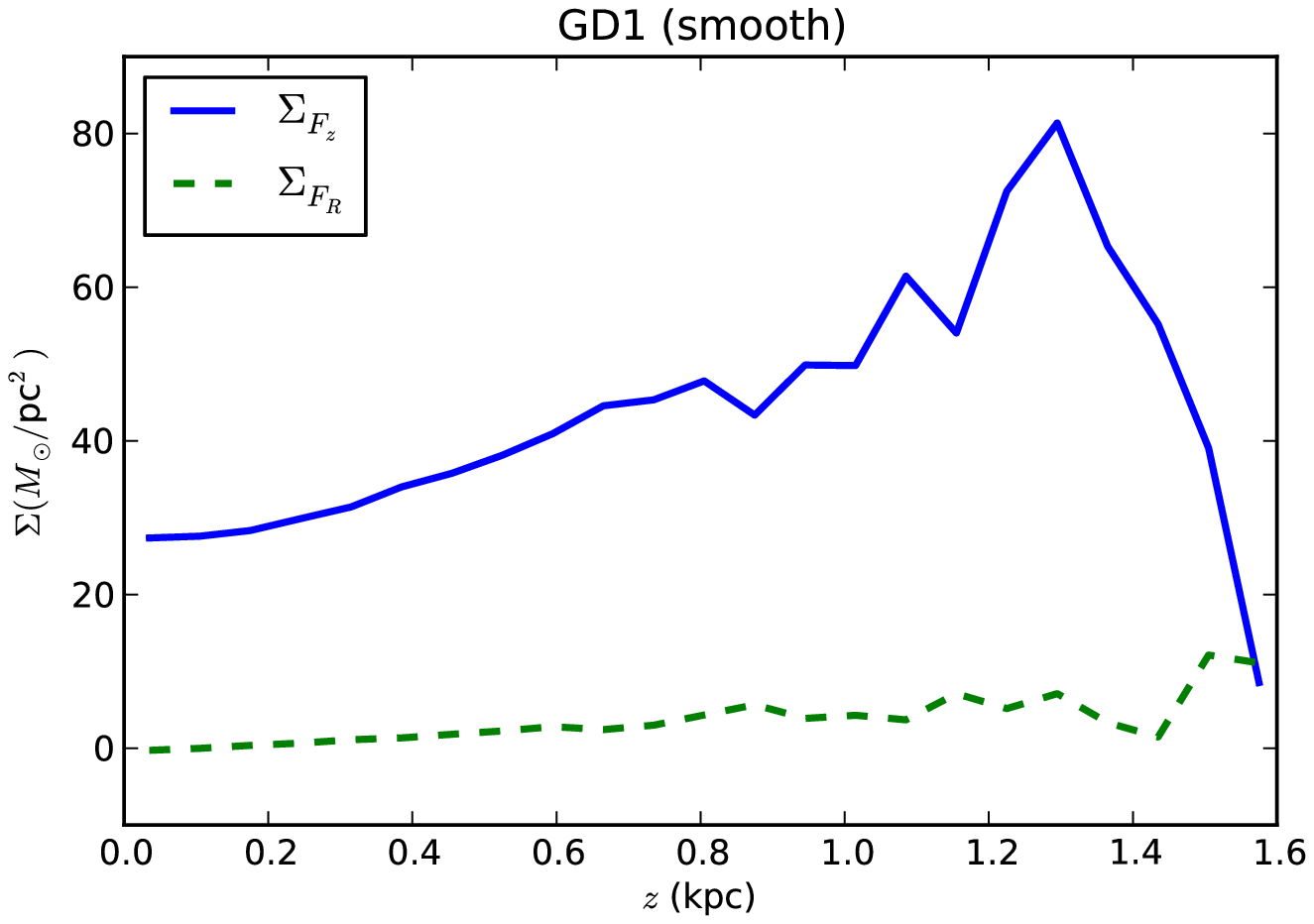} & \includegraphics[width=7.0cm]{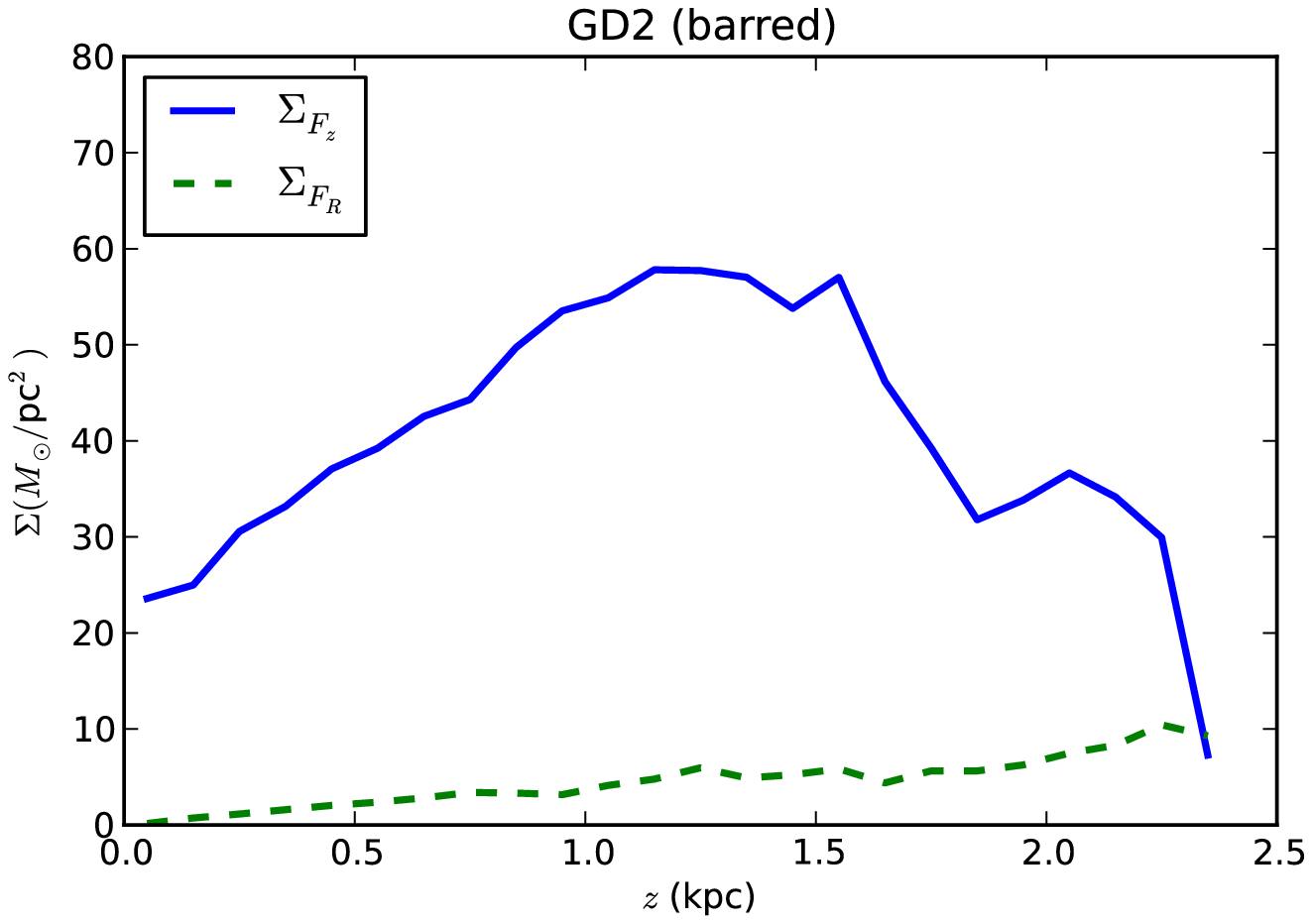}
\end{tabular}
\caption{The radial and vertical force terms of Eq. \ref{poissonvertical} as a function of height $z$, calculated using the stellar kinematics. The large radial terms in the MUGS models and MaGICC g1536 are primarily due to the large measured $h_U$ values relative to the $h_R$ values in these models.}
\label{surfdensforces}
\end{figure*}

\section{Results and analysis}
\label{results}
The fit parameters for each model are given in Table \ref{fitparams}, along with the $R^2$ value of each fit (in parenthesis). It is clear that most fits are good, except for those of the $\overline{UW}$ parameter and some parameters of the model without dark matter. There is substantial scatter around the value $\overline{UW} = 0$ in all of our models, and the fact that the best fit line is close to flat leads to very low $R^2$ values. We have considered variations of this parameter by determining $1\sigma$ values for the slope and checking the effect on the final calculated surface density of using these values, which is negligible (the maximum change at high $z$ is typically less than $10\%$).

The scale height and vertical velocity dispersion fit parameters for the dark matter-free model are also poor fits. Given that the model more closely resembles an isothermal slab model, we should not expect the assumption of an exponential profile in the vertical direction to be a good fit, but we will use this assumption in keeping with the approach employed for all other models. In the case of the vertical velocity dispersion, the flatness of this quantity in height leads to a low $R^2$ value for the fit line.

We also see that several assumptions of MB12 are not respected by our models, most notably the radial scale lengths of the velocity dispersions are all larger than that of the density, i.e. $h_U > h_R$ and $h_V > h_R$ for all models, and the radial dependence of $\overline{UW}$ is weak.

\begin{table*}
\begin{center}
\begin{tabular}{ l c c c c c c c c }
  \hline
  Model & $h_R$ & $h_z$ & $h_U$ & $\alpha_{W}$, $\beta_{W}$ & $h_{\sigma_V}$ & $\alpha_{UW}$ & $C_5 (\times 10^{-2})$ \\
  \hline                       
MUGS g1536 & 2.1 (0.99) & 2.0 (0.98) & 14.6 (0.80) & 6.0, 52.1 (0.93) & 14.4 (0.65) & 54.2 (0.27) & 3.9\\
MUGS g15784 & 2.5 (0.995) & 1.7 (0.97) & 8.5 (0.93) & 11.2, 55.8 (0.98) & 8.3 (0.82) & 93.2 (0.40) & 1.9\\
MaGICC g1536 & 2.5 (0.99) & 1.3 (0.97) & 11.2 (0.87)  & 5.3, 32.3 (0.93) & 6.6  (0.85) & 6.7 (0.02) & 2.9\\
MaGICC g15784 & 2.4 (0.999) & 0.9 (0.99) & 9.2 (0.98)  & 11.0, 36.2 (0.99) & 28.1 (0.65) & -3.7 (0.02) & 2.2\\
RaDES ``Selene'' & 3.2 (0.98) & 1.5 (0.94) & 5.1 (0.91) & 8.5, 51.7 (0.84) & 5.4 (0.83) & -18.6 (0.02) & -1.2\\
GD1 (smooth) & 3.0 (0.999) & 0.2 (0.96) & 3.7 (0.99) & 5.2, 13.3 (0.95) & 3.8 (0.98) & -2.1 (0.15) & -5.4\\
GD2 (barred) & 8.2 (0.97) & 0.3 (0.99) & 26.6 (0.73) & 6.1, 17.4 (0.93) & 10.4 (0.94) & -5.2 (0.35) & 1.9\\
No DM halo & 3.3 (0.98) & 7.8 (0.58) & 13.5 (0.79) & 0.3, 56.3 (0.05) & 5.2 (0.93) & 32.7 (0.02) & 2.8\\
  \hline
\end{tabular}
\end{center}
\caption{Fit parameters used in evaluation of the Jeans equation (with $R^2$ value for fit). The slope and intercept of the linear fit to $\sigma_W$ are $\alpha_W$ and $\beta_W$ respectively, and the slope of the linear fit to $\overline{UW}$ is $\alpha_{UW}$. The final column is the value of the coefficient to the integral in term (v), discussed in Section \ref{subsec:radvel}}
\label{fitparams}
\end{table*}

\subsection{Calculated surface density}
The calculated surface density is plotted along with the actual measured surface density in Fig.~\ref{surfdensresults}. The stellar surface density (and gas if present) is also shown. In addition we show the one-dimensional calculation of the surface density using only the vertical terms (viii) and (ix), as well as the full vertical force of Eq.~\ref{verticalforceterm} which includes the dependence on $\overline{UW}$, as utilised in BT12. For the cosmological models and the GD1 and GD2 thin disk models we show the effect of imposing the condition $h_U = h_R$ on the resulting calculated surface density, to be discussed further in Section~\ref{subsec:radvel}.

Due to the noise in the calculated curves, it is difficult to ascertain how well the analysis recovers the true surface density of the model. Therefore we have chosen to fit the following function to the $\Sigma^C$, $\Sigma^C_{Vert}$ and $\Sigma^C_{F_z}$ lines of Fig.~\ref{surfdensresults}:
\begin{equation}
\label{fittosurf}
f(z) = a \sqrt{z} + b
\end{equation}
where $a, b$ are adjustable parameters. This function, although an arbitrary choice, has the general behaviour expected of the surface density as a function of $z$: it bends down for small values of $z$ (although we allow for a vertical offset by including a non-zero intercept, $b$) and it flattens out for large values of $z$. To account for the noise arising due to low number statistics at high heights, we only choose to fit the function using $z$-bins for which the particle number is at least $>1\%$ of the total particle number in a vertical slice at the radius $R=8.2$~kpc (i.e. the sum of particle numbers in all $z$-bins at that radius). We will refer to this as our ``particle number'' condition. These fit lines are plotted in Fig.~\ref{surfdensresults} as a dotted line for $\Sigma^C$, a thin dashed purple line for $\Sigma^C_{Vert}$ and a thick dashed black line for $\Sigma^C_{F_z}$. The $1\sigma$ range on the fit parameters $a, b$ for $\Sigma^C$ are shown with the light grey bands. We label the fit line to $\Sigma^C$ as $\Sigma^C_{Fit}$ to distinguish the two quantities.

Our function fit may be interpreted as an indication of how high above the disk plane the method may be considered reliable in each model. For the case of the MaGICC models, the particle number condition includes all $z$-bins up to $2.5h_z$ ($3.3$~kpc) for MaGICC 1536, and $3h_z$ ($2.7$~kpc) for MaGICC 15784. The fit line shows a good match to the measured surface density, including when extrapolated beyond the region of the fit itself. Similarly the RaDES fit is within $1\sigma$ of the true surface density, using data up to $2.7h_z$ ($4.1$~kpc) in this case. The thin GD1 and GD2 disks also show reasonable fits, again extrapolating beyond the actual fit regions, which are all $z$-bins within $3h_z$ ($0.6$~kpc) and $2.9h_z$ ($0.9$~kpc) respectively. The result for the model with no DM halo is rather unusual, owing to the near-isothermal distribution of the particles, leading to an almost linear surface density curve. As such our choice of function is inappropriate in this case. Nevertheless the fit corresponds roughly to the trend of the true surface density, particularly at higher heights. The MUGS galaxies exhibit a more substantial overestimate at higher heights than in the other models, and so our fits must be constrained to no more than $h_z$ ($2.0$~kpc) and $1.5h_z$ ($2.6$~kpc) for the g1536 and g15784 models respectively, well below the height corresponding to our particle number condition. As the MaGICC models use a higher density threshold for star formation, this results in stars being born out of kinematically colder gas, which may contribute to the differences seen here. Furthermore, the halos of the MUGS galaxies have more substructure than the MaGICC galaxies, as the stronger feedback in the latter models decreases the number of satellites significantly. This additional substructure may lead to more external kinematic heating, and therefore less equilibriated disks. The assumption of a steady-state equilibrium test stellar population is of course central to the entire analysis, and so any sources of kinematic heating will introduce error. Further investigation of such effects is, however, beyond the scope of this paper.

\subsection{Relative importance of the terms in the Jeans equation}
\label{relimport}
The contribution of the various terms listed in Section \ref{listofterms} (multiplied by $1/(2\pi G)$) is plotted in Fig.~\ref{surfdenscontributions} as a function of $z$, taking into account the overall sign of the term. This is calculated by dividing each term by the total calculated surface density. Due to the presence of negative contributions this leads to some terms contributing more than $100\%$ of the total surface density at some heights, with the total contribution at each height summed over all terms being $100\%$. In all cases the vertical velocity component contributes by far the largest positive percentage of the total surface density, while the slope of the vertical velocity in $z$, term (ix), generally makes the largest negative contribution. In the MUGS models term (v) makes an appreciable positive contribution at higher $z$, and an especially large positive contribution in the dark matter-free model. It is also comparable to the (minor) contribution of term (vii), the mean azimuthal velocity term, in MaGICC g1536. In general, term (vii) is of only minor significance in all models, except the GD1 thin disk.

From the plots in Fig.~\ref{surfdenscontributions} we see that terms (i), (ii), (iii), (iv) and (vi) generally make only small contributions to the surface density calculation, increasing slightly for increasing $z$, but still contributing less than $\sim 10\%$ of the total at high $z$. This shows that for higher $z$ the vertical-only and full three-dimensional treatments do tend to diverge, as can be seen in the calculated surface densities of Fig.~\ref{surfdensresults}. As noted earlier, terms (iii) and (iv) do not appear in the analysis of MB12 as they were incorrectly neglected in the derivation. These terms use the radial behaviour of $\overline{UW}$, which is very close to flat for all of our models, thus these terms provide a negligible contribution to the surface density and may, in fact, be safely ignored. Term (vi) involves the radial behaviour of the azimuthal velocity dispersion. Again, for all models, this term is small, partly due to the fact that the radial scale length of $\sigma_V$ is rather large, sometimes as much as $\sim 10$ times larger than the radial scale length of the density. Imposing the $h_V = h_R$ condition reduces the surface density, by as much as $\sim 30\%$ at high $z$, and so this parameter must be reasonably well constrained for the analysis. The effect is less significant than for the $h_U$ parameter, however, so we will not focus on this quantity here.

The last two terms, (viii) and (ix), are those depending on the vertical velocity dispersion. In all models these are the dominant terms, particularly term (viii), as may be seen in Fig. \ref{surfdenscontributions}. The slope of $\sigma_w(z)$ controls the contribution of term (ix), and indeed one aspect of this analysis is that the vertical force does not vanish in the plane of the disk, due to the linear fit to $\sigma_w$ being a poor approximation at low $z$. This results in overestimates of the surface density for most models at very low $z$. 

This would suggest that, apart from the vertical terms (viii and ix), the radial velocity dispersion term (v) and the mean azimuthal velocity term (vi), all the other terms in the full three-dimensional approach are of minor significance. Term (vii) is obviously only of particular relevance in the case of a dark matter-dominated disk with an associated increasing rotation curve (as for the GD1 model), or a massive dark matter-free disk with a steeply decreasing rotation curve, neither of which appear to be true for the Milky Way. Term (v) is of significance only because of the sensitivity on $h_U$ (and $h_R$) as discussed in detail in Section \ref{subsec:radvel}.

\subsection{Comparison of the one-dimensional and three-dimensional approaches}
\label{comparison1d3d}
The vertical-only and complete vertical force formulations (i.e. $\Sigma^C_{Vert}$ and $\Sigma^C_{F_z}$) are very similar in most models, due to the negligible contribution of the `tilt' term, as we have already seen. The only model showing a significant disagreement between the two is the dark matter-free model, where the contribution of the first term in Eq.~\ref{verticalforceterm} is sufficiently large to lead to notable differences in the fit lines for the ``vertical'' formulations of the surface density.

Therefore, given our results in Section \ref{relimport} for the relative contributions of all terms, we see that in general the vertical formulations are competitive with the full three dimensional approach. There is, however, some disagreement for the MaGICC g1536 model, and significant differences for the dark matter-free model. We can understand these results when we look at the contributions from the various terms shown in Fig.~\ref{surfdenscontributions}. The two MUGS models and MaGICC g1536 have somewhat larger contributions arising from term (v), which depends on the radial velocity dispersion. In all these models this term contributes more at higher heights, with a particularly large contribution in MUGS g1536. This term therefore lifts the $\Sigma^C$ curve relative to the others. Note that the vertical-only and full formulation fit lines for the MUGS models, using data within approximately one scale height, do not differ a great deal, showing that the radial velocity dispersion is only relevant at higher heights. The large deviation of the $\Sigma^C$ line for $z > 2.5$~kpc in the MaGICC g15784 model is likely due to low number statistics (recall that our particle number condition has been applied to the fit lines). The importance of obtaining accurate measurements of the parameters that enter term (v) (i.e. $h_R$ and $h_U$) will be discussed in Section~\ref{subsec:radvel}.

The reasonable performance of the one dimensional formulations may also be seen in Fig.~\ref{sigmaw} where the thick green line is the vertical velocity dispersion derived from the true surface density under the assumption that only term (viii) in Eq.~\ref{jeansfinal} contributes to the surface density. This is an even less accurate approximation than the ``vertical-only'' formulation we have used, as we are now neglecting the derivative term (ix), which will act as a constant negative offset to the surface density in the case of a positive linear slope to the vertical velocity dispersion, as we have assumed earlier and is the case in almost all models. The thick green lines in Fig.~\ref{sigmaw} roughly correspond at least to the slope of the measured vertical velocity dispersion for almost all models. In the case of the stellar disk with no dark matter, our reconstruction of the velocity dispersion using only term (viii) leads to a vertically increasing $\sigma_w$, rather than the flat velocity dispersion measured in the model. This is because a vertically isothermal distribution implies a density distribution that goes like $\sech^2(z)$ rather than an exponential. Thus the assumption of an exponential distribution, which leads to the expression given in Eq.~\ref{jeansfinal} for term (viii), is a poor approximation in this case. Therefore, simply inverting Eq.~\ref{poissonvertical}, by using $F_z = -(1/h_z)\overline{W^2}$, to find $\sigma_w$ from the surface density leads to an erroneous result.


To compare with the discussion in Section \ref{analytic} we show the radial and vertical terms of Eq. \ref{poissonvertical} calculated for each model using the stellar kinematics in Fig. \ref{surfdensforces}. Almost all models exhibit a clearly dominant vertical term and flat or positive lines for the radial force term. The only model with an appreciable negative radial force term is the RaDES `Selene' model. For this model the fourth term in the curly brackets of Eq. \ref{radialforceterm}, involving an integral over $z$ of $\overline{UW}$, is the only significant positive contribution to the total radial force term, with all other terms (except for the radial derivatives of $\overline{UW}$ which are negligible) being negative contributions of similar magnitude. The large $h_U$ values for the MUGS models (especially g1536), MaGICC g1536 and the dark matter-free models lead to significant positive radial force terms. It is important to note, however, that although the GD2 model has an even larger value of $h_U$, the $h_R$ value is also large, and so the contribution of term (v) to the radial force is smaller than for the MUGS models.

The subdominance of the radial force term for almost all models is consistent with the demonstrated efficacy of the vertical-only formulation, especially for heights of order the scale height or lower. Furthermore, in all cases except RaDES `Selene' the contribution from the radial force term is positive, so the vertical-only treatment provides an underestimate of the surface density, as is also seen in Fig.~\ref{surfdensresults}.

Therefore, in almost all cases studied here, the one-dimensional approach is a reasonable approximation, especially at heights below the thick disk scale height. Far from the disk plane the effects of the additional terms can indeed become important, as demonstrated by the disagreements between the vertical-only and three-dimensional formulations at heights of order the vertical scale height in most models (see Fig. \ref{surfdensresults}), primarily due to the effect of term (v), that depends on the radial velocity dispersion. One must, however, keep in mind that low number statistics introduces noise at very high heights above the disk. The analytic results in Section \ref{analytic} are also supportive of the general applicability of the one-dimensional approach at heights up to the thick disk scale height, and even somewhat beyond.

\begin{figure*}
\centering
\begin{tabular}{cc}
\includegraphics[width=7.0cm]{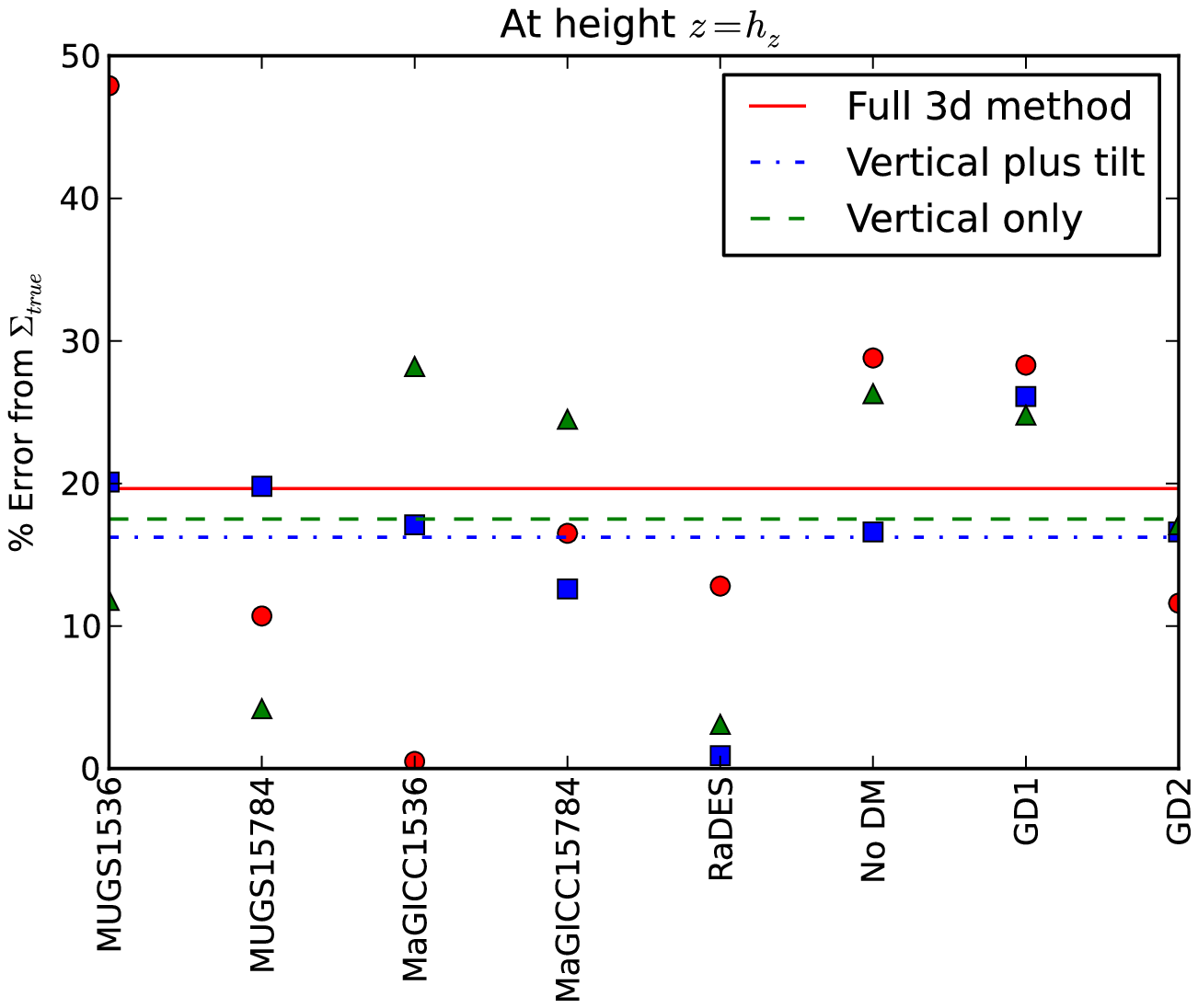} & \includegraphics[width=7.0cm]{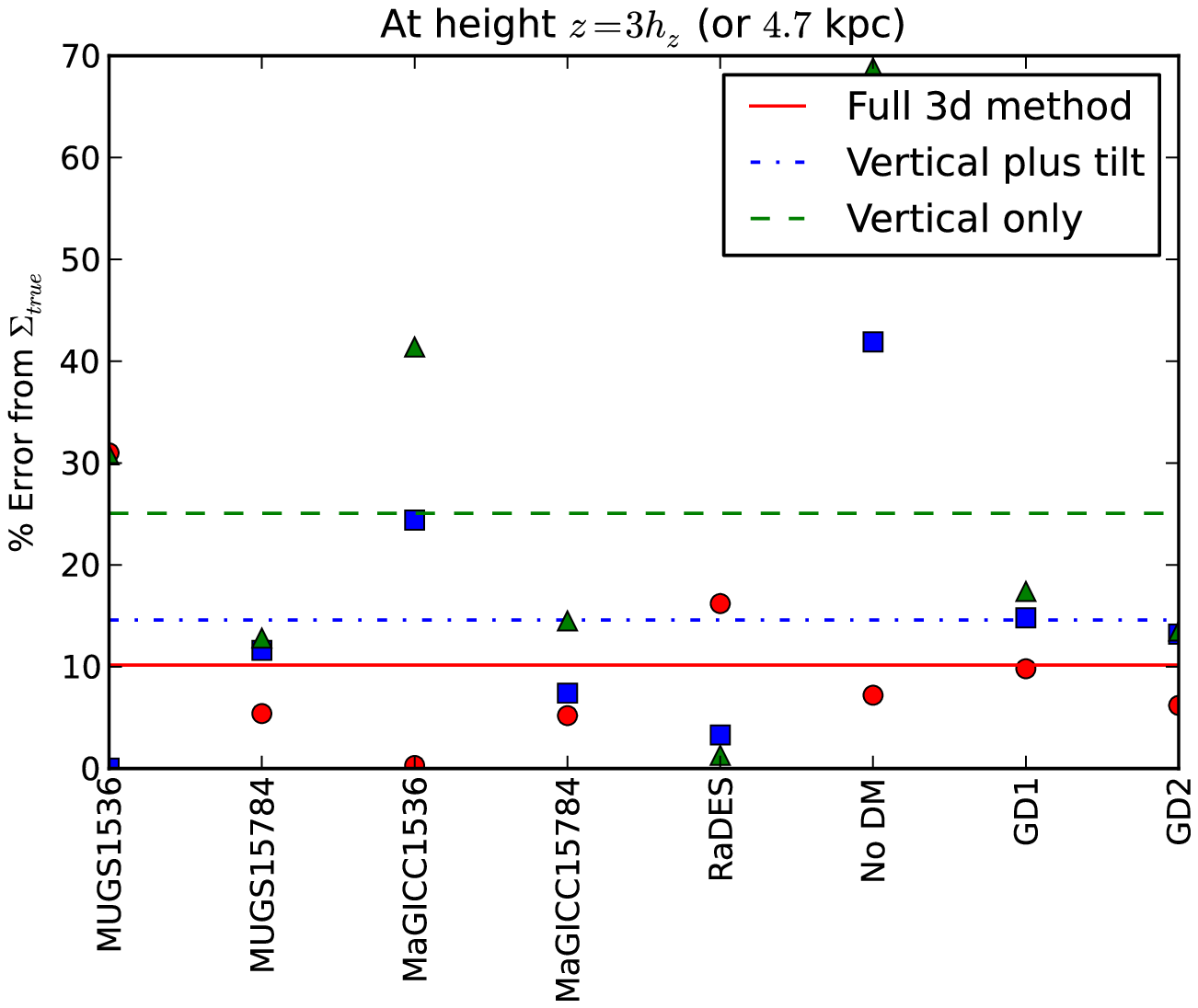}
\end{tabular}
\caption{The left panel shows the percentage error of the calculated surface densities at a height of $h_z$, i.e. the vertical scale height, and the right panel shows the same at a height of $3h_z$. The dashed horizontal lines are the mean percentage errors for each formulation across all models (triangles: vertical only; squares: vertical plus tilt; circles: full 3d).}
\label{comparisonmodels}
\end{figure*}

We now combine the results across all models to try to find a consistent picture of which approach is better. To do so, we calculate the percentage error of the calculated surface densities using the full three dimensional equation, the vertical force only (including the ``tilt'' term), and the vertical-only formulation (no ``tilt'' term): $\Sigma^C$, $\Sigma^C_{F_z}$ and $\Sigma^C_{Vert}$. The error is calculated with respect to the true surface density in the model $\Sigma_{true}$, for example
\begin{equation}
\Delta \Sigma = |\Sigma^C - \Sigma_{true}|/\Sigma_{true}.
\end{equation}
We perform this calculation at two heights: one scale height above the disk (i.e. $h_z$) and three scale heights above the disk (i.e. $3h_z$). For those models for which $3h_z$ is beyond $4.7$~kpc we choose the second height to be $4.7$~kpc. For the model without DM we choose the two heights as $1.1$~kpc and $4.7$~kpc due to the very large value of $h_z$ in this case. The results are shown in Fig.~\ref{comparisonmodels}, where it is clear that there is considerable variation across models. In Fig.~\ref{comparisonmodels} we also show the mean percentage error of each formulation as a horizontal dashed line. Although the differences are rather small, we can nevertheless see that the vertical force formulation, $\Sigma^C_{F_z}$ performs the best at one scale height above the disk, while the full three dimensional formulation performs better at three scale heights above the disk. This result, while not robust given the number of models considered, is expected: the one dimensional formulation provides reasonably good results when considering the kinematics of stars that are close to the disk plane, while the corrections from the additional terms in the three dimensional formulation are needed for data at higher heights. Such measurements are preferred for constraining the dark matter density as the halo becomes increasingly dominant at higher heights.


\subsection{Azimuthal dependence and low number statistics}
\label{azimuth}
\begin{figure*}
\centering
\begin{tabular}{cc}
\includegraphics[width=7.0cm]{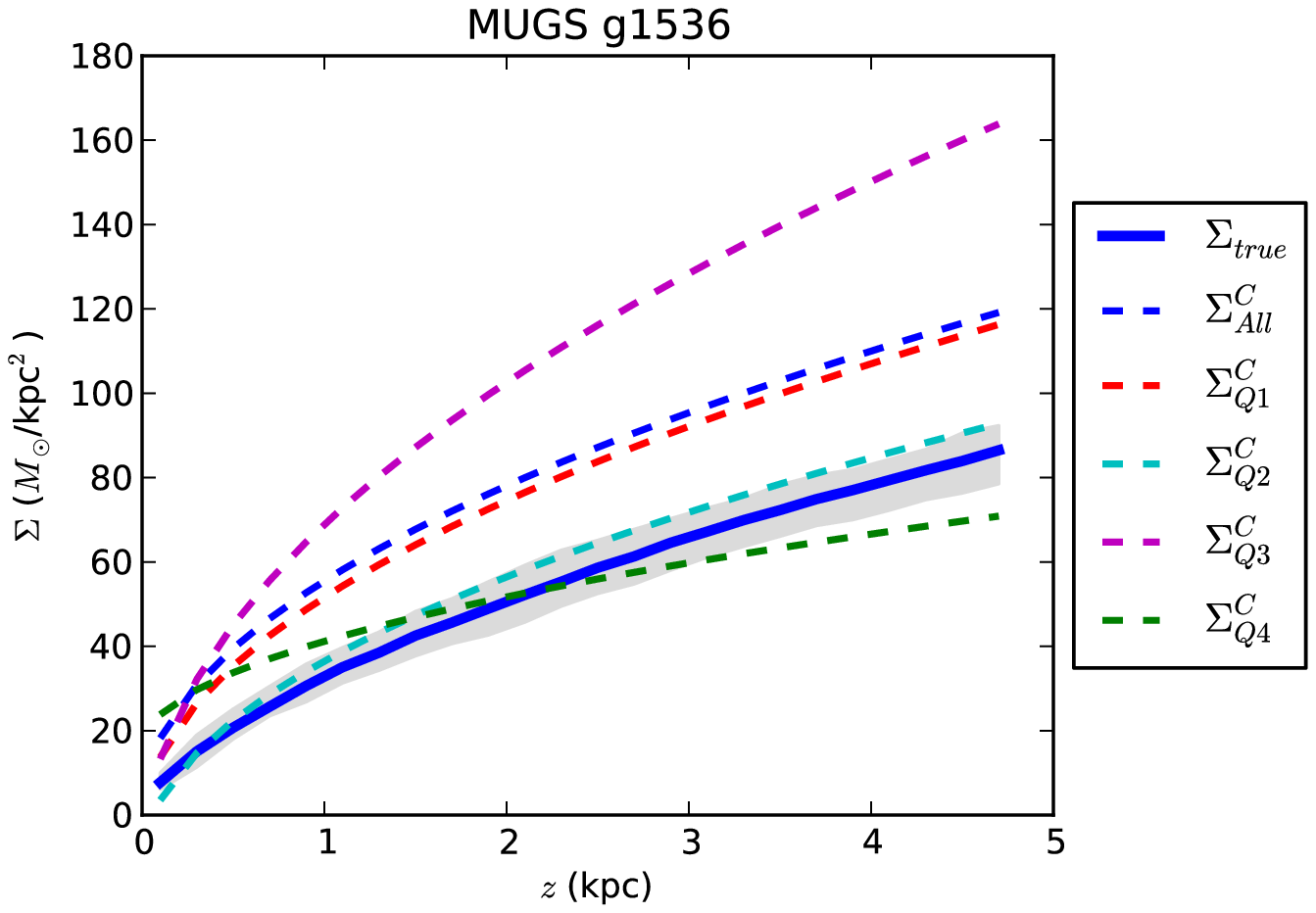} & \includegraphics[width=7.0cm]{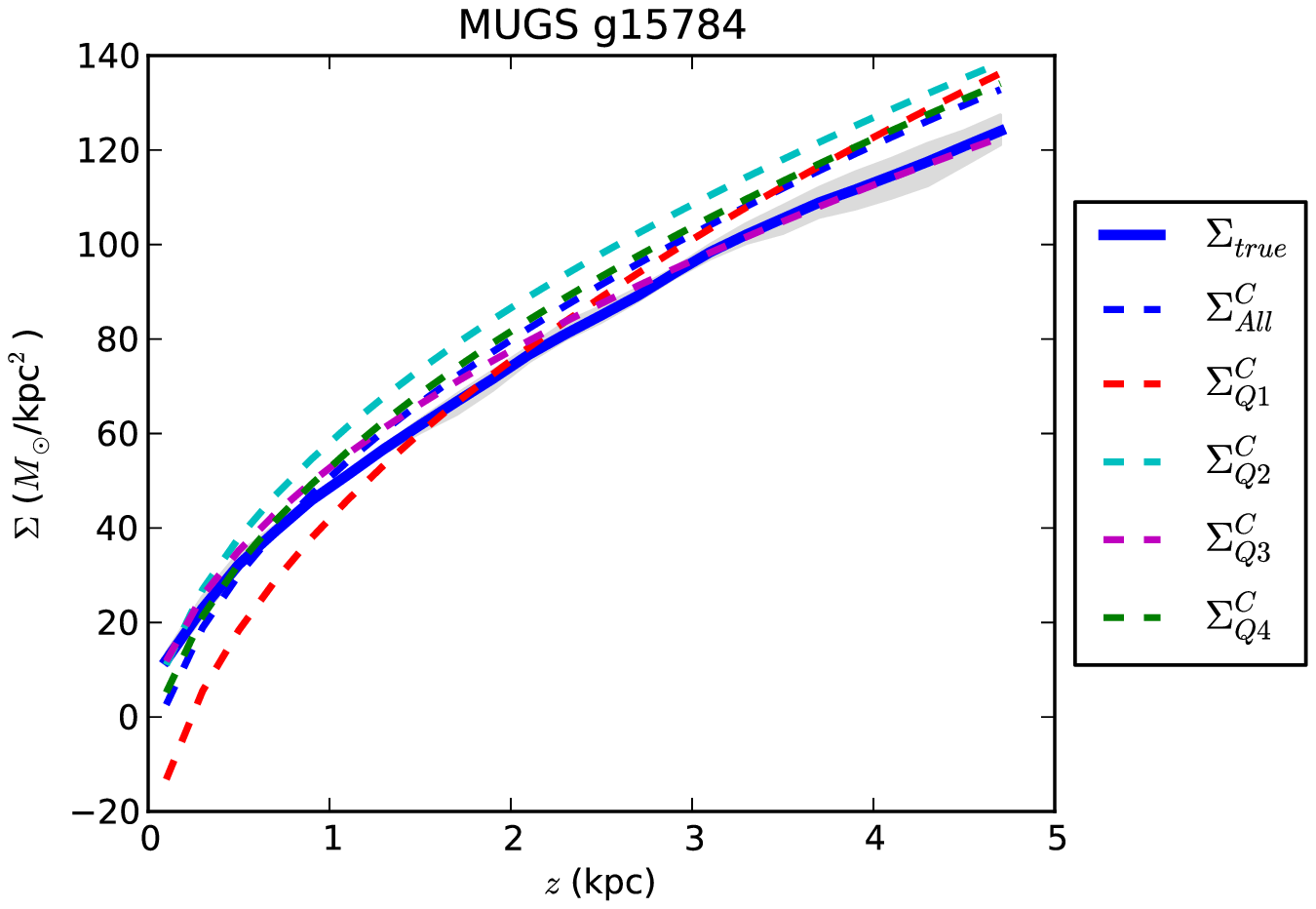} \\
\includegraphics[width=7.0cm]{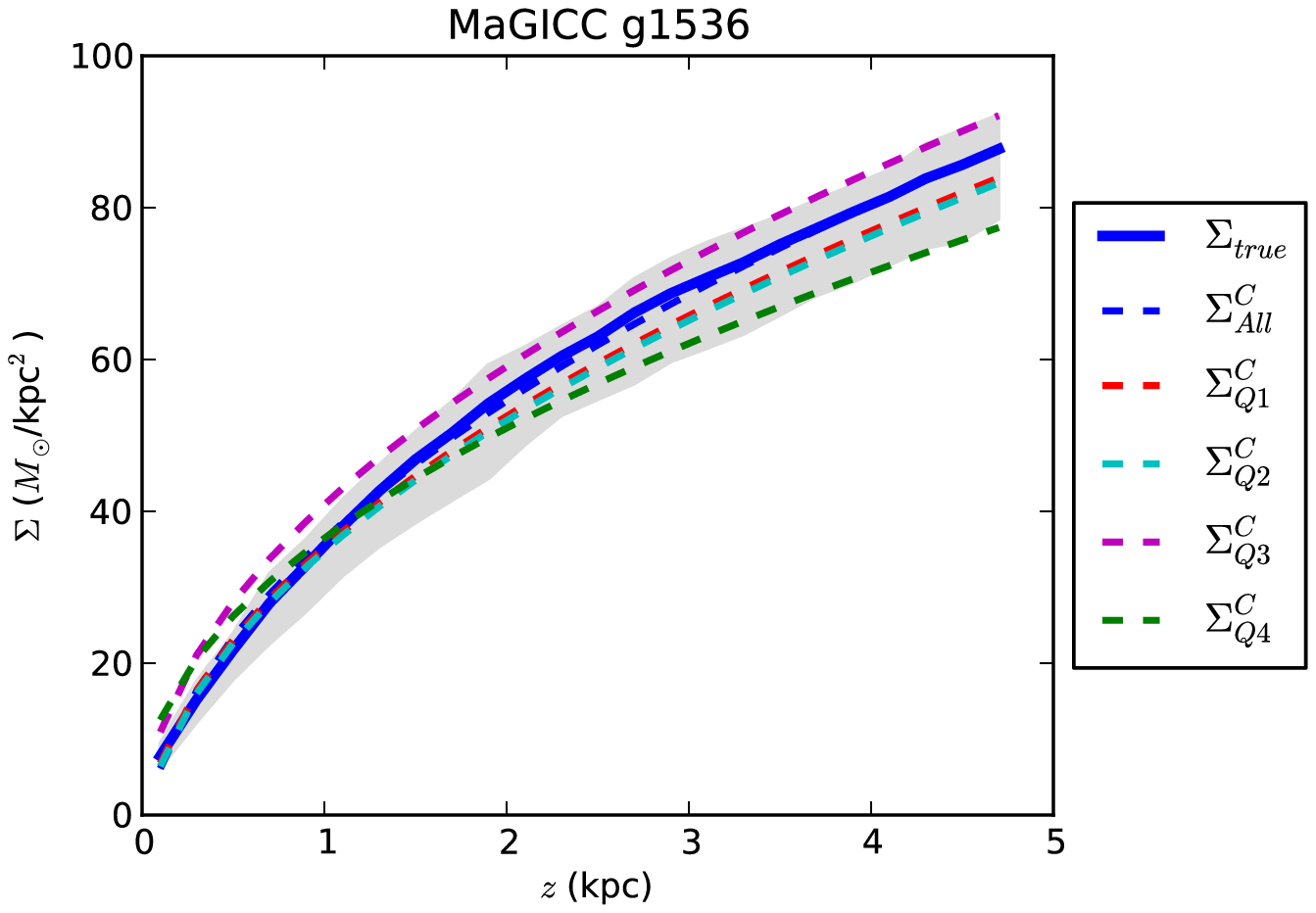} & \includegraphics[width=7.0cm]{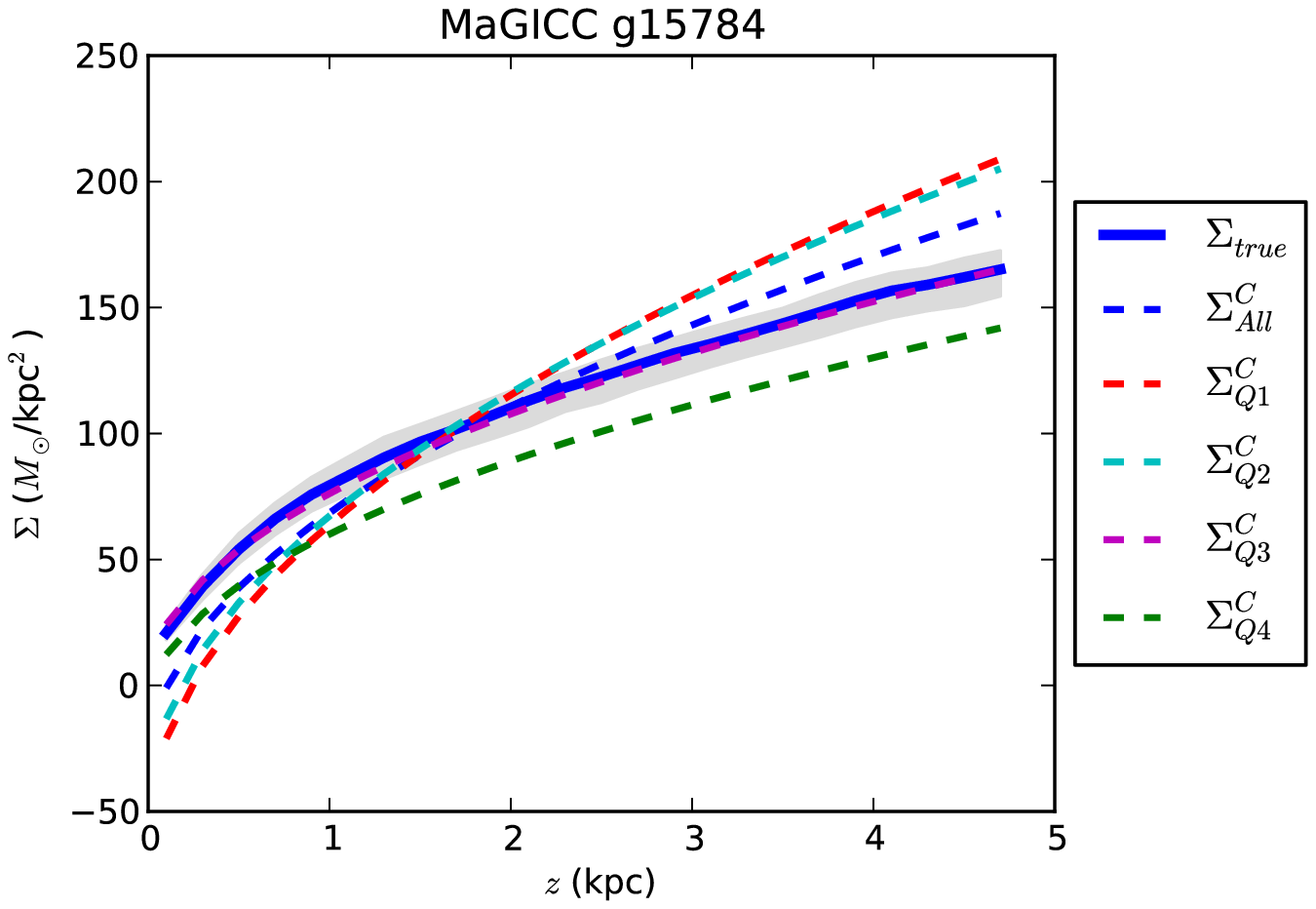} \\
\includegraphics[width=7.0cm]{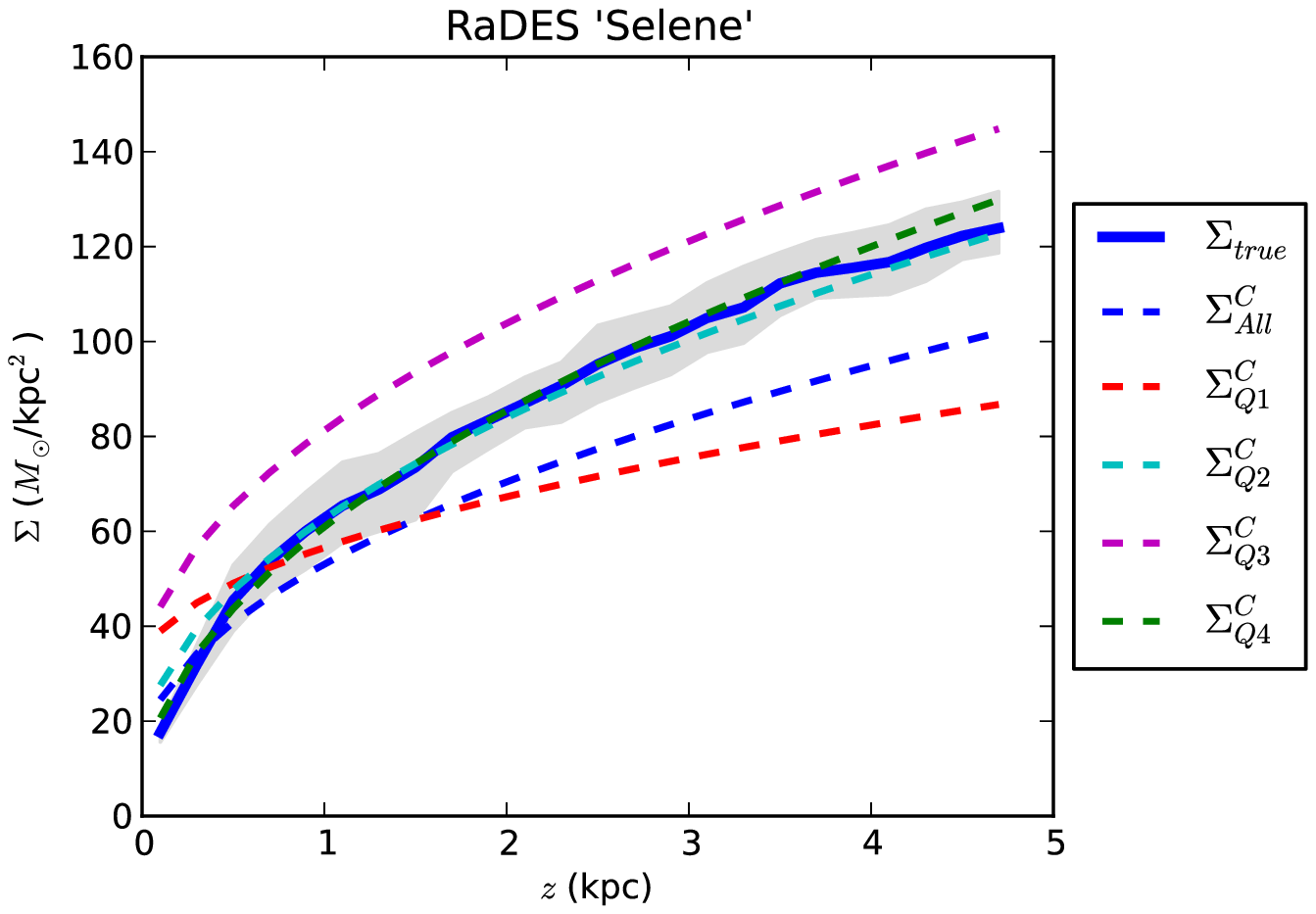} & \includegraphics[width=7.0cm]{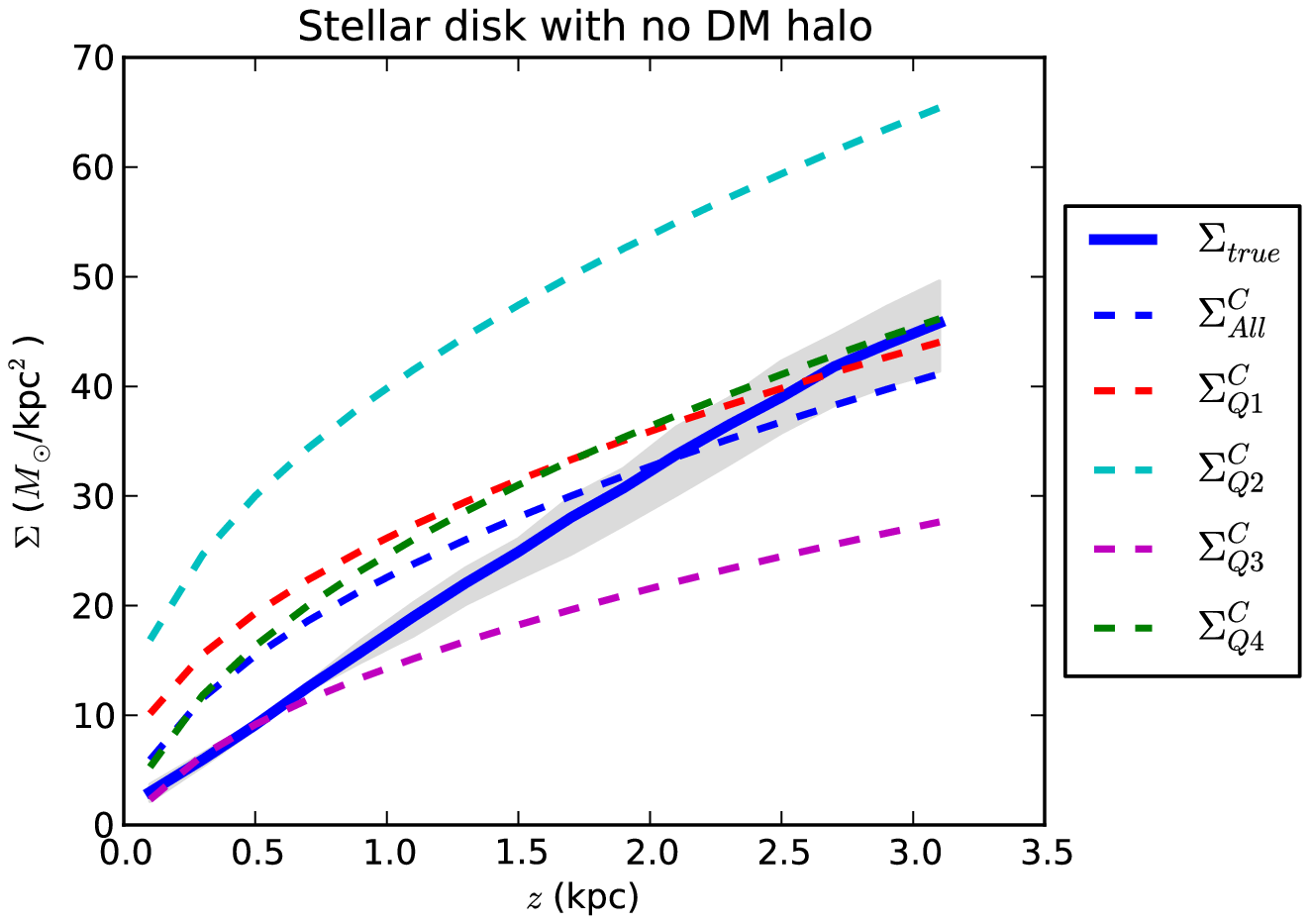} \\
\includegraphics[width=7.0cm]{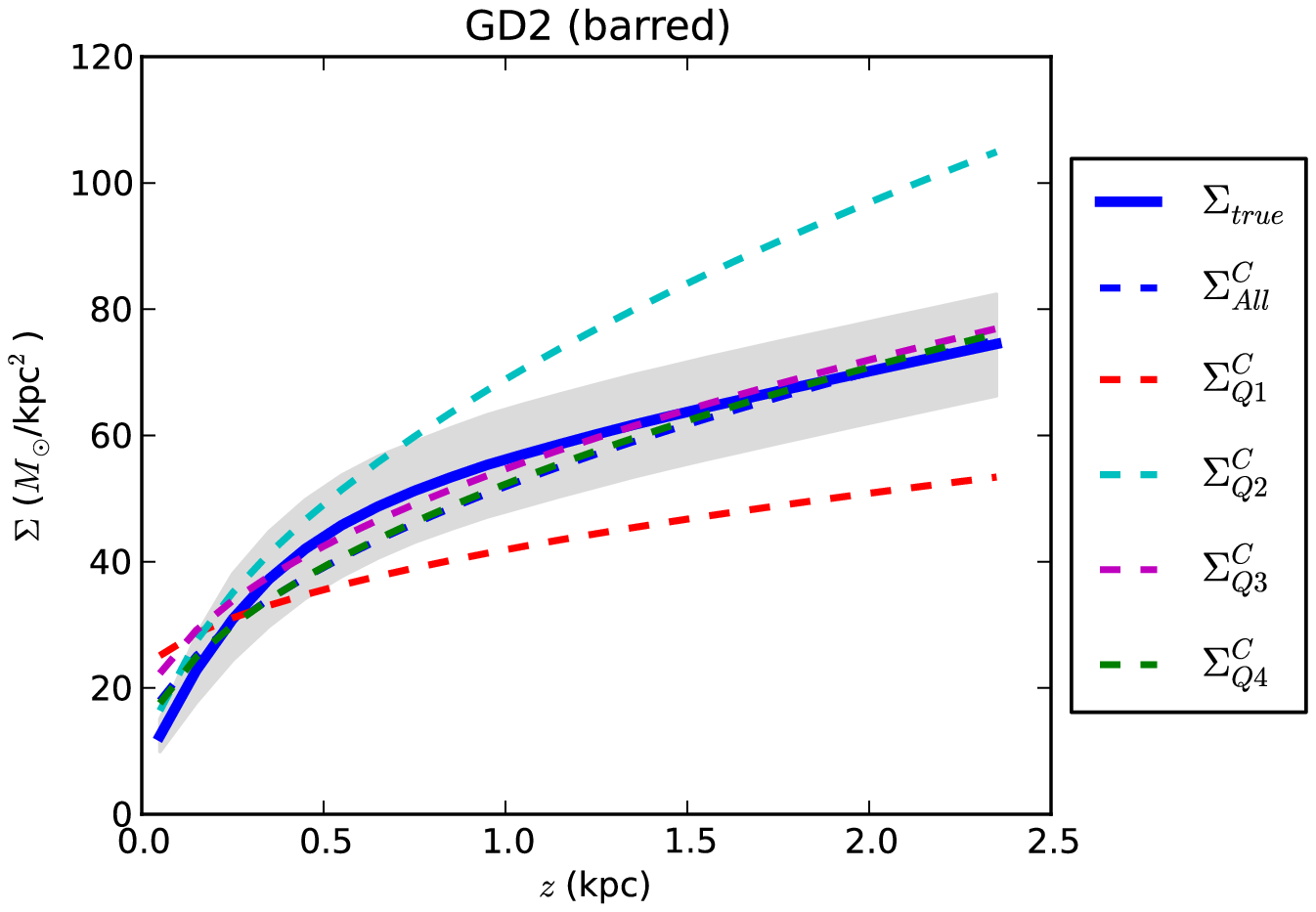} & \includegraphics[width=7.0cm]{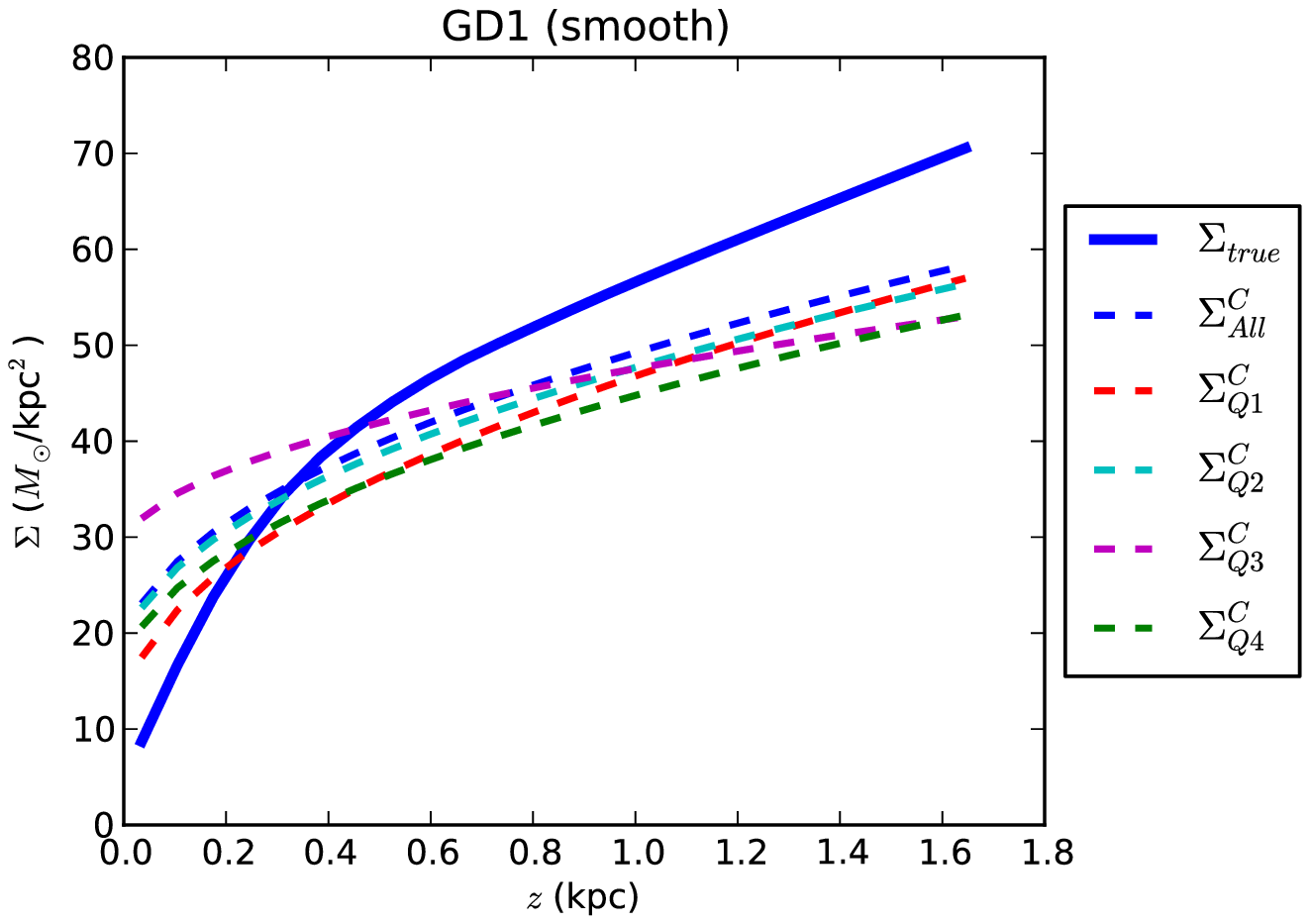}
\end{tabular}
\caption{Azimuthal dependence of the calculated and true surface densities, determined by considering each quadrant of the disk separately. The variation in the true surface density is illustrated by the grey band around the solid blue line (the surface density measured using the whole disk). The blue dashed line shows the calculated surface density using the whole disk, as in Fig.~\ref{surfdensresults}. The other dashed lines show the results of the analysis in each quadrant. All calculated surface densities use the three-dimensional formulation.}
\label{azimuthaldep}
\end{figure*}

To better compare our results with observations, we will now briefly consider how our calculated surface density (i.e. the fit line using Eq.~\ref{fittosurf}) compares with the true surface density using data that are restricted to lie within specific quadrants of the disk. Therefore, we create four subsets of the full data by selecting only the stars that lie in the following intervals of the azimuthal coordinate $\phi$: $[-\pi,-(1/2)\pi]$, $[-(1/2)\pi,0]$, $[0,(1/2)\pi]$ and $[(1/2)\pi,\pi]$. We then follow through with our analysis as described above, using each subset, until we derive a surface density curve using the three-dimensional formulation. Note that, when determining the various function fits necessary for calculating derivatives, we exclude all vertical bins where the particle number is less than approximately $10$ (the exact value varies between models due to differing particle numbers). This is to limit the impact of low number statistics on the fit parameters. The results for each quadrant in each model are shown in Fig.~\ref{azimuthaldep}. The variation of the measured (true) surface density is shown in this plot as a grey band around the surface density measured using all stars.
It is clear that the true surface density changes little as one moves around the disk, whereas the calculated surface density shows wider variation. The MUGS g15784 and MaGICC galaxies show reasonable agreement in all four quadrants. The GD1 model also shows good agreement in all four quadrants, albeit with a systematic underestimate of the surface density. The remaining models all show larger spreads. For the RaDES 'Selene' model and the disk without dark matter this may be partly related to noise arising from low number statistics, as the total number of particles used in the analysis (after restriction to the ``Solar'' radius) in each quadrant is of the order of $700$, whereas all other models (except MaGICC g1536) use at least $1200$ particles. It is also worth noting that the RaDES model disk exhibits some weak spiral structure. The dark matter-free disk presents particular problems for our analysis due to the assumption of an exponential vertical distribution being a poor fit to the data. This introduces errors in the estimation of $h_z$ which has a large impact on the final calculated surface density. In the case of the GD2 model the presence of a bar and weak spiral arms is likely to be the source of the azimuthal variation. Finally, the problematic MUGS g1536 model may exhibit azimuthal variation due to the increased substructure in the halo, as discussed earlier. It should be noted, however, that in most models the calculated surface density lines all track the \emph{trend} of the true surface density reasonably well, with errors in an overall offset. Thus the volume density inferred by integrating these lines between two heights will be similar in all cases. This is clearly less true for the MUGS g1536 and the GD2 model, which show more variation in the slopes of the lines.

\begin{figure}
\includegraphics[width=7.0cm]{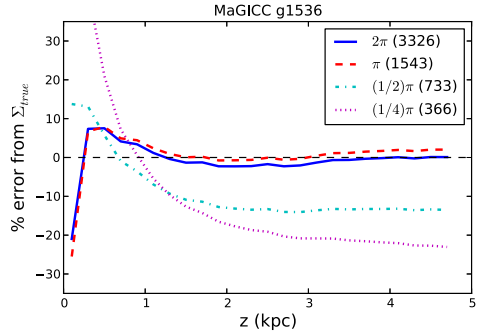}
\caption{The percentage error in the calculated surface densities for MaGICC g1536 as a function of height when compared with the true surface density (determined by azimuthally averaging around the whole disk). Each calculated surface density line is determined from a wedge region of the disk, with the size of the wedge and the associated number of star particles used in the analysis indicated in the legend.}
\label{lownumbers}
\end{figure}

In Fig.~\ref{lownumbers} we perform a similar experiment as shown in Fig.~\ref{azimuthaldep}, but this time we consider smaller regions of the disk to examine the effect of using fewer particles in the analysis. We have chosen the MaGICC g1536 model, as the calculated surface density when averaged around the whole disk is a good match for the true surface density. The number of star particles used in the analysis of each wedge region is shown in the plot, as well as the associated range of the azimuthal coordinate showing the size of each region ($2\pi$ is the whole disk, $\pi$ is half the disk, and so on). It is clear that the error increases substantially as we reduce the number of star particles. Note that, in order to fit the slope of the vertical velocity dispersion and the vertical scale height, we have only used bins that contain at least $20$ particles, otherwise the fits are heavily affected by noise.

The results of Figs.~\ref{azimuthaldep} and \ref{lownumbers} illustrate the challenge of correctly inferring the surface density using this method with more realistic data. Observations tend to probe a far smaller wedge of the disk than we are considering, and often do so with fewer stars, introducing two important sources of noise.

Returning to the specific terms of Eq. \ref{jeansfinal}, we now examine terms (vii) and (v) in more detail.

\subsection{Mean azimuthal velocity}
\label{meanazivel}
Term (vii) involves the mean azimuthal velocity $\overline{V}$ in the disk. This term was the focus of BT12 where $\partial_R \overline{V}$, poorly constrained by observations, was inferred from the requirement that the radial term of Eq. \ref{poissonvertical} vanishes, at least close to the plane of the disk. It was argued that one can safely ignore the radial term in Eq. \ref{poissonvertical} as this will be a subdominant correction to the vertical term, and that any deviation from this behaviour is argued to be in the form of a \emph{positive} radial term, leading to no more than a $20\%$ \emph{underestimate} of the local dark matter density when using the vertical term alone. As discussed extensively in MB15, and in Section \ref{analytic}, this is not consistent with all mass models, although an exponential disk requires a very short scale length to provide a negative radial contribution (with either a negligible amount of dark matter or a highly unlikely halo distribution that implies a radially decreasing rotation curve). In our simulations most of the models do indeed have subdominant \emph{positive} contributions from the radial term of Eq. \ref{poissonvertical}, consistent with the results in Section \ref{analytic} for exponential disks embedded in dark matter halos.

In BT12 it was suggested that neglecting term (vii) in MB12 led directly to the low dark matter density estimate. In MB15 this is shown to be incorrect, as the required \emph{positive} slope in $z$ of $\partial_R \overline{V}$ term is far too steep to be accomodated by the observational data. We will revisit this in Section \ref{subsub:cons_azi}. The only models for which term (vii) is significant (but still subdominant) are the GD1 thin disk and the MaGICC g1536 disk. The GD1 thin disk model shows the largest contribution from term (vii), although this is reduced by the negative contribution of term (v) to give a radial term in Eq. \ref{poissonvertical} that is small and only slightly increasing with height, as shown in Fig. \ref{surfdensforces}. The MaGICC g1536 disk shows a similar contribution from both term (vii) and term (v).

In all of our models term (vii) is therefore of only minor importance for the calculation of the surface density.

\subsection{Radial velocity dispersion}
\label{subsec:radvel}
\begin{figure}
\includegraphics[width=7.0cm]{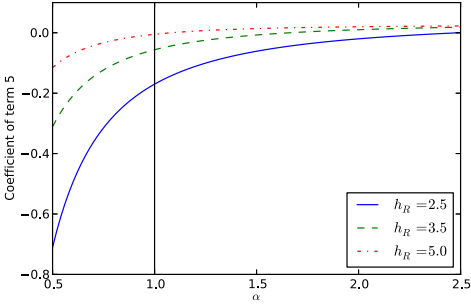}
\caption{Dependence of the coefficient of term (v) on varying values of $\alpha$, where $h_U = \alpha h_R$, for different disk scale lengths.}
\label{coeffsAlpha}
\end{figure}

We now turn to term (v). This term involves the radial velocity dispersion $\sigma_U$. From Fig.~\ref{surfdenscontributions} it is clear that term (v) is most important for the MUGS models and particularly the model without a DM halo. As stated earlier MaGICC g1536 also has an appreciable contribution from this term. Let us try to understand the importance of this term by looking at the coefficient of the integral shown in Section \ref{jeansfinal}:
\begin{equation}
\label{term5coeff}
C_5 = \frac{2}{Rh_U} + \frac{1}{Rh_R} - \frac{1}{h_U^2} - \frac{1}{h_Uh_R}.
\end{equation}
Setting $R = 8$~kpc and $h_U = \alpha h_R$, the function $C_5(\alpha)$ is plotted in Fig.~\ref{coeffsAlpha} for various choices of disk scale length $h_R$. In our simulations $1 \lesssim \alpha \lesssim 8$. We can see in Fig.~\ref{coeffsAlpha} that for larger scale lengths the dependence of the coefficient on $\alpha$ becomes weak for a wide range of $\alpha$ values, and therefore for a wide range of $h_U$ values. This illustrates that, for large $h_R$, a large uncertainty in the value of $\alpha$ (or equivalently $h_U$) does not significantly affect the contribution of term (v). This behaviour is clearly exhibited by the GD2 model, where $h_R = 8.2$ and $h_U = 26.6$. Setting $h_U = h_R$ in this model has almost no effect on the contribution from term (v), as seen in the coincidence of the $\Sigma^C$ and $\Sigma^C_{h_U}$ lines for this model in Fig.~\ref{surfdensresults}. For a smaller scale length disk, however, the dependence of the coefficient on $\alpha$ is stronger, especially for $\alpha \sim 1$ or less. It becomes possible to have a large negative contribution from term (v) with the imposition of $h_U = h_R$. This may be seen for the MUGS models in Fig.~\ref{surfdensresults}, where the $\Sigma^C_{h_U}$ line rapidly decreases with $z$ becoming large and negative, which is unphysical. These models have small $h_R$ but large $h_U$, and so a significant reduction of $h_U$ leads to a large change in the contribution of term (v). For the MaGICC models the difference between $h_R$ and $h_U$ is smaller, so the effect of imposing $h_U = h_R$ is less pronounced, but it still leads to a negative calculated surface density.

The values of the coefficient of term (v), $C_5$, are given in Table~\ref{fitparams}. For the MUGS and MaGICC models the contribution from term (v) is positive, and large in the case of the MUGS models. The RaDES `Selene' model has a slight negative contribution from term (v), as $h_R = 3.2$~kpc and $h_U = 5.1$~kpc, within the negative range of $C_5$ in Fig. \ref{coeffsAlpha}. The GD1 value of $C_5$ is much more negative than all other models, but this is cancelled by the positive contribution of term (vii) to give the almost flat radial force term of Fig. \ref{surfdensforces}. Therefore, we can characterise the discrepancies between the vertical-only and three-dimensional formulation results in Fig. \ref{surfdensresults} by considering the coefficient of term (v) alone, except for the GD1 model.

We now use Eq. \ref{term5coeff} to determine what scale length of disk leads to $C_5 = 0$, for a given value of $\alpha$. In the case of $\alpha = 1$, i.e. $h_U = h_R$, then the density scale length must be $h_R = 5.3$~kpc to ensure that term (v) does not give a negative contribution to the surface density (or the radial force term). For a larger radial velocity dispersion scale length, i.e. $\alpha = 2$, we find a minimum density scale length of $h_R = 3.0$ to ensure a non-negative term (v) contribution. A positive contribution from term (vii) will, of course, counteract any negative contribution from term (v), thus we do not strictly require $C_5 = 0$ to have a non-negative radial force term, and so lower values of $h_R$ would in that case be acceptable. For our Galaxy, the results of MB15 make it clear that the observations are not compatible with very large contributions from term (vii). Therefore, if the radial force term is to be non-negative (or only slightly negative), these estimates suggest that either the density scale length of the test population is longer than the $h_R = 3.8$~kpc choice of MB12 (and certainly far larger than the value $h_R = 2.0$~kpc of BT12) or $h_U$ is significantly larger than $h_R$. It is important to remember, however, that the radial force term in the Milky Way may well be more negative than exhibited by our models.

These results also suggest an approximate anticorrelation between $h_R$ and $h_U$ whereby a shorter disk scale length would imply a larger minimum value for the velocity dispersion scale lengths. For our models this is approximately true, with lower values of $h_R$ generally associated to larger values of $h_U$.

It is therefore necessary to measure these scale lengths as accurately as possible, especially given that the Milky Way density scale length appears to be in the region where a significant deviation in the value of $h_U$ leads to a large negative contribution from term (v) in the Jeans equation. Given the sensitivity of $\Sigma_{\text{calc}}$ on this parameter, it is certainly possible that inaccurate choices for the scale lengths have contributed to the result of MB12, where a surprisingly low dark matter density was reported.

\subsection{Revisiting MB12}
\label{revisitMB12}
The lessons from the simulations are that the most crucial extra terms in the three-dimensional formulation are those involving the radial velocity dispersion and the azimuthal velocity, although both are still subdominant. Furthermore, all models have $\partial_R \overline{UW} \approx 0$. We will now investigate what these lessons may imply for the analysis of MB12.

\subsubsection{Consequences for the mean azimuthal velocity}
\label{subsub:cons_azi}
To begin with, $\overline{UW}$ was assumed in MB12 to have a radial exponential dependence as for the squared velocity dispersions. Our simulation data do not respect this assumption, with the radial dependence of $\overline{UW}$ being remarkably flat in all models, at least at the solar radius. The analysis in MB12 also incorrectly neglected terms (ii) and (iv). When we include these extra terms using the parameter set of MB12 we find the surface density quickly becomes negative for increasing height. We can, however, restore the surface density to the result of MB12 by imposing $\partial_R \overline{UW} = 0$, a condition that is respected by all of our models. One further consequence of this condition is to imply that $\partial_R \overline{V}$ has only a weak dependence on $z$. To see this, we follow the logic employed by BT12 in arguing for the behaviour of term (vii) by appealing to the flatness of the Milky Way's circular velocity curve (ignoring the minor deviation from flatness at higher $z$), $\partial_R V_c = 0$. The expression for the asymmetric drift may be derived from the radial Jeans equation (see BT12) as:
\begin{equation}
\label{asymdrift}
V_c^2 - \overline{V}^2 = \sigma_V^2 + \sigma_U^2 \left[ R \left( \frac{1}{h_R} + \frac{1}{h_U} \right) - 1 \right] + \frac{R}{h_z} \overline{UW} - R \frac{\partial \overline{UW}}{\partial z}
\end{equation}
One may now differentiate this expression with respect to $R$ to find
\begin{equation}
\label{v2bargrad}
\begin{split}
-\overline{V}\frac{\partial \overline{V}}{\partial R} &= -\frac{1}{h_V}\sigma_V^2 - \frac{1}{h_U}\sigma_U^2 \left[ R \left(\frac{1}{h_R} + \frac{1}{h_U} \right) -1 \right] \\
&+ \sigma_U^2 \left(\frac{1}{h_R} + \frac{1}{h_U}\right) + \frac{1}{h_z}\overline{UW} + \frac{R}{h_z} \frac{\partial \overline{UW}}{\partial R} \\
&- \frac{\partial \overline{UW}}{\partial z} - R \frac{\partial^2 \overline{UW}}{\partial R \partial z}.
\end{split}
\end{equation}
where we have used the exponential dependence on $R$ of the squared velocity dispersions, as assumed throughout this study, and $\partial_R V_c = 0$. In BT12, using Eq. \ref{asymdrift} and Eq. \ref{v2bargrad} with the MB12 data, the vertical gradient of $\partial_R \overline{V}$ that is consistent with a flat circular velocity curve was found to be positive and significant: from $7$~km s$^{-1}$ kpc$^{-1}$ at $z = 0$~kpc up to $40$~km s$^{-1}$ kpc$^{-1}$ at $z = 3.5$~kpc. This was then shown in MB15 to be inconsistent with the few observations available for the Milky Way, and for external galaxies.

Let us now impose the alternative condition of $\partial_R \overline{UW} = 0$ in Eq. \ref{v2bargrad}. When we now include the MB12 data values we find that $\partial_R \overline{V}$ is much reduced, at all heights. Specifically, we find that it is equal to $5.4$~km s$^{-1}$  kpc$^{-1}$ at $z = 0$~kpc, and increases to just $14.5$~km s$^{-1}$  kpc$^{-1}$ at $z = 3.5$~kpc. Imposing a longer radial velocity dispersion, i.e. $h_U = 2h_R$, reduces further the values of $\partial_R \overline{V}$ and the vertical gradient: $3.2$~km s$^{-1}$  kpc$^{-1}$ at $z=0$ and $3.5$~km s$^{-1}$  kpc$^{-1}$ at $z = 3.5$~kpc. As stated in MB15 it is therefore only an approximation to set $\partial_R \overline{V} = 0$, but we find that it is a reasonable approximation given a very weak radial dependence in $\overline{UW}$, and we have also shown that it is roughly consistent with a flat circular velocity curve.

\subsubsection{Consequences for the surface density}
\label{mb12:surfdens}
\begin{figure}
\includegraphics[width=7.0cm]{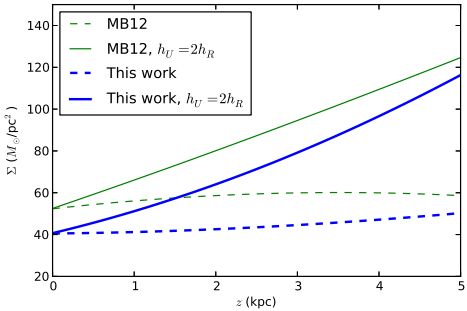}
\caption{The surface density calculated using the observational data and parameter choices of MB12, with the Jeans equation formulation of that work (green dashed line) and the current study (blue dashed line). These curves omit terms (ii) and (iv), and are without errors. The solid lines use $h_U = 2h_R$ in the formulation of MB12 (green solid line) and of the current study (blue solid line). The blue lines use $\partial_R \overline{UW} = 0$, as is approximately true in the simulation data.}
\label{mb12_curves}
\end{figure}

For the purposes of comparison we plot $\Sigma_{\text{calc}}(z)$ in Fig.~\ref{mb12_curves} (as the dashed green line) using the data of MB12 and the Jeans equation of that work. Therefore we neglect terms (ii), (iv) and (vii), and we impose an exponential dependence on $R$ in $\overline{UW}$ with scale length equal to the density, so that term (iii) becomes
\begin{equation}
\text{term 3} = \frac{2}{h_R}\overline{UW} \Big|_0^z = \frac{2}{h_R} (\gamma_{UW} z + \delta_{UW}) \Big|^z_0,
\end{equation}
with $\gamma_{UW}$ and $\delta_{UW}$ the slope and intercept of the linear fit line to $\overline{UW}(z)$. Note that the fit line used in MB12 for the $\overline{UW}$ data includes a non-zero intercept\footnote{The condition $F_z = 0$ in the plane of the disk implies that $\overline{UW}(z)$ must be an antisymmetric function of $z$, which is violated by the presence of a non-zero intercept term in the linear fit to the data. Presumably sufficiently accurate data would be consistent with an approximately zero intercept.} $\delta_{UW}$. The central data values of MB12 are used and we ignore errors.

Now we employ the Jeans equation terms given in Section \ref{listofterms}, with the exception of term (vii). The inclusion of terms (ii) and (iv), neglected in MB12, along with the imposition of $\overline{UW}(R) \sim \exp(-R/h_R)$ leads to a \emph{negative} surface density using the data of MB12, as stated earlier. Therefore we proceed as in Section \ref{subsub:cons_azi}, taking guidance from the simulation data, by setting $\partial_R \overline{UW} = 0$. While this is not precisely true for our simulations, we have seen in Section~\ref{results} that it is a good approximation. Thus, of the terms that depend on $\overline{UW}$, terms (iii) and (iv) do not contribute to the final result, leaving only term (ii):
\begin{equation}
\frac{1}{Rh_z} \int_0^z \overline{UW} = \frac{1}{Rh_z} \left( \frac{1}{2}\gamma_{UW}z^2 + \delta_{UW}z \right).
\end{equation}
The resulting $\Sigma_{\text{calc}}(z)$ is plotted in Fig.~\ref{mb12_curves} as the dashed blue line. The loss of term (iii) from the Jeans equation means that we no longer have the contribution from $\delta_{UW}$ at $z=0$, causing the $\Sigma_{\text{calc}}$ line to drop by approximately $10 M_{\odot}$ pc$^{-2}$. Clearly the surface density still barely increases with height, apparently indicating the presence of very little dark matter.

If we now relax the assumption that $h_U = h_R$, and instead choose $h_U = 2h_R$ (i.e. $\alpha = 2$ in the context of Section~\ref{subsec:radvel}, a value that is still considerably lower than that measured in all of the simulations except RaDES and the idealised GD1 model) we find the solid green and blue lines in Fig.~\ref{mb12_curves}, where the solid green line corresponds to the original MB12 treatment of the $\partial_R \overline{UW}$ term, while the solid blue line corresponds to our treatment of this term. As can be seen, the surface density is now an increasing function of $z$, with no flattening.

\begin{figure}
\includegraphics[width=7.0cm]{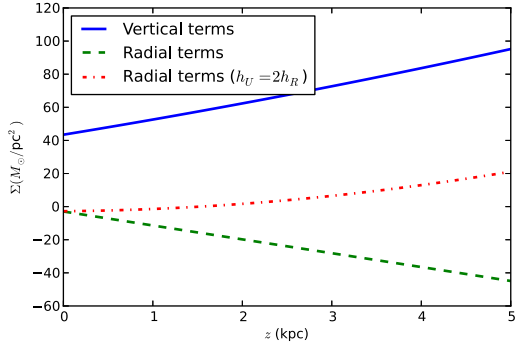}
\caption{Contributions to the calculated surface density from the vertical and radial force terms using the MB12 data, with the formulation of the Jeans equation used in this study and $\partial_R \overline{UW} = 0$. The radial force contribution is made positive with a rescaling of the $h_U$ parameter.}
\label{mb12forces}
\end{figure}

This increase of the surface density is, of course, entirely due to a change in the contribution from the radial force term of Eq. \ref{poissonvertical}, from negative to slightly positive, as can be seen in Fig. \ref{mb12forces}. This modified radial force curve is also more in line with the results of Section \ref{analytic} which demonstrated that even a dark matter-free exponential disk with a scale length of $h_R = 3.8$~kpc should have a positive radial force contribution to the surface density.

Let us now obtain the local dark matter volume density from these $\Sigma$ curves by calculating
\begin{equation}
\rho_{\text{DM}} = (\Sigma(z=4 \, \text{kpc}) - \Sigma(z=1.5 \, \text{kpc}))/(2 \times 2.5 \, \text{kpc})
\end{equation}
where we assume that the stellar surface density is constant at $z=1.5$~kpc and above. The additional factor of $2$ in the denominator arises due to the symmetric integral interval. For the original data in the original formulation of MB12 we obtain $\rho_{\text{DM}} = 4.7 \times 10^{-4} M_{\odot}$ pc$^{-3}$, ignoring errors. Setting $h_U = 2h_R$ in the original MB12 formulation we find instead that $\rho_{\text{DM}} = 7.3 \times 10^{-3} M_{\odot}$ pc$^{-3}$, fifteen times larger than the original estimate. Using our new formulation, with $h_U = h_R$, we find $\rho_{\text{DM}} = 1.1 \times 10^{-3} M_{\odot}$ pc$^{-3}$, a low estimate, as would be expected from the flat dashed blue line in Fig.~\ref{mb12_curves}, although this is over twice the estimate of the original formulation of MB12. Setting $h_U = 2h_R$ in our new formulation we find $\rho_{\text{DM}} = 7.9 \times 10^{-3} M_{\odot}$ pc$^{-3}$, seven times larger than the estimate with $h_U = h_R$. All of this strongly suggests that changes to the value of $h_U$ used in this method may well result in an estimate for the local dark matter volume density that, while somewhat lower than other estimates, is approximately consistent with the standard halo model value of $\rho_{\text{DM}} = 8.0 \times 10^{-3} M_{\odot}$ pc$^{-3}$ (\citealp{jungman}).

For a lower value of $h_R$ the estimated density decreases, possibly becoming negative and unphysical. For example, using $h_R = 2$~kpc and $h_U = 3.5$~kpc, as suggested by BT12, we find a negative density using either the formulation of MB12 or our new formulation. This is to be expected given the dependence of the contribution of term (v) on the scale lengths discussed in Section~\ref{subsec:radvel}. By scaling $h_U$ by a factor of $2$ the estimated dark matter density becomes positive: $\rho_{\text{DM}} \approx 8 \times 10^{-3} M_{\odot}$ pc$^{-3}$. While this scaling is arbitrary, it nonetheless suggests that, if this method is to reproduce other estimates of the local dark matter volume density, we would perhaps expect larger measured values of $h_U$ than used in either MB12 or BT12 (and a larger $h_R$ than that used in BT12), possibly values more consistent with those of our simulation data.

Furthermore, the value of $\alpha$ that is required to reproduce typical estimates of $\rho_{\text{DM}}$ would be reduced by the inclusion of the mean azimuthal velocity term, as the positive contribution from this term, discussed in Section \ref{meanazivel}, further increases the surface density.

\subsection{Comparisons with the Milky Way}
A central conclusion of our work, discussed in Section \ref{subsec:radvel} is that the scale lengths of the velocity dispersions, and particularly that of the radial velocity dispersion, must be very carefully measured if the full three-dimensional approach is to give a sensible result. We have demonstrated that all of our simulations show $h_U > h_R$ (and also $h_V > h_R$) and that simply assuming $h_U = h_R$ can lead to a catastrophic failure of the method and a completely unphysical result when calculating the surface mass density, especially for disks with a small scale length. Using the data of MB12, we have demonstrated (in Section \ref{revisitMB12}) that tension between the result of MB12 and other studies can be removed if the assumed value of $h_U$ in the analysis is scaled up by a factor entirely consistent with those measured in these simulations. Alternatively, if the condition $h_R = h_U$ is to be imposed, then the density scale length $h_R$ must be larger, possibly on the order of $h_R = 5$~kpc. Moreover, such a scaling would also modify the results of MB15: if we wish to impose a negligible radial force term, then a less negative contribution from the radial velocity dispersion implies a less positive gradient in $z$ of $\partial_R \overline{V}$, bringing this quantity in line with observations.

Our scaling of $h_U$ is, however, completely arbitrary as it stands. Without some observational constraint we cannot say that this definitely resolves the puzzle presented by the results of MB12. High-quality observational data for the scale lengths $h_R, h_U$ and, to a lesser extent, $h_V$, are therefore crucial for the robustness of this method. Observational studies have so far been unable to tightly constrain these scale lengths, although \cite{lewisfreeman} have determined the radial velocity dispersions to have an exponential profile in $R$, justifying assumptions 2 and 3 discussed earlier, and also reflected in our simulation data.

In Fig. 3 of MB12 several estimates for the vertical and radial disk scale lengths are shown. The spread in $h_R$ is large, with values from $1.5$~kpc to $6$~kpc being within the errors. As we have seen, values at the upper end of this range would lead to a significant increase in the local dark matter density volume measured from the MB12 dataset, even with $h_U = h_R$. In \cite{bovy2013} the favoured value is $h_R = 2.15 \pm 0.14$~kpc, considerably shorter than the $h_R = 3.8$~kpc used by MB12. If this value is accurate then our analysis would strongly suggest that the scale length of the radial velocity dispersion is likely to be much larger than the density, in order to avoid an unphysical surface density when the full three-dimensional Poisson equation is applied, using the Jeans equation to determine the forces.

Of course all of our conclusions must be counter-balanced with the caveat that none of the models that we have analysed are intended to reproduce our Galaxy. Therefore, there may well be significant differences between the properties of these models, particularly in the behaviour of the velocity dispersions, and the Milky Way. In particular, $\overline{UW}$ is proportional to the tilt angle (the inclination of the dispersion ellipsoids with respect to the Galactic axes). Measurements of the tilt angle indicate that the ellipsoid axis roughly points to the Galactic center at any height, i.e. the tilt increases with z. One would expect that this is roughly the same in the radial direction, the tilt slowly decreasing with $R$, and therefore $\overline{UW}$ decreasing also. The $\partial_R \overline{UW} \approx 0$ behaviour of our models is therefore rather unexpected for the Milky Way. Furthermore, the scant observations of the scale lengths that are available suggest that $h_U$ and $h_V$ are roughly comparable with $h_R$. If these statements are confirmed with further observations, then it appears very difficult to reconcile the results of MB12 with those of other studies.

\subsection{What of the other terms in the Jeans equation?}
If instead we assume that all data and parameters, including the scale lengths, of MB12 are correct, what does our analysis imply? Is it possible that term (v) is not the culprit, and we must look elsewhere?

Let us consider the ``tilt'' term $\overline{UW}$. All of our models have $\partial_R \overline{UW} \approx 0$ and a negligible contribution from the $\overline{UW}(z)$ terms in the Jeans equation. The observational situation with regards to the radial gradient of $\overline{UW}$ in our Galaxy is, unfortunately, rather uncertain. Again, accurate measurements of this quantity will allow us to conclusively determine whether this term can indeed be safely neglected in a Jeans equation analysis. For the MB12 dataset, a linear slope of $-65$~km$^2$ s$^{-2}$ kpc$^{-1}$ (assumed constant with height) would lead to a negative contribution from the sum\footnote{Term (iii) is a negative contribution to the surface density while term (iv) is positive, so they must be counted together.} of terms (iii) and (iv) of no more than $10\%$ of term (viii), the dominant vertical velocity dispersion term. For comparison, the MUGS g1536 model has the largest slope of $-53.4$~km$^2$ s$^{-2}$ kpc$^{-1}$. Note that the assumption of MB12 that the $\overline{UW}$ term has an exponential dependence on radius with the same scale length as the density corresponds to a slope in $R$ (at the disk plane) of $-159.7$~km$^2$ s$^{-2}$ kpc$^{-1}$, explaining why the surface density becomes negative if we use our formulation of the Jeans equation, including terms (ii) and (iv), with all the assumptions of MB12. 

It therefore seems that we would require a steep \emph{positive} radial gradient in $\overline{UW}$ in order to substantially increase the radial force contribution to the surface density with this term, something not exhibited by any of our models or the MB12 observational dataset. It is nonetheless conceivable that different galactic evolutionary histories will lead to different behaviour for the $\overline{UW}$ ``tilt'' term, and its radial gradient. Until more accurate data are available, however, the observations utilised in MB12 support the assertion that $\overline{UW}$ is not a significant contribution to the surface density, as do all of our models, provided the radial gradient is much less steep than for the density or velocity dispersions. The results in Section \ref{subsub:cons_azi} also suggest that the $\partial_R \overline{UW} = 0$ condition is consistent with both the observed low values of $\partial_R \overline{V}$ and a flat circular velocity curve.

Finally, we have had rather little to say about term (vi), the azimuthal velocity dispersion. This term gives a negative contribution to the surface density, that becomes increasingly significant at higher heights. In all simulations $h_V > h_R$, very roughly tracking the behaviour of $h_U$. For the MB12 data term (vi) is a negative contribution of approximately $25\%$ of the vertical velocity term at $z=5$~kpc. Increasing the radial scale length to $h_V = 2h_R$, in a similar manner to that discussed for the radial velocity dispersion, we find a negative contribution of $14\%$ at $z=5$~kpc. For comparison, setting $h_U = 2h_R$ changes term (v) from a negative contribution of $50\%$ of the vertical velocity term at $z=5$~kpc, to a positive contribution of $20\%$ at the same height. Therefore, while $h_V$ must also be observationally constrained for an accurate application of the three-dimensional method, uncertainties in this parameter do not lead to such drastic differences as in the case of the radial velocity dispersion.

\section{Conclusions}
\label{conclusions}
In this work, we have applied a fully three-dimensional method of reconstructing the disk surface density to data from simulated disk galaxies, formed in high resolution cosmological simulations that included state-of-the-art physics recipes for the various astrophysical feedback processes and star formation. These galaxies are known to satisfy many observed scaling relations and thus provide a realistic dataset for the purpose of attempting to reconstruct the surface mass density of the disk in the region of $R=8$~kpc. While we do not claim that these galaxies are simulated versions of our Milky Way, the fact that they have formed through cosmological processes, undergoing various secular and environmental effects throughout their evolution, provides a somewhat more realistic testing ground for the Jeans equation method, using dark matter dominated galaxies. In addition we have studied idealised thin disk galaxies embedded in analytic dark matter haloes, and a dark matter-free model. We then revisited the results of MB12, applying the lessons learned from the simulations to determine possible consequences for the Jeans equation analysis applied to our Galaxy.

As shown in Sections \ref{relimport} and \ref{comparison1d3d}, applying the full three-dimensional analysis to our simulated data, we find that the radial force term is usually subdominant at lower heights, and that the traditional vertical-only formulation is a good approximation, especially for vertical heights of order the thick disk scale height and below. Whenever there is more significant disagreement between the vertical-only and the full three-dimensional formulations, however, (such as for the MUGS 1536 model or the dark matter-free model) the main term responsible for this disagreement is the radial velocity dispersion, while the mean azimuthal velocity term is only a minor contribution in all models (see Sections \ref{meanazivel} and \ref{subsec:radvel}). At higher heights, however, the three-dimensional method is more accurate than the one-dimensional approximation. In Section \ref{azimuth} we have shown that restricting the analysis to a smaller subset of the data (as done in observational studies) by considering each quadrant of the disk separately leads to slightly more variation in the calculated surface densities than is present in the true surface density for some models, with more significant spread for those models with lower numbers of particles and/or structure in the disk. Thus low number statistics and disk structure will contribute further sources of error for this kind of analysis.

The quantity $\overline{UW}$, poorly constrained observationally, is a subdominant term in all models, and $\partial_R \overline{UW} \approx 0$ in all models, with no evidence for a strong dependence on radius. The use of the condition $\partial_R \overline{UW} = 0$, and imposing a flat circular velocity curve, implies small values of $\partial_R \overline{V}$ and a weak dependence on $z$, consistent with observations and the discussion of MB15, as shown in Section \ref{subsub:cons_azi}. This further demonstrates that this term alone cannot account for the ``missing dark matter'' result of MB12.

Taking guidance from the results of the simulations, we have therefore excluded the $\overline{UW}$ and $\partial_R \overline{V}$ terms as being capable of restoring the local dark matter volume density to results found elsewhere in the literature. Assuming that the kinematical dataset of MB12 is accurate, the results from the simulations lead us to the conclusion that the only mechanism for restoration of $\rho_{\text{DM}}$ to values seen in other studies is by increasing the radial scale length of $\sigma_U$, and possibly $\sigma_V$ and $\rho$ (see Section \ref{mb12:surfdens}). The increase of $h_U$ (and $h_V$) would also lead to a flat (or slightly positive) radial force term (as a function of $z$) in Eq. \ref{poissonvertical}, bringing the kinematical dataset of MB12 into line with the expectations (even in the absence of dark matter) for an exponential disk of scale length $h_R = 3.8$~kpc, as discussed in Section \ref{analytic}. Furthermore, if the scale length of the density is as low as $h_R = 2.15$, as claimed in \cite{bovy2013}, our analysis suggests we require an even larger value for the scale lengths of the velocity dispersions to reconcile $\partial_R V_c = 0$ with the kinematic data of MB12.

It must, however, be kept in mind that the entire analysis rests on the assumption that the test stellar population is indeed in equilibrium with the Galactic potential. Although we do not investigate this issue further in this study, it is worth noting that any kinematic heating of the stars will disrupt this equilibrium, leading to errors when using this method.

The use of the three-dimensional formulation for our Galaxy requires very accurate measurements of all parameters. Simplifying assumptions, such as $h_R = h_U$, are likely to lead to considerable error in estimates of the local dark matter volume density. Until accurate measurements are available, however, our study strongly suggests that the vertical-only formulation is as effective as the three-dimensional formulation up to the thick disk scale height. Using data at higher heights, where the gravitational potential of the dark matter halo begins to dominate, requires the inclusion of the radial force for an accurate result, and thus is subject to more sources of error. All approaches, however, require accurate measurments of all scale lengths, and accurate kinematics of at least several hundred stars at a range of heights in order to minimise errors, especially in the measurements of the vertical velocity dispersions. With more precise observations in the future, however, the use of this kind of three dimensional analysis will be helpful in constraining the local dark matter density and the kinematical properties of the Galactic disk.

\section*{Acknowledgments}
The generous allocation of resources from STFC's DiRAC Facility (COSMOS: Galactic Archaeology), the DEISA consortium, co-funded through EU FP6 project RI-031513 and the FP7 project RI-222919 (through the DEISA Extreme Computing Initiative), and the PRACE-2IP Project (FP7 RI-283493), are gratefully acknowledged. The support of our colleagues, Gareth Few, Chris Brook, and Greg Stinson, is likewise acknowledged. GNC acknowledges the support of FONDECYT grant 3130480, and partial support from the Center for Astrophysics and Associated Technologies CATA (PFB 06). CMB acknowledges the support of FONDECYT Regular project 1150060.


\end{document}